\documentclass[a4paper,11pt]{article}
\pdfoutput=1 

\usepackage{jheppub} 

\usepackage{graphicx}
\usepackage[figurename=Fig.]{caption}
\usepackage[tablename=Tab.]{caption}
\usepackage{color}
\usepackage{float}
\usepackage{amssymb}
\usepackage{amsmath}
\usepackage{textcomp}
\usepackage{cases}
\usepackage{makecell}
\usepackage{pdflscape}
\usepackage{subfigure}
\usepackage{pdfpages}
\usepackage{makeidx}
\usepackage{bm}
\usepackage{xstring}
\usepackage{lipsum}
\usepackage{slashed}
\usepackage{multicol}
\usepackage{setspace}
\usepackage{booktabs}
\usepackage[table]{xcolor}
\usepackage{multirow}
\usepackage{braket}
\usepackage{esvect}[e]
\usepackage[separate-uncertainty]{siunitx}
\usepackage[displaymath, mathlines]{lineno}

\newcommand{\eref}[1]{Eq.\,(\ref{#1})}
\newcommand{\sn}[2]{#1\times10^{#2}}
\newcommand{\w}{\omega}
\def\up{\mathrm}
\def\d{\mathrm{d}}

\def\i{\mathrm{i}}

\def\gms4{g_{\text{m}\sigma4}}
\def\gmp4{g_{\text{m}\pi 4}}
\def\gmv41{g_{\text{mv}41}}
\def\gmv43{g_{\text{mv}43}}
\def\gbs1{g_{\text{b}\sigma1}}
\def\gbp1{g_{\text{b}\pi1}}
\def\gbv1{g_{\text{bv}1}}

\renewcommand{\thetable}{\Roman{table}}
\newcommand{\vabs}[1]{|\vec #1\,|}
\renewcommand{\arraystretch}{1.2}

\newcommand{\comment}[1]{}

\graphicspath{{PDF/}}

\def\slash#1{\setbox0=\hbox{$#1$}           
   \dimen0=\wd0                                 
   \setbox1=\hbox{/} \dimen1=\wd1               
   \ifdim\dimen0>\dimen1                        
      \rlap{\hbox to \dimen0{\hfil/\hfil}}      
      #1                                        
   \else                                        
      \rlap{\hbox to \dimen1{\hfil$#1$\hfil}}   
      /                                         
   \fi}                                         %
\def\sl#1{\setbox0=\hbox{#1}
  \dimen0=\wd0
  \rlap{\hbox to \dimen0{\hss/\hss}}%
  #1}

\title{\boldmath Strong decays of $P_{\psi}^N(4440)^+$ and $P_\psi^N(4457)^+$ within the Bethe-Salpeter framework}

\author[1]{Qiang Li,}
\author[2,3]{Chao-Hsi Chang,}
\author[1]{Xin Tong,}
\author[4,5]{Xiao-Ze Tan,}
\author[6]{Tianhong Wang}
\author[7]{and Guo-Li Wang}

\affiliation[1]{School of Physical Science and Technology, Northwestern Polytechnical University, Xi'an 710072, China}
\affiliation[2]{CAS Key Laboratory of Theoretical Physics, Institute of Theoretical Physics, Chinese Academy of Sciences, Beijing 100190, China}
\affiliation[3]{School of Physical Sciences, University of Chinese Academy of Sciences, Beijing 100049, China}
\affiliation[4]{Deutsches Elektronen-Synchrotron DESY, Notkestr. 85, 22607 Hamburg, Germany}
\affiliation[5]{Department of Physics and Center for Field Theory and Particle Physics, Fudan University, Shanghai 200438, China}
\affiliation[6]{School of Physics, Harbin Institute of Technology, Harbin 150001, China}
\affiliation[7]{Department of Physics, Hebei University, Baoding 071002, China}

\emailAdd{liruo@nwpu.edu.cn}
\emailAdd{zhangzx@itp.ac.cn}
\emailAdd{tx2019301663@mail.nwpu.edu.cn}
\emailAdd{xiaoze.tan@desy.de}
\emailAdd{thwang@hit.edu.cn}
\emailAdd{wgl@hbu.edu.cn}

\abstract{
By combining the effective Lagrangian and Bethe-Salpeter framework,  we studied the mass spectra, wave functions, and strong decay widths of the two pentaquark states $P_\psi^N(4440)^+$ and $P_\psi^N(4457)^+$ reported by LHCb in 2019. Taking into account both the mass ordering and the decay widths, our results favor the interpretation of  $P_\psi^N(4440)^+$ and $P_\psi^N(4457)^+$  as the isospin-$\frac12$ $[\bar D^*\Sigma_c]$ molecular states with $J^P$ configuration $(\frac{3}{2})^-$ and $(\frac12)^-$, respectively.
We first calculate the one-boson-exchange interaction kernel of $[\bar D^*\Sigma_c]$ in the isospin-$\frac12$ configuration. Then we present the Bethe-Salpeter equation\,(BSE) and wave functions for the bound states of  a vector meson and a $\frac12$ baryon with $J^P={\frac12}^-$ and ${\frac32}^-$.  The obtained mass results for the $(\frac32)^-$ and $(\frac12)^-$ are $4.442$ and $4.457$\,GeV, respectively.  Combining the effective Lagrangians and the BS wave functions, we further calculate the strong decay channels $\bar D^{(*)0}\Lambda_c^+$, $J/\psi(\eta_c) p$, and $\bar D\Sigma_c^{(*)}$ for the two $P_\psi^N$ states. In the favored  $\frac32^-$ and $\frac12^-$ configuration, the obtained total widths are $21.8$\,\si{MeV} and $13.0$\,\si{MeV}, respectively, which are substantially consistent with the LHCb data. Our results suggest that $\bar D^{0}\Lambda_c^+$ and $\bar D^{(*)0}\Lambda_c^+$ are the dominant decay channels to detect $P_\psi^N(4440)^+$ and  $P_\psi^N(4457)^+$, respectively. 
}

\begin{document}
\maketitle
\flushbottom

\section{Introduction}\label{Sec-1}

The exotic pentaquark states are playing important roles in hadron physics. The LHCb collaboration first reported two exotic structures $P_c(4380)^+$ and $P_c(4450)^+$ in the $J/\psi p$ mass spectrum with high statistical significance in 2015\,\cite{LHCb2015-Pc}, which are interpreted as resonant states having minimal valence quark content $[c\bar c uud]$. The  detected masses and widths are $4380\pm30$\,MeV  and $205\pm 88$\,MeV for $P_c(4380)^+$, and are $4449.8\pm3.0$\,MeV and $39\pm20$\,MeV for the latter one, namely, $P_c(4450)^+$.  The preferred $J^P$ assignments are of opposite parity, with one state having spin $\frac32$ and the other $\frac52$. These two exotic charmonium-pentaquark states attracted great attention in particle physics community and inspired  lots of theoretical researches.

In 2019 the LHCb collaboration further resolved the $P_c(4450)^+$ into two different resonances $P_c(4440)^+$ and $P_c(4457)^+$\,\cite{LHCb2019-Pc}, and also identified a new $P_c(4312)^+$ state, which will be referred as  $P_\psi^N(4440)^+$, $P_\psi^N(4457)^+$, and $P_\psi^N(4312)^+$ respectively latter in this work following the naming scheme suggested by the LHCb in 2022\,\cite{LHCb2022-Naming}, where the superscript $N$ is used to denote the isotopic spin $I=\frac12$ and the subscript $\psi$ is used to represent the hidden charm flavor. The measured mass and total width are listed in \autoref{T-LHCb-Pc}.
\begin{table}[ht]
\caption{Mass and width\,(MeV) of the two pentaquark states reported by LHCb in 2019\,\cite{LHCb2019-Pc}.}
\label{T-LHCb-Pc}
\vspace{0.2em}\centering
\begin{tabular}{ c|cccccccccccc }
\toprule[1.5pt]
States	&Mass                       & Width \\
 \midrule[1.2pt]
$P_\psi^N(4440)^+$  & $4440.3\pm1.3^{+4.1}_{-4.7}$   & $20.6\pm4.9^{+8.7}_{-10.1}$\\ 
 \midrule[1.2pt]
$P_\psi^N(4457)^+$  &$4457.3\pm0.6^{+4.1}_{-1.7}$    & $6.4\pm2.0^{+5.7}_{-1.9}$\\
\bottomrule[1.5pt]
\end{tabular}
\end{table}
Although these two states have been discovered four several years, the inner structure, spin-parity and decay poperties are still not clear.
The proximity to the $\bar D^*\Sigma_c$ threshold of the two observed narrow peaks suggests that they may play an important role in the dynamics of  ${P_\psi^N(4440)^+}$ and $P_{\psi}^N(4457)^+$ states. The  $\bar D^*\Sigma_c$ molecular states picture is then a natural interpretation to these two exotic particles. In a previous work\,\cite{LiQ2023}, we studied the mass spectra, inner interaction kernel and strong decay behaviors of the  $P_\psi^N(4312)^+$ within the $\bar D\Sigma_c$ molecule picture based on the Bethe-Salpeter equation and effective Lagrangians. This work aims to further investigate the mass spectra and strong decay properties of the $P_\psi^N(4440)^+$ and $P_\psi^N(4457)^+$ states.
 
The hidden charm molecular pentaquark states have been proposed before the experimental confirmation\,\cite{WuJJ2010,WangWL2011,WuJJ2012,YangZC2012,LiXQ2014,Karliner2015,ChenR2015}.
After the LHCb discoveries, lots of literature explored these newly observed pentaquark states from different aspects within different approaches\,\cite{XiaoCJ2019,LiuMZ2019,ChenR2019,XiaoCW2019,HeJ2019,LinYH2019,ChenHX2019,Ali2019,MengL2019,Burns2019,Voloshin2019,GuoFK2020,KeHW2020,DuML2020,WangZG2020,Yamaguchi2020,XuH2020,Burns2022,WangXW2201,WangXW2205}. Although the properties of the two $P_\psi^{N+}$ states are most likely to be the $\bar D^*\Sigma_c$ molecules with $\ket{I,I_3}=\ket{\frac12,\frac12}$\,\cite{ChenR2019,ChenHX2019,LiuMZ2019,XiaoCJ2019,XiaoCW2019,HeJ2019,LinYH2019,XuH2020,Burns2022}, the possibilities of the compact pentaquark states\,\cite{Ruangyoo2021,Stancu2021} or kinematical effects\,\cite{Nakamura2103,Nakamura2109} still exist. On the other hand, within the $\bar D^*\Sigma_c$ molecule picture, there still exist ambiguities in spin-parity configuration, namely, ${\frac12}^-$ or $\frac32^-$.  Different interpretation pictures correspond to different inner structures and interaction kernels. The essence of these two pentaquark states is still an open problem.

Besides the mass spectrum, the strong decay properties may play more important roles in determining the nature of the pentaquark states. Several approaches are used to study the decay properties of these pentaquark states\,\cite{XiaoCJ2019,LinYH2019,Stancu2021,Sakai2019,DongYB2020,WangGJ2020,ChenHX2020,XuYJ2020,WangZG2020A}, including the effective Lagrangian methods\,\cite{XiaoCJ2019,LinYH2019}, the flavor-spin and heavy quark spin symmetry\,\cite{Stancu2021,Sakai2019}, the chiral
constituent quark model\,\cite{DongYB2020}, QCD sum rules\,\cite{WangZG2020A,XuYJ2020}, etc. Most of the previous studies are based on the nonrelativistic Schrodinger or Lippmann-Schwinger equation and the results are dependent on several introduced free parameters, especially the cutoff value in the form factors. These free parameters weaken the prediction power of the theories and bring ambiguities in interpreting the nature of the $P_\psi^{N+}$.  More studies on the decay behaviors of $P_\psi^{N+}$ can be important and helpful to explore their inner structures and dynamics. This work aims to calculate the strong decays of $P_\psi^N(4440)^+$ and $P_\psi^N(4457)^+$ to $J/\psi p$, $\bar D^{*0}\Lambda_c^+$, $\eta_c p$, $\bar D^0\Lambda_c^+$, $\bar D\Sigma_c$, and $\bar D\Sigma_c^*$ channels. Notice that $P_\psi^N(4457)^+$ may also strongly decay to $P_\psi^N(4312)^+\pi$, however, this  partial decay width is estimated to be $\sim100$\,keV\,\cite{LingXZ2021}.

In this work, we will calculate the strong decay widths of $P^N_\psi(4440)^+$ and $P_\psi^N(4457)^+$  by combing the Bethe-Salpeter\,(BS) framework  with the effective Lagrangians.  First within the heavy quark symmetry and the light quark chiral symmetry, we calculate the internal interaction kernel for the $\bar D^*\Sigma_c$ molecular states based on the one-boson exchange. Then by solving the corresponding Bethe-Salpeter euqation\,(BSE) to obtain the bound states mass spectra and wave functions. Finally, by combining the BS wave functions and the effective Lagrangians we calculate the main strong decay widths and then make comparison with the experimental data. The Bethe-Salpeter equation is a relativistic two-body bound state equation. The constructed BS wave functions only depend on the good quantum number spin-parity and the Lorentz covariance. The BS methods are widely used to deal with mass spectra of the doubly heavy baryons\,\cite{LiQ2020,LiQ2022},  producing the observed molecular  pentaquarks\,\cite{XuH2020} and the fully heavy tetraquark $ T_{QQ\bar Q\bar Q}$ states\,\cite{LiQ2021}, and also the hadronic transitions and decays\cite{Chang2005,WangZ2012A,WangT2013,LiQ2016,LiQ2017A,LiQ2019A}.

This paper is organized as follows. After the introduction, we start with the Bethe-Salpeter equation for $P_\psi^N$ as the molecular state of a vector meson and a baryon, including  the interaction kernel and the relevant Salpeter wave functions (Sect. \ref{Sec-2}), then we calculate the strong decay widths of $P_\psi^N(4440)^+$ and $P_\psi^N(4457)^+$ to the eight channels aforementioned\,(Sect.\,\ref{Sec-3}). We finally present the numerical results, discussion and summaries in Sect.\,\ref{Sec-4}. The complicated Lagrangian expressions and calculation details are all collected in the appendix.

\section{$P^N_{\psi}$ as the $\bar D^* \Sigma_c$ molecular states}\label{Sec-2}

In this section, we first review the Bethe-Salpeter equation for bound states consisting of a vector meson and a spin-$\frac12$ baryon. Then we calculate the pentaquark interaction kernel based on the one-boson exchange. The BS wave functions of the $J^P=(\frac12)^-$ and $(\frac32)^-$ $P_\psi^N$ states will be constructed and solved numerically to prepare for the following strong decays.
\begin{figure}[htpb]
\centering
\includegraphics[width = 0.7\textwidth, angle=0]{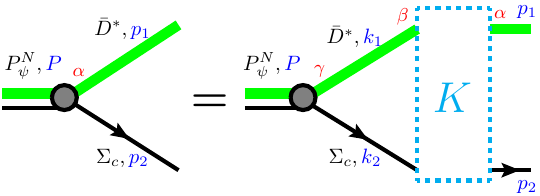}
\caption{Bethe-Salpeter equation of the molecular states consisting of a vector meson and a spin-$\frac12$ baryon.  The red letters denote the Lorentz indices; the blue letters $P,~p_1(k_1),~p_2(k_2)$ denote the momenta of the pentaquark, constituent meson and baryon, respectively.}\label{Fig-BSE-Pc-1-N}
\end{figure}

\subsection{Bethe-Salpeter equation of a spin-1 meson and a spin-$\frac{1}{2}$ baryon}

The Bethe-Salpeter equation for the molecular state of a meson and a baryon is schematically depicted in \autoref{Fig-BSE-Pc-1-N}, and can be expressed as
\begin{gather} \label{E-BSE-Pc}
\Gamma^\alpha(P,q,r) =\int \frac{\up d^4 k}{(2\pi)^4} (-\i)K^{\alpha\beta}(k,q)  [S(k_2) \Gamma^\gamma(P,k,r) D_{\gamma\beta}(k_1)] ,
\end{gather}
where $\Gamma(P,q,r)$ denotes the effective vertex for the pentaquark, constituent meson and baryon; the symbols $P$, $q$, and $r$ represent the pentaquark total momentum, inner relative momentum, and spin state, respectively; the inner relative momentum $q$ and $k$ are defined as $q\equiv\alpha_2p_1-\alpha_1p_2$, $k\equiv\alpha_2k_1-\alpha_1k_2$, with $\alpha_{1(2)}\equiv \frac{m_{1(2)}}{m_1+m_2}$, $k_{1(2)}$ denoting the momentum of the constituent meson\,(baryon), and $m_{1(2)}$ is the corresponding mass; $S(k_2)=\frac{\i}{\slashed k_{\!2} - m_2}$ is the free Dirac propagator of the baryon; the effective propagator for a spin-1 constituent meson reads
\begin{gather}
D^{\beta\gamma}(k_1)=\i\frac{\left(-g^{\beta\gamma} + \frac{k_{1\perp}^{\beta} k_{1\perp}^{\gamma}}{m_1^2}\right)}{k_1^2-m_1^2}.
\end{gather}
Here we introduce the abbreviation $x_\perp\equiv x- x_P \hat P$ for any four-vector $x$ to represent the corresponding spacelike variable  with $x_P \equiv x\cdot \hat P$,  $\hat P\equiv \frac{P}{M}$ and $M$ denoting the pentaquark mass, then $p_{1\perp}$ is implied.
The $(\i \epsilon)$ should be implied in all the propagators. 

In this work both the constituents $\bar D^*$ and $\Sigma_c$ contain a heavy charm quark, the velocity of the relative motion would be small. The interaction kernel $K(k,q)$ is assumed to be instantaneous and is not dependent on the time component of the exchanged momentum $(k-q)$, namely, $K(k,q)\sim K(k_\perp-q_\perp)$. Throughout this work, this instantaneous approximation is assumed for the pentaquark kernel.

The four-dimensional Bethe-Salpeter wave function is  defined as
\begin{gather} \label{E-wave-4D}
\psi_\alpha (q)=S(p_2) \Gamma^\beta(q) D_{\beta\alpha}(p_1),
\end{gather}
where the dependence on $P$ and $r$ is omitted for simplicity.
Under the instantaneous approximation of interaction kernel $K(k_\perp-q_\perp)$, the integral over the time component of $q$ can be directly absorbed into the wave function by defining the Salpeter wave function as
\begin{gather}
 \varphi_\alpha(q_\perp) \equiv - \i \int \frac{\d q_{P}}{2\pi} \psi_\alpha(q).
\end{gather}
where the convention factor $(-\i)$ is for later convenience.

By using the instataneous approximation and the Salpeter wave function, we can reduce one-dimensional integral of the BS equation. Performing the contour integral over $q_P$ on both sides of \eref{E-wave-4D}, we obtain the Salpeter equation (SE) for meson-baryon bound state\,\cite{XuH2020},
\begin{align} \label{E-SE-0}
\varphi_\alpha(q_\perp)=\frac{1}{2w_1}\left[ \frac{\Lambda^+(p_{2\perp}) }{M-w_1-w_2} +  \frac{\Lambda^-(p_{2\bot})  }{M+w_1+w_2}  \right] d_{\alpha\beta}(p_{1\perp})\Gamma^\beta(q_\perp),
\end{align}
where 
\begin{gather}
d_{\alpha\beta} = \left(-g_{\alpha\beta} + \frac{p_{1\perp\alpha} p_{1\perp\beta}}{m_1^2}\right)
\end{gather}
is the numerator of the vector propagator; $w_{i}=\left({m_{i}^2-p_{i\perp}^2}\right)^{1/2}~(i=1, 2)$ denotes the kinetic energy of the constituent meson and baryon respectively; the BS vertex $\Gamma^\beta(q_\perp)$ is expressed by a three-dimensional integral of the defined Salpeter wave function, 
\begin{align} \label{E-Gamma-3D}
\Gamma^\beta(q_\perp) =\int \frac{\up d^3 k_\perp}{(2\pi)^3}  K^{\beta\gamma}(k_\perp-q_\perp)  \varphi_\gamma(k_\perp).
\end{align}
The projector operators $\Lambda^\pm(p_{2\perp})$ are defined as 
\begin{gather}
\Lambda^\pm(p_{2\perp})=\frac{1}{2}\left[1 \pm {H}_2(p_{2\perp})\right]\gamma_0,
\end{gather}
where the dimensionless operator $H_2(p_{2\perp})$ is the usual Dirac Hamiltonian divided by the kinetic energy $w_2$,
\begin{gather}
H_2(p_{2\perp})=\frac{1}{w_2} \left(p^\alpha_{2\perp}\gamma_\alpha+m_2 \right)\gamma^0.
\end{gather}
Using the projector operator, we can further define the positive and negative energy wave functions as $\varphi^{\pm} \equiv \Lambda^\pm  \gamma^0 \varphi $, and we also have $\varphi=\varphi^+ + \varphi^-$. The Salpeter equation above can also be rewritten as the following  type
\begin{gather} \label{E-VS-SE}
M \varphi_\alpha =  (w_1+w_2) H_2(p_{2\perp})\varphi_\alpha +\frac{ 1 }{2w_1} d_{\alpha\beta}\gamma_0 \Gamma^\beta(q_\perp) .
\end{gather}

The  Salpeter  \eref{E-VS-SE} is in fact an eigenvalue equation about the Salpeter wave function $\varphi_\alpha(q_\perp)$, where the pentaquark mass $M$ behaves as the eigenvalue. The three-dimensional BSE, namely, \eref{E-VS-SE}, indicates that the mass of the pentaquark state consists of two parts, the kinetic energy and the potential energy. 

The normalization of the BS wave function is generally expressed as
\begin{align*}
&-\i\int \int \frac{\up{d}^4 q}{(2\pi)^4}\frac{\up{d}^4 k}{(2\pi)^4} \bar{\psi}^\alpha(P,q,\bar r) \frac{\partial}{\partial P^0} I_{\alpha\beta}(P,k,q) \psi^\beta(P,k,r)  = 2M \delta_{r\bar r},
\end{align*}
where the integral kernel in the normalization condition reads,
\begin{align*}
I_{\alpha\beta}(P,q,k)&=(2\pi)^2 \delta^4(k-q) S^{-1}(p_2)D^{-1}_{\alpha\beta}(p_1) + iK_{\alpha\beta}(P,k,q).
\end{align*}
Notice in this work, under the instantaneous and on-shell approximation, the interaction kernel has no dependence on $P^0$ and $q_P$, namely, $K_{\alpha\beta}(P,k,q)\sim K_{\alpha\beta}(k_\perp-q_\perp)$. Then the normalization would only involve the inverse of the two propagators. The inverse of the vector propagator reads,
\begin{gather} 
D^{-1}_{\alpha\beta}(p_1)=\vartheta_{\alpha\beta}D^{-1}(p_1),
\end{gather}
where $\vartheta^{\alpha\beta}=-g^{\alpha\beta}-{p^\alpha_{1\perp} p^\beta_{1\perp}}/{w_1^2}$ and  fulfills $\vartheta^{\alpha\beta}d_{\beta\gamma}=\delta^{\alpha}_{\gamma}$.
Notice now  there is no $P^0$ dependence in the numerator of $D^{\alpha\beta}(p_1)$.
Inserting the inverse of the propagators, we further obtain the normalization condition for the general Salpeter wave function,
\begin{align}\label{E-Norm-D1}
\int \frac{\up{d}^3 q_\perp}{(2\pi)^3} 2w_1  \vartheta^{\alpha\beta}  \bar\varphi_\alpha(q_\perp,\bar r) \gamma^0   \varphi_\beta(q_\perp,r)=2M\delta_{r\bar r}.
\end{align}

\subsection{Interaction kernel based on the one-boson exchange}

The $P_{\psi}^N(4440)^+$ and $P_\psi^N(4457)^+$ are consistent with the  $\bar D^* \Sigma_c$ molecular state with  isospin $\ket{I,I_3}=\ket{\frac12,+\frac12}$. We will extract the interaction kernel by calculating the $\bar D^* \Sigma_c$ scattering under the isospin symmetry,
\begin{gather}
 \braket{\textstyle\frac12,+\frac12|H_\up{eff}|\textstyle\frac12,+\frac12}.
\end{gather}
Since the isospin eigenstate does not coincide with the flavor eigenstate, it is more convenient to work in the uncoupled representation of isospin,
\begin{gather}
\textstyle \ket{\frac12, \frac12} =\frac{ \sqrt{2}}{ \sqrt{3}} \ket{1,+1}\ket{\frac12,-\frac12} - \frac{1}{\sqrt3} \ket{1,0}\ket{\frac12,\frac12}=\frac{ \sqrt{2}}{ \sqrt{3}} \ket{\Sigma_c^{++}}\ket{D^{*-}} - \frac{1}{\sqrt3} \ket{\Sigma_c^+}\ket{\bar D^{*0}}.
\end{gather}
In the molecular state scenario, the interaction between the two constituents $\Sigma_c$ and $\bar D^*$ can be realized by the one-boson exchange. 
Different from the $P_\psi^N(4312)^+$ consisting of the $\bar D\Sigma_c$ where the usual one-pion exchange is not possible for the parity, we need to take into account both the light (pseudo)scalar and vector meson exchange.

The effective hamiltonian is responsible for $\bar D^*\Sigma_c$ scattering by the possible light meson exchanges, namely,
\begin{gather}
H_\up{eff} = -\i \sum L_{\bar D^*\bar D^* E_i} L_{\Sigma_c\Sigma_c E_i}
\end{gather}
with $E_i=\sigma,\pi,\eta,\rho $ and $\omega$ representing the exchanged light mesons. Since both the constituent meson $\bar D^*$ and baryon $\Sigma_c$ contains a heavy $c$ quark, we use the effective Lagrangians under the chiral and heavy quark symmetry, which are collected in the appendix. The involved Lagrangians for meson sector behaves
\begin{equation}
\begin{aligned}
L_{\bar D^*\bar D^*\sigma} &= g_{\text{m}\sigma4}\bar D^{*\alpha\dagger} \bar D^*_{\alpha} \sigma,\\
L_{\bar D^*\bar D^*\Sigma}&= -\i g_{\text{m}\pi 4} \frac{1}{\i M_{D^*} } \epsilon^{\alpha\beta\mu\nu} \partial_\beta\bar D^{*\dagger}_{\nu}\partial_\alpha \Sigma \bar D^*_{\mu},\\
L_{\bar D^*\bar D^* V} &= -\i g_{\text{mv}41} \bar D^{*\alpha\dagger} ( \partial_\alpha  V_\beta -\partial_\beta  V_\alpha) \bar D^{*\beta} +\i g_{\text{mv}43}  \frac{1}{\i M_{\bar D^*}} \partial_\beta \bar D^{*\alpha\dagger}  V^\beta \bar D^{*}_{\alpha},
\end{aligned}
\end{equation}
where $\Sigma$ represents  the $3\times3$ traceless hermitian matrix consisting of eight pseudoscalar meson fields, 
\begin{gather}\label{E-Sigma}\renewcommand{\arraystretch}{1.8}
\Sigma =
\begin{bmatrix}
\frac{\pi^0}{\sqrt2}+ \frac{\eta}{\sqrt6} & \pi^+  & K^+ \\
\pi^-  & -\frac{\pi^0}{\sqrt2}+ \frac{\eta}{\sqrt6} & K^0 \\
K^-  & \bar K^0 & -\frac{2}{\sqrt6} \eta
\end{bmatrix};
\end{gather}
and  the symbol ${V}$ denotes the $3\times3$ matrix consisting of the 9 light vector meson fields
\begin{gather}\label{E-V}\renewcommand{\arraystretch}{1.8}
{V}=
\begin{bmatrix}
 \frac{(\rho^0+ \omega) }{\sqrt2} & \rho^+  & K^{*+ }\\
\rho^-  & - \frac{(\rho^0 - \omega) }{\sqrt{2}}   & K^{*0} \\
K^{*-}  & \bar K^{*0}  & \phi
\end{bmatrix}.
\end{gather}
The involved Lagrangians for the baryon sector behaves
\begin{equation}
\begin{aligned}
L_{\bar B B\sigma} &= -g_{\text{b}\sigma1}\bar B  B \sigma,\\
L_{\bar B B\Sigma}&= -\i g_{\text{b}\pi 1}  \bar B \Sigma \gamma_5 B,\\
L_{\bar BB V} &= -\i g_{\text{bv}1} \bar B \gamma_\alpha V^\alpha B,
\end{aligned}
\end{equation}
where $B$ denotes the symmetric baryon sextet and is specified in \eref{E-B6}. The coupling constants $g_{\text{m}\sigma4}$, $g_{\text{m}\pi4}$, $g_{\text{mv}43}$, and $g_{\text{b}\sigma1}$, $g_{\text{b}\pi1}$, $g_{\text{bv}1}$ are related to the chiral coupling constants and will be specified numerically later, 
\begin{equation}
\begin{aligned}
g_{\text{m}\sigma4} &= 2g_s M_{D^*},      &\gbs1 &=l_s,\\
g_{\text{m}\pi 4} &= 2M_{D^*} \frac{g}{f}, &\gbp1&=2 M_{\bar D^*}\frac{g_1}{f},\\
g_{\text{mv}43} &= \sqrt2 g_V\beta M_{D^*}, &\gbv1&=\frac{1}{\sqrt2} g_V\beta_S.
\end{aligned}
\end{equation}
It can be seen that the 8 chiral coupling constants $g_s$, $g$, $l_s$, $f$, $g_1$, $g_V$, $\beta$, and $\beta_S$ determine the strength of the interaction kernel. However, the numerical values of  these chiral coupling constants are just very  rough estimations; and on the other hand, the heavy quark approximation also takes non-negligible uncertainties into the interaction strength. 

Using these effective Lagrangians under the chiral and heavy quark symmetry, we can calculate the interaction kernel of  $\bar D^*\Sigma_c$ in isospin-$\frac{1}{2}$ based on the one-boson exchange model as
\begin{gather}\label{E-kernel}
K_{\alpha\beta} (s) = K^\Sigma_{\alpha\beta} +K^\sigma_{\alpha\beta} +  K^V_{\alpha\beta} = -K_1 \i   \epsilon^{s \hat P \alpha\beta} + K_2 g_{\alpha\beta} + K_3 \gamma_0 g_{\alpha\beta},
\end{gather}
where the spacelike $s = (k-q)_\perp$ represents the the exchanged momentum in the inner interaction;  $K^\Sigma$, $K^\sigma$, and $K^V$ denote the contributions from $\pi(\eta)$, $\sigma$, and $\rho(\omega)$ exchange, respectively,
\begin{equation}\label{E-As}
\begin{aligned}
K_1 &= - A_s^2\gbp1\gmp4 \Big(D_\pi - \frac16D_\eta\Big),\\
K_2 &=  A_s^2\gbs1\gms4   D_\sigma,\\
K_3 &= A_s^2\gbv1\gmv43 \Big(D_\rho - \frac12D_\omega\Big),
\end{aligned}
\end{equation}
where $D_\eta=(s^2 - m^2_\eta)^{-1}$ denotes the propagator of the inter-mediator $\eta$ meson, and then $D_\pi$, $D_\sigma$, $D_\rho$ and $D_\w$ are implied; $A_s$ is an parameter to adjust the relative interaction strength overall. It is obvious to see that $\pi$ and $\eta$ would made destructive contribution, and so for $\rho$ and $\omega$.  The contribution from pseudoscalar meson exchange is suppressed by the exchange momentum $s$ at low momentum range. Notice this interaction kernel is spin-independent since the spin-dependent items are incorporated in the construction of wave functions within the BS framework.

Notice the interaction kernel from $\rho(\omega)$ exchange share the same form with the potential from the one-gluon exchange in the diquark-quark model of baryons\,\cite{LiQ2020,LiQ2022}.  Since neither  mesons nor baryons  are point-like particles, it is necessary to introduce a form factor in the interaction vertex of hadrons to describing this non-pointlike effect. There is no general method to choose the form factor functions. In the previous work of the doubly heavy baryons\,\cite{LiQ2020,LiQ2022}, we calculated the $(cc)$ and $(bc)$ diquark form factors based on the BS framework, which can be parameterized by the exponential functions. In this work, we also choose the following exponential form factor
\begin{gather}\label{E-form-factor}
F(s^2) =\exp\left(\frac{s^2}{m_\Lambda^2}\right),
\end{gather}
and then the interaction kernel will behave $K_i \to F^2(s^2) K_i$ with $i=1,2,3$ representing the pseudoscalar, scalar and vector meson exchanges, respectively. Notice $s$ is spacelike and $s^2$ ranges from $0$ to negative infinity.

The cutoff parameter  $m_\Lambda$ is introduced to characterize this regulator function, and will be  determined together with $A_s$ by fitting the bound state mass to the experimental data, which is found to be $m_\Lambda=0.64\,\si{GeV}$ for $P_\psi^N(4440)^+$  and  $P_\psi^N(4457)^+$.  In the limit  ${s}^2\to 0$, the heavy hadron is seen by the inter-mediator mesons as a point-like particle, and hence the regular function is normalized to 1.  The cutoff value $m_\Lambda$ is usually believed to be much larger than the typical energy scale  $\sqrt{2\mu\epsilon}\sim0.05\,\si{GeV}$ for $P_\psi^N(4440)^+$ and $\sim0.007\,\si{GeV}$ for $P_\psi^N(4457)$, where $\mu=\frac{m_1m_2}{m_1+m_2}$ is the reduced mass of the two-hadron system and $\epsilon=(m_1+m_2-M)$ denotes the bound energy\,\cite{GuoFK2018,LinYH2019}. Our determined cutoff value is consistent with this universal estimation. 
The obtained $K_i$ for isospin-$\frac{1}{2}$ are displayed graphically in \autoref{F-Ki}.

\subsection{BS wave function for the $J^P=\frac{1}{2}^-$ and $\frac32^-$ pentaquark states}

From a $1^-$ vector meson and a $\frac12^+$ baryon, we can form bound states with $J^P=\frac12^-$ and $\frac32^-$. According to the parity properties, the constructed wave functions behave as vector-spinors under the space transformations. On the other hand, the $\frac12$ state can be expanded by the Dirac spinor $u_{(r)}(P)$ with the spin state $r=\pm\frac12$, while the $\frac32$ state can be constructed based on the Rarita-Schwinger vector-spinor $u^\alpha_{(r)}(P)$ with the spin state $r=\pm\frac32,\pm\frac12$.  
Then according to the spin-parity properties, and also considering the proper Lorentz structures,
the ${\frac{1}{2}}^-$ Salpeter wave function formed by the $1^-$ meson and $\frac{1}{2}^+$ baryon can be written  as
\begin{gather} \label{E-wave-1-2N}
\varphi_{\alpha}(P,q_\perp,r) =  A_\alpha(x) u(P,r) ,
\end{gather}
with
\begin{gather} \label{E-A-alpha}
A_\alpha= \left( g_1+ g_2  \slashed x  \right)(\gamma_\alpha -\hat P_\alpha)  + \left( g_3+ g_4  \slashed x \right)x_{\alpha},
\end{gather}
where we introduce a spacelike dimensionless variable $x_{\alpha}= \frac{q_{\perp\alpha}}{ |\vec q\,|}$ to simplify the expression; $u(P,r)$ represents the Dirac spinor with momentum $P$ and spin state $r$; the radial wave function $g_i(\vabs{q})~(i=1,\cdots,4)$ just depends on $\vabs{q}$ explicitly. To see the partial wave components more clearly, we express the above wave function  in terms of the usual spherical harmonics $Y_l^m$, namely,
\begin{equation} \label{E-1-2-Ylm}
\begin{aligned}
A_\alpha &=C_0  Y_0^0 \left(g_1 \xi_{\alpha} - g_4\frac{ \gamma_\alpha}{3} \right)  + C_1\left(Y_1^0\Gamma_{\alpha3}-Y_1^{-1} \Gamma_{\alpha+} -  Y_1^{1}\Gamma_{\alpha-}  \right)  \\
&+C_2g_4\left[Y_2^{-2}g_{\alpha+} \gamma_+ -Y_2^{-1}\frac{ (g_{\alpha3}\gamma_+  + g_{\alpha+}\gamma_3 )}{\sqrt{2}}  +Y_2^0 \frac{(g_{\alpha+}\gamma_-+g_{\alpha-}\gamma_++2 g_{\alpha3}\gamma_3)}{\sqrt{6}} \right. \\
& \left. - Y_2^1 \frac{(g_{\alpha3}\gamma_-+g_{\alpha-} \gamma_3 )}{\sqrt{2}} + Y_2^2 g_{\alpha-} \gamma_- \right],
\end{aligned}
\end{equation}
where the coefficients are $C_0=2\sqrt{ \pi} $, $C_1=\sqrt{\frac13}C_0$, and $C_2=\sqrt{\frac{2}{15}}C_0$; and we define the following abbreviations, $\xi_\alpha=(\gamma_\alpha-\hat P_\alpha)$, $\Gamma_{\alpha i} = (g_2 \gamma_i \xi_{\alpha} + g_3 g_{\alpha i})$, and
\begin{equation}\notag
\gamma_{\pm} = \mp\frac{1}{\sqrt{2}} (\gamma_1\pm i \gamma_2), ~~g_{\alpha\pm} = \mp\frac{1}{\sqrt{2}} (g_{\alpha1}\pm i g_{\alpha2}), ~~
\Gamma_{\alpha{\pm}} = \mp \frac{1}{\sqrt{2}}(\Gamma_{\alpha1}\pm i \Gamma_{\alpha2})
\end{equation}
with $i=1,2,3$ and $g_{\alpha\beta}$ denoting the Minkowski metric tensor; $Y_l^m$ is the usual spherical harmonics. Then it is clear to see that $g_1$ item corresponds to the $S$-wave component, $g_{2(3)}$ item corresponds to the $P$-wave components, and $g_4$ contributes to both the $S$ and $D$ partial waves.

Notice that the transformation properties of the above wave function depend on both the involved momenta, the Dirac  matrices and the spinor, and then the $P$-wave component can also make contribution in the $\frac12^-$ structure, which is different from the usual Schrodinger framework. It can be checked that each item in above wave function behave as a vector-spinor under the space transformation.
It can be seen that the BS wave function depends only on the good quantum number $J$ and parity while not the spin and orbit numbers. 

Inserting \eref{E-wave-1-2N} into \eref{E-Norm-D1}, the normalization behaves as
\begin{align}
\int \frac{\up{d}^3 q_\perp}{(2\pi)^3} 2 w_1  \vartheta^{\alpha\beta} \left(\bar u_{\bar r} \bar A_{\alpha} \gamma_0 A_{\beta} u_r \right) = 2M \delta_{r\bar r}.
\end{align}
Eliminating the polarization states and expand $A_\alpha$ explicitly, we obtain the following normalization
\begin{align*}
\int \frac{\up{d}^3 q_\perp}{(2\pi)^3} 2w_1  \left[
2 \left(g_1^2+g_2^2 \right) + o_0(g_4-g_1)^2+o_0(g_3+g_2)^2\right]
&=1
\end{align*}
with $o_0={m_1^2}/{w_1^2}$.

For two-body bound states with $J^P=\frac{3}{2}^-$, the Salpeter wave function can be expressed as
\begin{align} \label{E-wave-3-2N}
\varphi_{\alpha}(P,q_\perp,r)
&= A_{\alpha\beta} \gamma_5 u^\beta (P,r),
\end{align}
 where $ u^\beta (P,r)$ is the Rarita-Schwinger spinor, and $r=\pm\frac{3}{2},~\pm\frac{1}{2}$ denotes the polarization state; and
\begin{gather}
A_{\alpha\beta}(x) = \left( h_1+ h_2  \sl x \right ) g_{\alpha\beta}+ \left( h_3+ h_4 \slashed x \right)(\gamma_\alpha +\hat P_\alpha) x_{\beta}  +\i \epsilon_{\alpha\beta \hat P x}(h_5 + h_6\sl x)\gamma_5 + \left(h_7+h_8 \sl x \right) x_{\alpha} x_{\beta},
\end{gather}
where the radial wave function $h_i(|\vec q\,|)$ just depends on $|\vec q\,|$ explicitily.
Both the Salpeter wave functions constructed in \eref{E-wave-3-2N} and \eref{E-wave-1-2N} also fulfill the Lorentz condition $P^\alpha \varphi_\alpha=0$. It can also be checked that each item above behave as a vector-spinor under the space transformation. The decomposition of above wave function in terms of the spherical harmonics $Y_l^m$ is quite similar with that for the $\frac12^-$. We can also see that $h_1$, $h_{2(3,5)}$, $h_{4(6,7)}$, and $h_8$ mainly represent the contribution from $S$, $P$, $D$, and $F$-wave components, respectively.

For pentaquark states with $J^P=\frac{3}{2}^-$, the normalization is expressed as
\begin{align}
-\int \frac{\up{d}^3 q_\perp}{(2\pi)^3} 2 w_1  \vartheta^{\alpha\beta} \left(\bar u^\nu_{\bar r}\gamma_5 \bar A_{\alpha\nu} \gamma_0 A_{\beta\mu} \gamma_5 u^\mu_r \right) = 2M \delta_{r\bar r},
\end{align}
which can be further expanded explicitly as
\begin{equation}
\begin{aligned}
\int& \frac{\up{d}^3 q_\perp}{(2\pi)^3} \frac{2w_1}{3}   [(o_0+2) (h_1^2+h_2^2+h_3^2+h_4^2)\\
& -o_0 (2 h_1 h_7+2 h_2 h_8+2 h_3 h_8-2 h_4 h_7-h_7^2-h_8^2)  \\
&-2 (1-o_0) (h_2 h_3-h_1 h_4)-2 (-h_1 h_6+h_2 h_5-h_3 h_5-h_4 h_6-h_5^2-h_6^2)
 ]=1,
\end{aligned}
\end{equation}
where  the following completeness relation of the Rarita-Schwinger spinor\,\cite{Behrends1957} has been used,
\begin{equation}\label{E-RS}
u^\alpha(P,r) \bar u^\beta(P,r)=(\sl P+M)\left[-g^{\alpha\beta}+\frac{1}{3}\gamma^\alpha\gamma^\beta - \frac{P^\alpha\gamma^\beta-P^\beta\gamma^\alpha}{3M}+\frac{2P^\alpha P^\beta}{3M^2} \right].
\end{equation}

Inserting the Salpeter wave function \eref{E-wave-1-2N} for $J^P=\frac12^-$ into the Salpeter equation \eref{E-VS-SE}, eliminating the spinor, and then calculating the possible different traces, we can obtain four coupled eigenvalue equations with the bound state mass $M$ as the eigenvalue and the radial wave functions $g_{1-4}$ as the eigen functions (also see refs.\,\cite{LiQ2020,LiQ2022,XuH2020} for details to solve the BSE). The obtained eigenvalue equation for $\frac12^-$ state behaves 
\begin{equation}\label{E-w-K}
M \begin{pmatrix}
g_1(q) \\
g_2(q) \\
g_3(q) \\
g_4(q)
\end{pmatrix}
=
\begin{bmatrix}
-w_{m} & w_q & 0 &0 \\
w_q &w_m & 0 & 0 \\
0&0& w_m &-w_q \\
0& 0& -w_q & -w_m
\end{bmatrix}
\begin{pmatrix}
g_1(q) \\
g_2(q) \\
g_3(q) \\
g_4(q)
\end{pmatrix}
+\frac{1}{2w_1} \int \frac{\d^3 k_\perp}{(2\pi)^3}K(s)
\begin{pmatrix}
g_1(k) \\
g_2(k) \\
g_3(k) \\
g_4(k)
\end{pmatrix},
\end{equation}
where the kernel matrix $K$ reads
\begin{gather}
\begin{bmatrix}
K_{2-3} & -\frac12(s_k+s_q)K_1 &0 & -\frac12 s^2_\theta K_{2-3} \\
s_q K_1    & -c_\theta K_{2+3} & 0 & \frac12(c_\theta s_k-s_q) K_1 \\
-s_q K_1   &-   c_\theta q_m^2 K_{2+3} & -c_\theta w_{1m}^2K_{2+3}  & -\frac12(c_\theta s_k-s_q) K_1 \\
-q_m^2 K_{2-3} &[ (c_\theta s_q - s_k)(q^2_m-\frac32) -s_k]K_1 &0 & (c_\theta^2w_{1m}^2 -\frac12s^2_\theta)K_{2-3}
\end{bmatrix},
\end{gather}
where the involved symbols are defined as
\begin{gather*}
w_m=(w_1+w_2)m_2/w_2,~~w_q=(w_1+w_2)q/w_2,~~,K_{2\pm3}=K_2\pm K_3,\\ s_q=(\bm{q}\cdot \bm{s})/q, ~~ w_{1m} = w_1/m_1,~~s_k=\bm{k}\cdot \bm{s}/k, ~~q_m=q/m_1,~~ c_\theta=\cos\theta, ~~s_\theta=\sin\theta
\end{gather*}
with $\theta$ denoting the angle between $\bm{q}$ and $\bm{k}$. From above eigenvalue equation, it can also been seen that the $P$-wave component $g_2$ and $g_3$ but not the usual $S$-wave $g_1$ play the dominant roles in the  positive energy eigenvalues for $\frac12^-$ states. The first $4\times4$ matrix in \eref{E-w-K} represents the contribution from the kinetic energy parts. All the information of the interaction including the spin-spin and spin-orbit coupling has all been represented by the above $4\times4$ kernel matrix $K(s)$. We can also see that the scalar and vector meson exchange make constructive contribution to the molecular bound states, while the contribution from the pseudoscalar meson exchange is suppressed by the factor $s$. The contribution from the pseudoscalar meson exchange is difficult to analyze from above equation, but the numerical calculations also reveal that it makes positive contribution to bind the molecular states. 

Solving the eigenvalue equations numerically,  we obtain the corresponding mass spectra and numerical wave functions, which are also graphically displayed in \autoref{F-wave}. The BSE for $J^{P}=\frac{3}{2}^-$ pentaquark bound state can be solved similarly. The obtained eigenvalue equations show that the wave function of the ${\frac12}^-$ state is dominated by the $P$-wave $g_2$ and $g_3$, while the ${\frac32}^-$ state is dominated by the $S$-wave $h_1$.

\section{Strong decays of $P_{\psi}^N$  with the BS wave function}\label{Sec-3}

In this section, we first present the relevant effective Lagrangians for $P_\psi^N$ strong decays; then we give the decay amplitude by using the BS wave function combining with the effective Lagrangians; finally, the expressions of the partial decay widths are presented in terms of the relevant form factors. 

In this work, we interpret $P_\psi^N(4440)^+$ and $P_\psi^N(4457)^+$ as the molecular states $P_{\psi1/2}^N$ and $P_{\psi3/2}^N$  consisting of $\bar D^*\Sigma_c$ constituents, where the particular spin configuration are not assumed.  We will calculate the following eight decay channels, $J/\psi p$, $\bar D^{*0}\Lambda_c^+$, $\eta_c p$, $\bar D^0\Lambda_c^+$, $D^-\Sigma_c^{++}$, $\bar D^0\Sigma_c^+$, $D^-\Sigma_c^{*++}$ and $\bar D^0\Sigma_c^{*+}$. According to the final products, we categorize the main strong decay modes of $P_\psi^N$ as three ones: a pseudoscalar meson and a spin-$\frac12$ baryon ($P+B$), a vector meson and a spin-$\frac12$ baryon ($V+B$), and a pseudoscalar meson and a spin-$\frac32$ baryon $(P+B^*)$.

For the $(V+B)$ decay mode, the mainly decay channels are $J/\psi p$ and $\bar D^{*0}\Lambda_c^+$. The $J/\psi p$ channel can be realized by a $D$ or $D^*$ exchange, while the $\bar D^* \Lambda_c$ channel can be realized by a $\pi$ or a $\rho$ exchange. The $(P+B)$ decay mode mainly contains the $\eta_c p$, $\bar D^0\Lambda^+_c$ and $\bar D\Sigma_c$, where the last one consists of the $\bar D^0\Sigma_c^+$ and $D^-\Sigma_c^{++}$ two decay channels. For the $(P+B^*)$ decay mode, the main strong decay channel is $\bar D\Sigma_c^*$ which can be realized by a $\pi$ or $\rho\,(\omega)$ exchange. The interactions mentioned above can be represented by the corresponding effective Lagrangians under the hadronic level. All the relevant Lagrangians have been collected in the appendix \ref{A-Ls}.

\subsection{$P_\psi^N\to (V+B)$}

$P_\psi^N$ as the $\bar D^*\Sigma_c$ molecular state can strongly decay to a vector meson and a baryon by exchanging either a vector meson or a pseudo-scalar meson. We first calculate the $J/\psi p$ channel amplitude and then the similar channel $\bar D^{*0}\Lambda_c^+$  will be given directly. $P_\psi^N[\bar D^* \Sigma_c]$ molecular state can decay to $J/\psi p$ by  exchanging either a  $D$ or a $D^*$ virtual meson, and the total amplitude is the sum of the two.

\begin{figure}[htpb]
\centering
\includegraphics[width = 0.45\textwidth, angle=0]{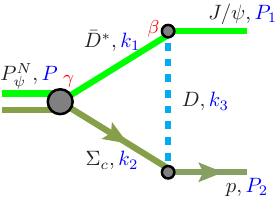}
\includegraphics[width = 0.45\textwidth, angle=0]{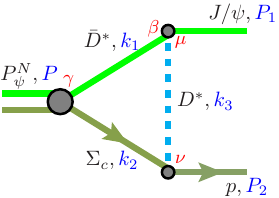}
\caption{Strong decay of $P_\psi^N[\bar D^*\Sigma_c]^+$ to the $J/\psi p$ by exchanging a virtual mediator $D$\,(left panel) and $D^*$\,(right panel). $P,~k_1,~k_2, ~P_1,~P_2$ denote the momenta of $P_\psi^N$, constituent meson,  constituent baryon, the final $J/\psi$, and the final $p$ respectively.}\label{Fig-Pc-psi+p-D}
\end{figure}

\subsubsection{$P_\psi^N\to J/\psi p$ by $D$ exchange}
We take the $P_\psi^N\to J/\psi p$ by $D$ exchange as an example to show the strong decay calculation details by combining the effective Lagrangians and BS wave functions, and other calculations will be listed in the \ref{App} for simplicity.
The effective Lagrangians responsible for the $D\bar D^*J/\psi$ and $\bar p \Sigma_cD$  interactions read
\begin{equation} \label{E-L-11}
\begin{aligned}
{L}_{ D\bar D^{*}\psi} =& +g_{  D \bar D^*\psi}\frac{1}{M_\psi} \epsilon^{\alpha\beta\mu\nu} \partial_\alpha D \bar D^*_\beta \partial_\mu \psi^\dagger_\nu, \\
L_{\bar p\Sigma_cD} =& -g_{ND\Sigma_c}\i \left( \bar p \gamma_5 \Sigma_c^{++} D^{+\dagger} - \frac1{\sqrt2} \bar p \gamma_5\Sigma_c^+ D^{0\dagger}\right). 
\end{aligned}
\end{equation}
Given above Lagrangians, the amplitude for$P_{\psi1/2}^N\!\to\! J/\psi p$ by exchanging one $D$ can be expressed by the BS wave vertex as
\begin{equation}\label{E-A-psi+p-D}
\begin{aligned} 
\mathcal{A}_{11;1/2}
&=  - \i^3g_{N\!D\Sigma_c} g_{D\bar D^*\psi} \bar u_2 \gamma_5 \int \frac{\d^4k}{(2\pi)^4} [S(k_2) \Gamma^\gamma(k,r) D_{\gamma \beta}(k_1)] F^2 D(k_3) \frac{ \epsilon^{\alpha\beta P_1 \nu}}{M_1}  e^*_{1\nu}  k_{3\alpha} ,
\end{aligned}
\end{equation}
where $u_2$ is short for $u(P_2,r_2)$ with $r_2$ representing the proton spin state; $e^*_{1}$ is short for $e^*_{(r_1)}(P_1)$ representing the polarization vector of the final $J/\psi$ with $P_1$ denoting the $J/\psi$ momentum and $r_1=0,\pm 1$ representing the 3 possible polarization states, which fulfills the Lorentz condition $e_{1}^\alpha  P_{1\alpha} =0$. The momentum of the exchanged virtual charmed meson is denoted as $k_3=(k_1-P_1)$.
We will use $M_1$ and $M_2$ to denote the masses of the final meson and baryon (here $J/\psi$ and proton) respectively. To calculate this decay amplitude, we strip off the triangle amplitude involved the integral over $k$ as
\begin{gather}
T_{11}^\nu u(P,r) = \i \frac{ \epsilon^{\alpha \beta P_1\nu } }{M_1} \gamma_5 \int \frac{\d^4k}{(2\pi)^4}  [S(k_2) \Gamma^\gamma(k,r) D_{\gamma \beta}(k_1)] D(k_3) k_{1\alpha}. 
\end{gather}
Then the decay amplitude by exchanging a $D$ can be expressed as
\begin{gather}
\mathcal{A}_{11;1/2} = G_{11} e^{*\nu}_{1} \bar u_2 T_{11 \nu} u(P,r) .
\end{gather}
where we also define a dimensionless constant
\begin{gather}
G_{11} = g_{ND\Sigma_c} g_{D\bar D^*\psi}
\end{gather}
to denote the strong interaction strength for $\bar D^{*-}\Sigma_c^{++}\to J/\psi p$ scattering by one $D$ exchange. 

Now the main task is to calculate the above triangle integral. 
To express the amplitude by the Salpeter wave functions, we need perform a contour integral over $k_P$. First we split off the time component of all the items involved $k$, 
\begin{gather}
D(k_1) = \i \frac{1}{2w_1}\left( \frac{1}{k_P -\zeta_1^+ +\i \epsilon} - \frac{1}{k_P -\zeta_1^- -\i \epsilon} \right), \\
S(k_2) = (-\i) \left( \frac{\Lambda^+(k_\perp)}{k_P -\zeta^+_2 -\i \epsilon} + \frac{\Lambda^-(k_\perp)}{k_P -\zeta^-_2 +\i \epsilon}  \right), \\
D(k_3) =  \i \frac{1}{2w_3}\left( \frac{1}{k_P -\zeta_3^+ +\i \epsilon} - \frac{1}{k_P -\zeta_3^- -\i \epsilon} \right),
\end{gather}
where the six poles are defined as
\begin{equation}
\begin{aligned}
\zeta_1^\pm  = -\alpha_1 M \pm w_1 ,~~
\zeta_2^\pm  = +\alpha_2 M \mp w_2, ~~
\zeta_3^\pm  =E_1-\alpha_1M \pm w_3.
\end{aligned}
\end{equation}
Using above expression we obtain the following result,
\begin{align} \label{E-Duv-k1-kP}
\int \frac{\d k_P}{2\pi} [S(k_2) \Gamma_\gamma(k,r) D_{\gamma \beta}(k_1)] D(k_3)k_{1\alpha} &=  \frac{1}{2w_3}\left( a_{1\alpha} \Lambda^+  + a_{2\alpha} \Lambda^- \right) \gamma_0\varphi_\beta.
\end{align}
where the two variables $a_1$ and $a_2$ behave
\begin{equation}\label{E-a1-a2}
\begin{gathered}
a_{1\alpha} = c_1 k_{11\alpha} + c_3 k_{13\alpha} + c_5 k_{15\alpha}, \\
a_{2\alpha} = c_2 k_{12\alpha} + c_4 k_{14\alpha} + c_6 k_{16\alpha}, 
\end{gathered} 
\end{equation}
where $k_{1i} = k_1(k_P=k_{Pi})$ with $i=1,\cdots 6$, and $k_{Pi}$s are defined as
\begin{gather}
k_{P1}=\zeta_1^+,~k_{P2}=\zeta_1^-,~ k_{P3}=\zeta^+_2,~k_{P4}=\zeta_2^-,~ k_{P5}=\zeta_3^+,~ k_{P6}=\zeta_3^-.
\end{gather}
Here we need to express $k_1=\alpha_1 P + k$.
The coefficients $c_i$s $(i=1,\cdots 6)$ are defined as 
\begin{equation} \label{E-cn}
\begin{aligned}
c_{1(2)} &= \mp\frac{1}{E_1\mp(w_1+w_3)},\\
c_{3(4)}   &= \mp\frac{1}{E_2\mp(w_2+w_3)}, \\
c_{5(6)}   &= \pm\frac{M\mp(w_1+w_2)}{[E_1\pm(w_1+w_3)][E_2\mp(w_2+w_3)]}.
\end{aligned}
\end{equation}
Inserting above result of the contour integral, we can express $T_{11}^\nu u$ by the Salpeter wave function as
\begin{align} \label{E-A-psi+p-D}
 {T}^\nu_{11} u&=  \i \frac{ \epsilon^{\alpha\beta P_1\nu} }{M_1} \gamma_5 \int \frac{\d k^3_\perp}{(2\pi)^3} \frac{1}{2w_3}F^2\left( a_{1\alpha} \Lambda^+  + a_{2\alpha} \Lambda^- \right) \gamma_0\varphi_\beta (k_\perp ).
\end{align}

Now $T_{11}$ has been expressed by the integral over the three-dimensional Salpeter wave function $\varphi_\beta$.
Inserting the Salpeter wave functions \eref{E-wave-1-2N} for $P_{\psi1/2}^N$ into \eref{E-A-psi+p-D}, we can further simplified the amplitude $T_{11\alpha}$ in terms of two form factors as
\begin{gather} \label{E-ff-s11}
{T}_{11\alpha;1/2} = (s^1_{11} \gamma_\alpha  + s^1_{12} \hat P_\alpha ).
\end{gather}
where the following identity of the Levi-Civita symbol are used,
\begin{gather}
\gamma_\alpha \epsilon^{\alpha\beta\gamma\delta} =\i (\gamma^\beta \gamma^\gamma \gamma^\delta - \gamma^\delta g^{\beta\gamma} +\gamma^\gamma g^{\beta\delta} - \gamma^\beta g^{\gamma\delta})\gamma^5.
\end{gather}
The amplitude  then behaves
\begin{gather}
\mathcal{A}_{11;1/2}= G_{11}  e^{*\alpha}_{1}   \bar u_{2}  \left(s^1_{11} \gamma_\alpha  + s^1_{12} \hat P_\alpha \right) u.
\end{gather}
Here we introduce a symbol $s^n_{m(o)}$ to represent the strong decay form factor, where the first subscript $m\,(1,\cdots,8)$ distinguishes the eight decay channels, namely, $J/\psi p$, $\bar D^{*0}\Lambda_c^+$, $\eta_c p$, $\bar D^0\Lambda_c^+$, $D^-\Sigma_c^{++}$, $\bar D^0\Sigma_c^+$, $D^-\Sigma_c^{*++}$ and $\bar D^0\Sigma_c^{*+}$, respectively; the superscript $n$ is used to denote the different exchanged particle, with 1 for $D(\pi)$, 2 for $D^*(\rho)$, and $4$ for $\omega$ exchange;  the third one $o$ is used when the possible form factor number is more than one.

For $P_{\psi3/2}^N$ to $J/\psi p$ channel by $D$ exchange, the calculation process is quite similar and the only difference is to make a replacement
\begin{gather}
T_{11}^\alpha u \to T_{11}^{\alpha\beta} u_\beta.
\end{gather}
Namely, the decay amplitude is expressed as
\begin{gather}
\mathcal{A}_{11;3/2} =G_{11}  e^{\alpha*}_{1} \bar u_2 T_{11{\alpha\beta}} u^\beta,
\end{gather}
where  the triangle amplitude is expressed by the integral of the Salpeter wave function as
\begin{align} \label{E-A-psi+p-D-4457}
 {T}^{\nu\gamma}_{11} u_\gamma&=  \i \frac{ \epsilon^{\alpha\beta P_1\nu} }{M_1} \gamma_5 \int \frac{\d k^3_\perp}{(2\pi)^3} \frac{1}{2w_3}F^2\left( a_{1\alpha} \Lambda^+  + a_{2\alpha} \Lambda^- \right) \gamma_0A_{\beta\gamma}u^\gamma.
\end{align}
After inserting the BS wave function \eref{E-wave-3-2N} into above expression, we further express $T_{11{\alpha\beta}}$ as the form factors,
\begin{gather}\label{E-ff-t11}
T_{11{\alpha\beta}} =t_{11}^1\i\epsilon_{\alpha\beta\hat P\hat P_1}+\left(t_{12}^1g_{\alpha\beta} + t_{13}^1 \gamma_\alpha \hat P_{1\beta} + t^1_{14} \hat P_{\alpha}\hat P_{1\beta}\right)\gamma_5,
\end{gather}
where we used $t^n_{m(o)}$ to denote the form factor for $P_{\psi3/2}^N$ decay with  $m,n,o$ representing the same meanings as those in $s^n_{m(o)}$ for $P_{\psi1/2}^N$.

In the appendix \ref{App} we collect the specific expressions of the two form factors in terms of the Salpeter wave functions $g_1$, $g_2$, $g_3$ and $g_4$, and the numerical values can then be calculated by performing the three-dimensional integral over $k_\perp$ numerically.

\subsubsection{$P_\psi^N\to J/\psi p$ by $D^*$ exchange}
The effective Lagrangian responsible for the $J/\psi D^* \bar D^*$ interaction is
\begin{equation}\label{E-L-12}
\begin{aligned}
{L}_{D^* \bar D^*\psi} =
& - g_{ D^*\bar D^*\psi} \i({\partial}_\alpha D^{*}_{\beta} \bar D^{*\alpha}    \psi^{\dagger\beta}+ 2   D^{*}_\alpha {\partial}_\beta \bar D^{*\alpha}  \psi^{\dagger\beta} +   D^{*\alpha} \bar D^*_\beta   {\partial}_\alpha \psi^{\dagger\beta}),\\
L_{\bar p\Sigma_cD^{*}} =
& - g_{N\Sigma_cD^*} \left( \bar p \gamma^\alpha \Sigma_c^{++} D^{*+\dagger}_\alpha - \frac1{\sqrt2} \bar p \gamma^\alpha\Sigma_c^+ D^{*0\dagger}_\alpha \right).
\end{aligned}
\end{equation}
Then we can express the invariant amplitude for $P_{\psi1/2}^N\!\to\! J/\psi p$ by exchanging one $D^*$ as
\begin{align}
\mathcal{A}_{12;1/2}=&G_{12} \bar u_2 \gamma_\nu \int \frac{\d^4 k}{(2\pi)^4} [S(k_2) \Gamma_\gamma(k,r) D^{\gamma\beta}(k_1)] F^2D^{\mu\nu}(k_3) e^{*\alpha}_{1}  O_{\alpha\beta\mu\rho} (k_1+k_3)^\rho,
\end{align}
where the propagator of the exchanged $D^*$ meson behaves as
\begin{gather}
D_{\mu\nu}(k_3) = \i \frac{-g_{\mu\nu} + {k_{3\mu} k_{3\nu}}/{m_3^2}}{k_3^2 - m_3^2 + \i \epsilon}
\end{gather}
with the propagator mass $m_3=M_{D^*}$; and 
\begin{gather}
O_{\alpha\beta\mu\rho} \equiv   - (g_{\alpha\mu} g_{\beta\rho} -\frac12\iota_2 g_{\beta\mu}g_{\alpha\rho}+ \iota_3 g_{\alpha\beta} g_{\mu\rho}  ),
\end{gather}
where we introduced  $\iota_{2(3)}$ for later convenience, and obviously here $\iota_2=2$ and $\iota_3=1$.
Also as before we define the dimensionless constant
\begin{gather}\notag
G_{12} = g_{ND^*\Sigma_c} g_{ D^*\bar D^*\psi}
\end{gather}
to denote the strong interaction strength for $\bar D^{*-}\Sigma_c^{++}\!\!\to\!\! J/\psi p$ scattering by one $D^*$ exchange. Finally, by a similar calculation (\ref{A-2-1})  as before we express the decay amplitude  as
\begin{align*}
\mathcal{A}_{12;1/2}
=& G_{12}  (e_{1}^{\alpha})^*  \bar u_2 T_{12\alpha} u = G_{12}  (e_{1}^{\alpha})^*  \bar u_2  (s^2_{11} \gamma_\alpha  + s^2_{12} \hat P_\alpha )u.
\end{align*}
where $s^2_{11}$ and $s^2_{12}$ are the form factors expressed by the integral of the wave functions.

For $P_{\psi3/2}^N\to J/\psi p$ by exchanging $D^*$ meson, the decay amplitude is similar with that for $P_{\psi1/2}^N$ and the only difference is to make replacement
\begin{gather}
T_{12\alpha}u  \to T_{12\alpha\gamma} u^{\gamma};
\end{gather}
and then replace the Salpeter wave function of $P_{\psi1/2}^N$ to $P_{\psi3/2}^N$,
namely, 
\begin{gather}
\varphi_\alpha=A_\alpha u \to \varphi_\alpha = A_{\alpha\gamma} u^\gamma.
\end{gather}
After performing the numerical integration, we express $T_{12{\alpha\beta}}$ by the form factors as
\begin{gather} \label{E-ff-t12}
T_{12{\alpha\beta}} =  t^2_{11}\i \epsilon_{\alpha\beta \hat P \hat P_1} + \left(t_{12}^2g_{\alpha\beta} + t_{13}^2 \gamma_\alpha \hat P_{1\beta} + t_{14}^2 \hat P_{\alpha}\hat P_{1\beta}\right)\gamma_5.
\end{gather}

Combing the two amplitudes from $D$ and $D^*$ exchange together, we obtain the full invariant amplitude for $P_{\psi1/2}^N \to J/\psi p$ decay by two form factors,
\begin{gather} \label{E-A}
\mathcal{A} [{P_{\psi1/2}^N\!\to \!J/\psi p}] =  e^{*\alpha}_{1}  \bar u_{2} \left(   s_{11} \gamma_\alpha + s_{12} \hat P_\alpha \right) u.
\end{gather}
where $s_{11}$ and $s_{12}$ are related to the corresponding strong interaction constants and are expressed as 
\begin{equation}\label{E-s1-s2}
\begin{aligned}
s_{1i}&=G_{11}s_{1i}^1 +G_{12} s_{1i}^2.
\end{aligned}
\end{equation}

The full amplitude for $P_{\psi3/2}^N\to J/\psi p$ is expressed by three form factors,
\begin{align}
\mathcal{A}[{P^N_{\psi3/2}\!\to\!\! J/\psi p}]
=& e_{1}^{*\alpha}  \bar u_2 \left( \i t_{11} \epsilon_{\alpha\beta \hat P \hat P_1} + t_{12}g_{\alpha\beta}\gamma_5 + t_{13} \hat P_{1\beta} \gamma_\alpha \gamma_5 + t_{14} \hat P_{\alpha}\hat P_{1\beta} \gamma_5 \right) u^\beta,
\end{align}
where $t_{1i}~(i=1,2,3,4)$ is related to the coupling constants by
\begin{equation}\label{E-tn}
\begin{aligned}
t_{1i}&=G_{11} t_{1i}^1 + G_{12} t_{1i}^2.
\end{aligned}
\end{equation}
In above equations, the specific expressions of $s_{11}^2$ and $s_{12}^2$ can be obtained by inserting the Salpeter wave functions  and performing the integral numerically. 

\subsubsection{$P_\psi^N \to \bar D^{*0}\Lambda_c^+$}

The decay channel for $P_\psi^N\to \bar D^{*0}\Lambda_c^+$ is quite similar with that of $P_\psi^N\to J/\psi p$. The only difference is the final vector meson $J/\psi$ replaced by $\bar D^{*0}$, the final proton baryon replaced by the $\Lambda_c^+$ baryon, the propagator $D$ and $D^*$ replaced by the $\pi$ and $\rho$ in the pseudoscalar and vector exchange respectively. Then we can obtain the amplitude directly from the $J/\psi p$ channel.  

From \eref{E-L-DPD} and \eref{E-L-AMB} we obtain the relevant Lagrangians with $\pi$ exchange as
\begin{equation}\label{E-L-21}
\begin{aligned}
L_{\bar D^{0*}\bar D^{*-} \pi} & = + g_{\bar D^{*0}\bar D^{*-} \pi}\frac{1}{M_{\bar D^{*0}}} \epsilon^{\alpha\beta\mu\nu} \partial_\alpha \pi^+ \bar D_\beta^{*-} \partial_\mu \bar D^{*0\dagger}_\nu,\\
L_{\Lambda_c^+\Sigma_c^{++}\pi} &= +g_{\Lambda_c^+\Sigma_c^{++}\pi}\i \bar \Lambda_c^+ \gamma_5 \pi^- \Sigma_c^{++},
\end{aligned}
\end{equation}
where we have divided a  $M_{\bar D^{*0}}$ in the first Lagrangian to keep the coupling constant dimensionless; and the coupling constants now behaves
\begin{equation}
\begin{aligned}
g_{\bar D^{*0}\bar D^{*-}\pi} &=M_{\bar D^{*0}}\frac{2g}{f},\\
g_{\Lambda^+_c\Sigma_c^{++}\pi} &= \frac{1}{\sqrt3}(M_{\Lambda_c^+} + M_{\Sigma_c^{++}}) \frac{g_4}{f}.
\end{aligned}
\end{equation} 
For the $\rho$ exchange, the relevant Lagrangians have been given in \eref{E-L-DVD} and \eref{E-L-AMB}, which can be further expressed as 
\begin{equation}\label{E-L-22}
\begin{aligned}
L_{\bar  D^{*0}\bar D^{*-}\rho} =
&  g_{\bar  D^{*0}\bar D^{*-}\rho}  \i(\partial^\alpha \rho^{+\beta} \bar D^{*-}_{\alpha}  \bar D^{*0\dagger}_\beta + \rho^{+\alpha}\partial^\beta \bar D^{*-}_{\alpha} \bar D_\beta^{*0\dagger} +A_\up{r} \rho^{+\alpha} \bar D^{*-\beta} \partial_\alpha \bar D_\beta^{*0\dagger}),\\
L_{\bar \Lambda_c^+  \Sigma_c^{++}\rho} = & -g_{\Lambda_c^+ \Sigma_c^{++}\rho} \bar \Lambda_c^+  \rho_\alpha \gamma^\alpha  \Sigma_c^{++},
\end{aligned}
\end{equation}
where the two coupling constants read
\begin{equation}
\begin{aligned}
g_{\bar D^{*0}\bar D^{*-}\rho} &=R_{\braket{\bar D^{*0}\bar D^{*-}}} \sqrt8(\lambda g_V),\\
g_{\Lambda_c^+ \Sigma_c^{++}\rho} &=\frac{\sqrt2}{\sqrt3}(M_{\Lambda_c^+} + M_{\Sigma_c^{++}})(\lambda_I g_V).
\end{aligned}
\end{equation}
In this work we calculate all the decay widths under the isospin symmetry, then the contributions from other Lagrangians, such as $L_{\bar D^{0*}\bar D^{*0}\pi}$ and $L_{\Lambda_c^+\Sigma_c^{+}\pi}$, will be induced into the final isospin factors.

Comparing the above Lagrangians with the ones in $J/\psi p$ channel, we find the Lagrangians involved in the two decay modes share the same structure, and then we can directly express the strong decay amplitude for $P_\psi^N$ to $\bar D^{*0}\Lambda_c^+$ by the form factors as
\begin{align}
\mathcal{A}[P_{\psi1/2}^N\!\to\! \bar D^{*0}\Lambda^+_c] &=   e_{1}^{*\alpha} \bar u_2 \left(s_{21} \gamma_\alpha + s_{22}\hat P_\alpha\right) u,\\
\mathcal{A}[P_{\psi3/2}^N\!\to\! \bar D^{*0}\Lambda^+_c] &=   e_{1}^{*\alpha} \bar u_2 \left(
t_{21}\i \epsilon_{\alpha\beta \hat P \hat P_1} + (t_{22}g_{\alpha\beta} + t_{23} \gamma_\alpha \hat P_{1\beta} + t_{24} \hat P_{\alpha}\hat P_{1\beta})\gamma_5
\right) u,
\end{align}
where the form factors $s_{2i}\,(i=1,2)$ and $t_{2j}\,(j=1,2,3,4)$ are defined as 
\begin{align}
s_{2i} &= G_{21}s_{2i}^1 + G_{22}s_{2i}^2, \\
t_{2j} &= G_{21}t_{2j}^1 + G_{22}t_{2j}^2,
\end{align}
where we define two dimensionless constants
\begin{equation}\notag
\begin{aligned}
G_{21}&=g_{\Lambda_c^+ \Sigma_c^{++}\pi } g_{\bar D^{*0} \bar D^{*-} \pi},\\
G_{22}&= g_{\Lambda_c^+ \Sigma_c^{++}\rho} g_{\bar D^{*0}\bar D^{*-}\rho}
\end{aligned}
\end{equation}
to represent the strong interaction strength for $\bar D^{*-}\Sigma_c^{++}\to \bar D^{*0}\Lambda_c^+$ scattering by one pion and one $\rho$ exchange respectively.

The form factors $s_{2i}^1$, $s_{2i}^2$, $t_{2j}^1$, and $t_{2j}^2$ share exactly the same expressions with $s_{1i}^1$, $s_{1i}^2$, $t_{1j}^1$, and $t_{1j}^2$ of $P_\psi^N\to J/\psi p$ decays, respectively, while the only difference is to change the masses of $m_3$, $M_1$, and $M_2$ from $M_D(M_{D^*})$, $M_{\psi}$, and $M_p$ to $M_\pi(M_\rho)$, $M_{\bar D^{*0}}$, and $M_{\Lambda_c}$, respectively, namely,
\begin{equation}
\begin{aligned}
s_{2i}^1 &= s_{1i}^1[m_3\to M_\pi, M_1\to M_{\bar D^{*0}}, M_2\to M_{\Lambda_c^+}],\\
t_{2j}^1 &= t_{1j}^1[m_3\to M_\pi, M_1\to M_{\bar D^{*0}}, M_2\to M_{\Lambda_c^+}],\\
s_{2i}^2 &= s_{1i}^2[M_{1}\to M_{\bar D^*}, M_{2}\to M_{\Lambda^+_c}, m_3\to M_{\rho},  \iota_2\to1,\iota_3\to A_{\up{r}}], \\
t_{2j}^2 &= t_{1j}^2[M_{1}\to M_{\bar D^*}, M_{2}\to M_{\Lambda^+_c}, m_3\to M_{\rho},  \iota_2\to1,\iota_3\to A_{\up{r}}],
\end{aligned}
\end{equation}
where $s_{2i}^1\,(i=1,2)$ and $t_{2j}^1\,(j=1,2,3,4)$ represent the contribution from pion exchange, while $s_{2i}^2$ and $t_{2j}^2$ represent the contribution from $\rho$ exchange.

\subsection{$P_\psi^N \to (P+B)$}
For the decay mode of $P_\psi^N\to(P+B)$ decay mode, namely, a pesudoscalar meson and a spin-$\frac12$ baryon, the involved decay channels are $\eta_c p$, $\bar D^0\Lambda_c^+$, and $\bar D\Sigma_c$. The $\eta_c p$ channel can be realized by a $D$ or $D^*$ exchange with $P$-wave barrier, while last two can be realized by the $\pi$ or $\rho(\omega)$ exchange. 

\subsubsection{$P_\psi^N \to \eta_c p$}
Different from the $P_\psi^N(4312)^+[\bar D\Sigma_c]$ which can only decay to $\eta_c p$  by exchanging a $D^*$, $P_{\psi}^N[\bar D^*\Sigma_c]$ can decay to $\eta_c p$ through both the $D$ and $D^*$ mode. From \eref{E-L-etac}, the involved effective Lagrangians for $D\bar D^*\eta_c$ interaction reads
\begin{equation}
\begin{aligned}
{L}_{D\bar D^*\eta_c}  
&= - \i g_{D \bar D^*\eta_c}  D  \bar D^{*\alpha}   \partial_\alpha \eta^\dagger_c,
\end{aligned}
\end{equation}
and the Lagrangian for $\bar p\Sigma_cD$ interaction has been given in \eref{E-L-11}.

The corresponding Feynman diagram is similar with that for the decay to $J/\psi p$. The decay amplitude for $P_{\psi1/2}^N\to \eta_c p$ behaves as
\begin{align}
\i  \mathcal{A}_{31;1/2} & = \i^3\bar u_2  (g_{N\!D\Sigma_c} )\gamma_5 \int \frac{\d^4 k}{(2\pi)^4} [S(k_2) \Gamma^\gamma(k,r) D_{\beta\gamma}(k_1)] F^2D(k_3) (g_{ D\bar D^*\eta_c})(\i P_1^\beta ).
\end{align}
Performing the integral over $k$\,(see \ref{A-A31} for details) as before,  we obtain the amplitude for decay to $\eta_c p$ by form factor $s_{3}^1$ with a simple form
\begin{gather} \label{E-Am-etac}
\i \mathcal{A}[P_{\psi1/2}^N\!\xrightarrow{}\!\eta_c p(D)]  =G_{31} \bar u_2 s_{3}^1 \gamma_5 u,
\end{gather}
where we define 
\begin{gather}\notag
G_{31} =  g_{\!N\!D\Sigma_c} g_{D\!\bar D^*\eta_c} 
\end{gather}
to denote the strong interaction strength for $\bar D^{*-}\Sigma_c^{++}\to \eta_c p$ by one $D$ exchange.
The corresponding decay amplitude for $P^N_{\psi3/2} $ can then be expressed as
\begin{gather} \label{E-Am-4457-etac}
\i \mathcal{A}[P_{\psi3/2}^N\!\to\!\eta_c p(D)]  =G_{31}  \bar u_2  T_{31\gamma} u^\gamma (P,r)=G_{31}  \bar u_2  \left(  t_{3}^1 \hat P_{1\gamma}  \right) u^\gamma (P,r).
\end{gather}

The $D^*$ exchange involves the effective Lagrangian $D^*\bar D^*\eta_c$, which behaves
\begin{gather}
L_{D^*\bar D^*\eta_c^\dagger} =- g_{ D^*\bar D^*\eta_c} \frac{1}{M_{\eta_c}}\epsilon^{\alpha\beta\mu\nu}  {\partial}_\alpha D^*_\beta \bar D^*_\mu{\partial_\nu \eta^\dagger_c};
\end{gather}
while the Lagrangian for $\bar p\Sigma_c^{++}D^*$ has been given in \eref{E-L-11}.
The decay amplitude by $D^*$ exchange is then expressed as
\begin{align}
\i \mathcal{A}[P_{\psi1/2}^N\!\to\!\eta_cp(D^*)]  
 =& -\i G_{32} \bar u_2   \gamma^\nu  \int \frac{\d^4 k}{(2\pi)^4} [S(k_2) \Gamma^\gamma(k,r) D_{\beta\gamma}(k_1)] D_{\mu\nu}(k_3) \frac{\epsilon^{k_3\beta \mu P_1}}{M_1},
\end{align}
where we define
\begin{gather}\notag
G_{32} = g_{N\!D^*\Sigma_c}g_{ D^*\bar D^*\eta_c}
\end{gather}
to denote the strong interaction strength for $\bar D^{*-}\Sigma_c^{++}\to \eta_c p$ by one $D^*$ exchange.
By a similar calculation\,(see \ref{A-A32}), we express this amplitude by one form factor $s^2_{32}$ as 
\begin{gather}
\i\mathcal{A}[P_{\psi1/2}\to \eta_c p(D^*)] = G_{32}\bar u_2 T_{32} u(P,r) = G_{32}\bar u_2  (s_{3}^2 \gamma_5) u(P,r).
\end{gather}

The corresponding amplitude for $P_{\psi3/2}^N \!\to\! \eta_c p$ can be obtained by making a replacement
\begin{gather}
T_{32}u(P,r) \to T_{32\gamma} u^\gamma(P,r).
\end{gather}
And the by inserting the Salpeter wave function $\varphi_\beta[P_{\psi3/2}^N]=A_{\beta\gamma}u^\gamma$ and performing the integral, we obtain
\begin{gather}\label{E-ff-t32}
\i\mathcal{A}[P_{\psi3/2}\to \eta_c p(D^*)] = G_{32}\bar u_2 T_{32\gamma} u = G_{32}\bar u_2 ( t_{3}^2 \hat P_{1\gamma}) u.
\end{gather}

Combining the amplitudes for $D$ and $D^*$ exchange, we finally obtain
\begin{align}
\i\mathcal{A}[P_{\psi1/2}^N\!\to \eta_c p]  &= s_3  \bar u_2   \gamma_5 u, \\
\i\mathcal{A}[P_{\psi3/2}^N\!\to \eta_c p]  &= t_3   \bar u_2  \left(\hat P_{1\alpha} \right) u^\alpha,
\end{align}
where the form factors $s_3$ and $t_3$ behave
\begin{equation}
\begin{aligned}
s_3  &= G_{31}  s_{3}^1 + G_{32}  s_{3}^2,\\
t_3  &=  G_{31} t_{3}^1 +  G_{32}  t_{3}^2.
\end{aligned}
\end{equation}

\subsubsection{$P_\psi^N\to \bar D^0\Lambda_c^+$}

The $\bar D^0\Lambda^+_c$ decay channel is almost the same compared with the $\eta_cp$ channel, while the only differences are $\eta_c$ replaced by $\bar D^0$, $p$ replaced by $\Lambda^+_c$, and the propagators $D$ and $D^*$ replaced by the $\pi$ and $\rho$ in the pseudoscalar and vector exchange respectively.  From the Lagrangians in \eref{E-L-DPD} and \eref{E-L-DVD}, we can rewrite the relevant effective Lagrangians for $\bar D\pi\bar D^*$ and $\bar D\rho D^*$ interactions as 
\begin{equation}\label{E-L-4}
\begin{aligned}
L_{\bar D^0 \bar  D^{*-}\pi} &= -g_{\bar D^0 \bar D^{*-}\pi } \i \partial_\alpha \bar D^{0\dagger}   \pi \bar D^{*-\alpha}, \\
{L}_{\bar D^0\bar D^{*-}\rho } &= + g_{\bar D^0\bar D^{*-}\rho} \frac{\epsilon^{\alpha\beta\mu\nu}}{M_{\bar D^0}}  \partial_\alpha \rho_\beta \bar D^{*-}_\mu \partial_\nu \bar D^{0\dagger},
\end{aligned}
\end{equation}
where the coupling constants behave 
\begin{align*}
g_{\bar D^0 \bar D^{*-}\pi }&=(M_{\bar D^0} M_{\bar D^{*-}})^{\frac12} \frac{2g}{f},\\
g_{\bar D^0\bar D^{*-}\rho} &=R_{\braket{\bar D^0\bar D^{*-}}} M_{\bar D^0}\sqrt8 (\lambda g_V).
\end{align*}
The Lagrangians for the baryon sector, namely, $\Lambda^+_c\pi(\rho)\Sigma_c^{++}$ have been given in \eref{E-L-21} and \eref{E-L-22}. Also the contribution from other similar Lagrangians will be induced into the isospin factor. 

Comparing the above Lagrangians with those of the $\eta_cp$ channel,  we can express the the amplitude for $P_{\psi1/2}^N$ and $P_{\psi3/2}^N$ to $\bar D^0\Lambda^+_c$  as
\begin{gather}  \label{E-A4}
\i \mathcal{A}[P_{\psi1/2}^N \to \bar D^0\Lambda_c^+]  =  \bar u_2  \left( s_4 \gamma_5\right) u,\\
\i \mathcal{A}[P_{\psi3/2}^N\to \bar D^0\Lambda_c^+]  =   \bar u_2  \left(t_4 \hat P_{1\alpha} \right)  u^\alpha,
\end{gather}
where the two form factors are expressed as
\begin{equation}\label{E-s4-t4}
\begin{aligned}
s_4 &= G_{41}s_{4}^1 + G_{42}s_{4}^2,\\
t_4  &= G_{41}t_{4}^1 + G_{42}t_{4}^2,
\end{aligned}
\end{equation}
and we define two dimensionless constants 
\begin{align*}
G_{41} &= g_{\bar D\Sigma\bar D^*} g_{\Lambda_c^+\Sigma_c^{++}\pi} ,\\
G_{42} &= g_{\bar D^{0}\rho\bar D^{*-}} g_{\Lambda^+_c\rho\Sigma_c^{++}}
\end{align*}
to denote the strong interaction strength for $\bar D^{*-}\Sigma_c^{++}\to \bar D^0\Lambda_c^+$ scattering by one $\pi$ and one $\rho$ exchange respectively.

The form factors $s_{4}^1$ and $t_{4}^1$ denote the contribution from pion exchange, while $s_{4}^2$ and $t_{4}^2$ represent the contribution from $\rho$ exchange for $P_{\psi1/2}^N$ and $P_{\psi3/2}^N$ decays, respectively, which be obtained from the results of $P_\psi^N\to\eta_c p$ channel by making the following replacements,
\begin{equation}
\begin{aligned}
s_{4}^1&=s_{3}^1[M_1= M_{\bar D^0}, M_2= M_{\Lambda^+_c}, m_3= M_\pi],\\
t_{4}^1&=t_{3}^1[M_1= M_{\bar D^0}, M_2= M_{\Lambda^+_c}, m_3= M_\pi],\\
s_{4}^2 &= s_{3}^2[M_1=\bar D, M_2=M_{\Lambda_c}, m_3=M_\rho],\\
t_{4}^2  &= t_{3}^2[M_1=\bar D, M_2=M_{\Sigma_c}, m_3=M_\rho].
\end{aligned}
\end{equation}

\subsubsection{$P_\psi^N\to \bar D\Sigma_c$}
The $P_{\psi}^N$ can decay to both $ D^-\Sigma_c^{++}$ and $\bar D^0\Sigma_c^+$ channels. We will use $\bar D\Sigma_c$ to denote the sum of these two channels. We first consider the $D^-\Sigma_c^{++}$ channel. Besides the contributions from the $\pi$ and $\rho$ exchange, the $\omega$ exchange can also contribute in this channel compared with $\bar D^0\Lambda_c^+$ channel. The involved   Lagrangian  for meson sector read
\begin{equation}\label{E-L5M}
\begin{aligned}
L_{D^- D^{*-}\pi} &= +g_{D^- \bar D^{*-}\pi } \i \partial_\alpha D^{-\dagger}   \pi \bar D^{*-\alpha}, \\
{L}_{D^-D^{*-}\rho } &= - g_{D^- D^{*-}\rho} \frac{\epsilon^{\alpha\beta\mu\nu}}{M_{D^-}}  \partial_\alpha \rho_\beta D^{*-}_\mu \partial_\nu   D^{-\dagger},\\
{L}_{D^-D^{*-}\omega } &= + g_{D^- D^{*-}\omega} \frac{\epsilon^{\alpha\beta\mu\nu}}{M_{D^-}}  \partial_\alpha \omega_\beta D^{*-}_\mu \partial_\nu   D^{-\dagger},
\end{aligned}
\end{equation}
where the coupling constants behaves
\begin{align*}
g_{D^-  D^{*-}\pi }&=(M_{D^-} M_{D^{*-}})^{\frac12} \frac{\sqrt2g}{f},\\
g_{D^-  D^{*-}\rho} &=R_{\braket{D^-D^{*-}}} M_{D^-}2 (\lambda g_V),\\
g_{D^-  D^{*-}\omega} &=g_{D^-  D^{*-}\rho}.
\end{align*}
The involved Lagrangian for baryon sector read
\begin{equation} \label{E-L5B}
\begin{aligned}
L_{\bar\Sigma_c^{++}\Sigma_c^{++}\pi} &= - g_{\Sigma_c^{++}\Sigma_c^{++}\pi} \i  \bar \Sigma_c^{++}\gamma_5\pi\Sigma_c^{++},\\
L_{\bar\Sigma_c^{++}\Sigma_c^{++}\rho} & =-g_{\Sigma_c^{++}\Sigma_c^{++}\rho}\bar \Sigma_c^{++} \gamma_\alpha \rho^\alpha \Sigma_c^{++},\\
L_{\bar\Sigma_c^{++}\Sigma_c^{++}\omega} & =-g_{\Sigma_c^{++}\Sigma_c^{++}\omega}\bar \Sigma_c^{++} \gamma_\alpha \omega^\alpha \Sigma_c^{++},\\
\end{aligned}
\end{equation}
where the coupling constants read
\begin{equation}
\begin{aligned}
g_{\Sigma_c^{++}\Sigma_c^{++}\pi} & =\frac{1}{\sqrt2} (M_{\Sigma_c^{++}} + M_{\Sigma_c^{++}})\frac{g_1}{f},\\
g_{\Sigma_c^{++}\Sigma_c^{++}\rho} & = \frac{1}{2} g_V\beta_S,\\
g_{\Sigma_c^{++}\Sigma_c^{++}\omega} & = g_{\Sigma_c^{++}\Sigma_c^{++}\rho}.
\end{aligned}
\end{equation}
Again the contribution from other similar Lagrangian will be induced into the isospin factor. From isospin analysis, the isospin factor behaves $C_5=\frac{2\sqrt2}{\sqrt3}$ for $\pi$ and $\rho$ exchange,  and for the $\omega$ exchange the isospin factor is $\frac{\sqrt2}{\sqrt3}$. It is clear to see  from above Lagrangian that the $\rho$ and $\omega$ would make destructive interference in this decay.

Compared with the $\bar D^0\Lambda_c^+$ channel, the difference in $D^-\Sigma_c^{++}$ channels are the replacements of $\Lambda_c^+$ by $\Sigma_c^{++}$ for $\pi$ and $\rho$ exchanges. Then the amplitude for $P_\psi^N\to D^-\Sigma_c^{++}$ can be expressed as
\begin{equation}
\begin{aligned}
\i \mathcal{A}[P_{\psi1/2}^N\to D^-\Sigma_c^{++}] &=C_5 \bar u_2 (s_{5}\gamma_5) u(P,r),\\
\i \mathcal{A}[P_{\psi3/2}^N\to D^-\Sigma_c^{++}] &=C_5 \bar u_2 (t_{5}\hat P_{1\alpha}) u^\alpha(P,r),
\end{aligned}
\end{equation}
where the two form factors are
\begin{equation}
\begin{aligned}
s_{5}&=G_{51}s_{5}^1 + G_{52}s_{5}^2 - \frac12G_{54}s_{5}^4,\\
t_{5}&=G_{51}t_{5}^1\hspace{1pt} +  G_{52}t_{5}^2  \hspace{1.5pt}- \frac12G_{54}t_{5}^4,
\end{aligned}
\end{equation}
where $\frac12$ comes from the relative isospin factor aforementioned.
As usual we also define three dimensionless constants
\begin{equation}\notag
\begin{aligned}
G_{51} &= g_{\Sigma_c^{++}\Sigma_c^{++}\pi} g_{D^{-} D^{*-}\pi}, \\
G_{52} &= g_{\Sigma_c^{++}\Sigma_c^{++}\rho} g_{D^{-} D^{*-}\rho},\\
G_{54} &= g_{\Sigma_c^{++}\Sigma_c^{++}\omega} g_{D^{-} D^{*-}\omega},
\end{aligned}
\end{equation}
to denote the strong interaction strength for $\bar D^{*-}\Sigma_c^{++}\to D^-\Sigma_c^{++}$ by one pion, one $\rho$ and one $\omega$ exchange, respectively, and $G_{54}=G_{52}$ in the chiral limit. 

The form factors $s_{5}^1$, $t_{5}^1$ denote the contributions from pion exchange while $s_{5}^2$ and $t_{5}^2$ from $\rho$ exchange, and $s_{5}^4$ and $t_{5}^4$ for $\omega$ exchange for $P_{\psi1/2}^N$ and $P_{\psi3/2}^N$ decay, respectively, which can be obtained by making the following replacements 
\begin{equation}
\begin{aligned}
s_{5}^1 &=s_{3}^{1}[M_1\to M_{D^-}, M_2\to M_{\Sigma_c^{++}}, m_3\to M_\pi],\\
t_{5}^1 &=t_{3}^{1}[M_1\to M_{D^-}, M_2\to M_{\Sigma_c^{++}}, m_3\to M_\pi];\\
s_{5}^2 &=s_{3}^{2}[M_1\to M_{ D^-}, M_2\to M_{\Sigma_c^{++}}, m_3\to M_\rho],\\
t_{5}^2 &=t_{3}^{2}[M_1\to M_{ D^-}, M_2\to M_{\Sigma_c^{++}}, m_3\to M_\rho];\\
s_{5}^4 &=s_{3}^{2}[M_1\to M_{ D^-}, M_2\to M_{\Sigma_c^{++}}, m_3\to M_\omega],\\
t_{5}^4 &=t_{3}^{2}[M_1\to M_{ D^-}, M_2\to M_{\Sigma_c^{++}}, m_3\to M_\omega].
\end{aligned}
\end{equation}

From an isospin analysis, we can see that the decay amplitude for $\bar D^0\Sigma_c^+$ channel is just the same as that for $\bar D^-\Sigma_c^{++}$ channel except for the isospin factor $C_6[\bar D^0\Sigma_c^+]=\frac1{\sqrt2}C_5[\bar D^-\Sigma_c^{++}]$. And then within the isospin symmetry, the decay width for  $P_\psi^N$ to $\bar D^0\Sigma_c^{++}$ channel would be $\frac12$ of that for $D^-\Sigma_c^{+}$ channel.

\subsection{$P_\psi^N\to (P+B^*)$}
$P_{\psi}^N[\bar D^*\Sigma_c]$ can also decay to $ \bar D\Sigma_c^{*}$ channel with the final particles being a pseudoscalar meson and a $\frac32$ baryon, which can be realized by one $\pi$, $\rho$ or $\omega$ exchange. The isospin factor for $P_\psi^N\to \bar D\Sigma_c^*$ is exactly the same as that for $\bar D\Sigma_c$ channel, namely, $C_7[D^-\Sigma_c^{*++}]=\sqrt2C_8[\bar D^0\Sigma_c^{*+}]=C_5$.

\subsubsection{$P_\psi^N\to \bar D\Sigma_c^*$}

We first consider the $D^-\Sigma_c^{*++}$ channel and then the result for $\bar D^0\Sigma_c^+$ channel can be obtained by isospin factor directly.
The effective Lagrangian for $D^- D^{*-}\pi$ has been given in \eref{E-L5M}, while the Lagrangian for $\Sigma_c^{*++}\Sigma_c^{++}\pi$ reads
\begin{gather}
L_{\bar\Sigma_c^{*++}\Sigma_c^{++} \pi}=+g_{\bar\Sigma_c^{*++}\Sigma_c^{++} \pi}  \frac{1}{M_{\Sigma^*_c}}
\bar \Sigma_c^{*++\alpha}\partial_\alpha\pi \Sigma_c^{++},
\end{gather}
where we divided a $M_{\Sigma_c^*}$ to keep the coupling constant dimensionless, and now
\begin{align}\notag
g_{\bar\Sigma_c^{*+}\Sigma_c^{++} \pi} = \frac{\sqrt3}{2\sqrt2} \frac{g_1}{f}M_{\Sigma_c^*}
\end{align}
Again contributions from other Lagrangian, such as $\bar\Sigma_c^{*+}\pi\Sigma_c^{++}$ will be collected into the isospin factor. Then by a similar calculation\,(\ref{A-A71}) as before, the amplitude by $\pi$ exchange can be expressed by the form factor as
\begin{gather}
\i\mathcal{A}_{71;1/2} = G_{71} \bar u^\alpha T_{71\alpha} u =G_{71} \bar u^\alpha  ( s_{7}^1  \hat P_\alpha) u,  
\end{gather}
where we defined 
\begin{gather}\notag
G_{71} = g_{\Sigma_c^{*++}\Sigma_c^{++}\pi} g_{D^-D^{*-}\pi}
\end{gather}
to denote the strong decay strength for $\bar D^{*-}\Sigma_c^{++}\!\to\! D^{-}\Sigma_c^{*++}$ scattering by one pion exchange.

For $P_{\psi3/2}^N \to  D^-\Sigma_c^{*++}$ by $\pi$ exchange, the amplitude can be represented by three form factors as
\begin{gather}
\i\mathcal{A}_{71;3/2}= G_{71} \bar u_\alpha T_{71}^{\alpha\gamma} u_\gamma =  G_{71} \bar u_\alpha\left( \i t_{71}^1 {\epsilon^{\alpha\beta \hat P\hat P_1}} +  ( t_{72}^1 g^{\alpha\beta}  +t_{73}^1\hat P^\alpha \hat P_1^\beta  ) \gamma_5  \right) u_\gamma.
\end{gather}

The $D^-\Sigma_c^{*++}$ channel by $\rho$ and $\omega$ exchange also involves the $\Sigma_c^*\Sigma_c\rho(\omega)$ interaction which can be described by the effective Lagrangian
\begin{equation}
\begin{aligned}
L_{\Sigma_c^{*++}\Sigma_c^{++}\rho} &=+ \i g_{\Sigma_c^{*++}\Sigma_c^{++}\rho}   \frac{1}{M_{\Sigma_c^{*}}} \bar \Sigma_c^{*++\beta} \gamma^\alpha\gamma^5(\partial_\alpha \rho_\beta -\partial_\beta \rho_\alpha) \Sigma_c^{++},\\
L_{\Sigma_c^{*++}\Sigma_c^{++}\omega} &=+ \i g_{\Sigma_c^{*++}\Sigma_c^{++}\omega}   \frac{1}{M_{\Sigma_c^{*}}} \bar \Sigma_c^{*++\beta} \gamma^\alpha\gamma^5(\partial_\alpha \omega_\beta -\partial_\beta \omega_\alpha) \Sigma_c^{++},
\end{aligned}
\end{equation}
where we have divided a $M_{\Sigma_c^*}$ to keep the coupling constant $ g_{\Sigma_c^{*++}\Sigma_c^{++}\rho(\omega)}$ dimensionless, and then 
\begin{align}
g_{\Sigma_c^{*++}\Sigma_c^{++}\rho} =g_{\Sigma_c^{*++}\Sigma_c^{++}\omega} = \frac{1}{\sqrt{12}} M_{\Sigma_c^*} g_V \lambda_S.
\end{align}
The Lagrangian for $D^- D^{*-}\rho(\omega)$ has been presented in \eref{E-L5M}. Now we first consider the $\rho$ exchange and $\omega$ exchange can be obtained by a simple replacement.
The corresponding decay amplitude can then be expressed by one form factor as\,(see \ref{A-A72})
\begin{gather}
\i\mathcal{A}[P_{\psi1/2}^N\to\bar D^-\Sigma_c^{*++}(\rho)] =G_{72}\bar u^{\alpha}(P_2,r_2) T_{72\alpha} u =G_{72}\bar u^{\sigma}(P_2,r_2)\left(   s_{7}^2\hat P_\alpha  \right) u. 
\end{gather}
As usual we define two dimensionless constants
\begin{equation}\notag
\begin{aligned}\notag
G_{72} &=  g_{\Sigma_c^{*++}\Sigma_c^{++}\rho} g_{D^- D^{*-}\rho},\\
G_{74} &=  g_{\Sigma_c^{*++}\Sigma_c^{++}\omega} g_{D^- D^{*-}\omega}
\end{aligned}
\end{equation}
to represent the strong interaction strength for $D^{*-}\Sigma_c^{++}\to  D^-\Sigma_c^{*++}$ by one $\rho(\omega)$ exchange.  

For $P_{\psi3/2}^N \to \bar D^-\Sigma_c^{*++}$ decay by $\rho$ exchange, the calculations are very similar and we just need make a replacement
\begin{gather}
T_{72}^\alpha u \to T_{72}^{\alpha\beta} u_\beta.
\end{gather}
And then  the amplitude can be further express by three form factor as
\begin{gather}
\i\mathcal{A}[P^N_{\psi3/2}\to \bar D^-\Sigma^{*++}_c(\rho)] =G_{72} \bar u_\alpha \left(\i t_{71}^2 {\epsilon^{\alpha\beta \hat P\hat P_1}} + t^2_{72} g^{\alpha\beta}\gamma_5 +t_{73}^2\hat P^\alpha \hat P_1^\beta \gamma_5 \right) u_\beta.
\end{gather}

The form factors for $\omega$ exchange can be obtained from those for $\rho$ exchange by making replacement $m_\rho\to m_\omega$.
Finally, combining the contributions from the $\pi$,  $\rho$ and $\omega$ exchanges, we obtain the amplitude for $P_\psi^N$ to $D^-\Sigma_c^{*++}$ channel as
\begin{gather}
\i\mathcal{A}[P^N_{\psi1/2}\to D^-\Sigma_c^{*++}] = \bar u^\alpha_2 (s_{7}\hat P_\alpha)u, \\
\i\mathcal{A}[P^N_{\psi3/2}\to D^-\Sigma_c^{*++}] = \bar u_{2\alpha} \left(\i t_{71} \epsilon^{\alpha\beta \hat P\hat P_1} + t_{72} g^{\alpha\beta}\gamma_5 +t_{73}\hat P^\alpha \hat P_1^\beta \gamma_5 \right)u_\beta, 
\end{gather}
where $s_7$ and $t_{7i}\,(i=1,2,3)$ behaves
\begin{equation}
\begin{aligned}
s_7 &  = G_{71}s_{6}^1 \hspace{1.5pt}+ G_{72}s_{7}^2  \hspace{1.5pt}-\frac12 G_{74} s_{7}^4 , \\
t_{7i}& = G_{71} t_{7i}^1 + G_{72} t_{7i}^2 -\frac12 G_{74} t_{7i}^4.
\end{aligned}
\end{equation}
with the form factors from $\omega$ exchange reading
\begin{equation}
\begin{aligned}
s_{7}^4 &= s_{7}^2[m_3 \to m_\omega] , \\
t_{7i}^4 & = t_{7i}^2[m_3\to m_\omega].
\end{aligned}
\end{equation}

\subsection{Partial decay widths}

To obtain the partial decay widths, we first square the amplitude and then sum all the polarization states. Then the two-body partial decay width is expressed as
\begin{gather}
\Gamma[P_\psi^N \!\to \! M_{P(V)}B^{(*)}] = \frac{|\bm{P}_1|}{8\pi M^2}  C^2_i\frac{1}{2J+1} \sum |\mathcal{A}|^2 ,
\end{gather} 
where $J$ denotes spin of the initial $P_\psi^N$ state;  the three-momentum of the final meson is given by
\begin{gather}
|\bm{P}_1| =\frac{1}{2M}\sqrt{[(M^2-(M_1+M_2)^2] [M^2-(M_1-M_2)^2)]};
\end{gather}
$C^2_{i}$ denotes the isospin factor, which is $C^2_1=C^2_2=C^2_3=C^2_4=\frac32$ and $C^2_5=C^2_7=\frac83$, and $C_6^2=C_8^2=\frac43$;  the summation is over all the spin states of the involved initial and final particles. The specific forms of $\sum |\mathcal{A}|^2$ depend on both the initial $P_\psi^N$ state and the final decay mode, which are collected in the appendix \ref{A-3}.

\section{Numerical results and discussions}\label{Sec-4}
\subsection{Numerical parameters}
Before giving the decay widths, we first summarize the numerical values used in this work, including the hadron masses, decay constants and coupling constants in effective Lagrangians involved. The hadron masses involved are listed in \autoref{T-Hadron-Mass}.
\begin{table}[ht]
\caption{Hadron mass used in this work in units of GeV\,\cite{PDG2022}.}
\label{T-Hadron-Mass}
\vspace{0.2em}\centering
\begin{tabular}{ ccccccccccccc }
\toprule[1.5pt]
$P_\psi^N(4440)$	&$\bar D^0$    &$\psi$  &$\eta_c$  &$p$  &$\Lambda_c^+$  &$\Sigma_c^{++}$    & $\Sigma_c^+$  &$\Sigma_c^{*++}$  &$\Sigma_c^{*+}$ \\
4.440  & 1.865 & 3.097 & 2.983 & 0.938  &2.286  & 2.454& 2.453   &2.518 &2.517\\ 
 \midrule[1.2pt]
$P_\psi^N(4457)$  &$\bar D^{*0}$  &$D^-$  &$D^{*-}$  &$\pi^0$ &$\pi^+$  &$\sigma$  &$\eta$  &$\rho$ &$\omega$\\
4.457 &2.007 &1.868 & 2.010  & 0.135 &0.140   &$0.45$ &$0.548$ &0.775 &0.783\\
\bottomrule[1.5pt]
\end{tabular}
\end{table}
While the masses of the mediator mesons are always isospin-averaged. 

The pion decay constant we used is $f=0.132\,\si{GeV}$, while the $J/\psi$ decay constant $f_\psi=0.416\,\si{GeV}$ is estimated from the dilepton decay width\,\cite{PDG2022}.
The chiral parameters between the singly heavy hadron and the light bosons are obtained under the heavy quark symmetry and the chiral symmetry\,\cite{Casalbuoni1997,YangZC2012,LiuYR2012,ChenR2015,ChenR2019}, which are listed in \autoref{T-chiral-c}. Notice the coupling constant is directly related to the specific form of the Lagrangian used. 
\begin{table}[ht]
\caption{The chiral constants used are from the previous literatures\,\cite{Casalbuoni1997,YangZC2012,LiuYR2012,ChenR2015,ChenR2019}.}
\label{T-chiral-c}
\vspace{0.2em}\centering
\begin{tabular}{ ccccccccccccc }
\toprule[1.5pt]
$g_s$  &$g$  &$g_V\beta $  &$g_V\lambda $  &$l_S$ &$g_1$  &$g_V\beta_S$  &$g_V\lambda_S$  &$g_4$ &$g_V\lambda_I $\\
0.76 &0.59  &5.22 &3.25\,GeV$^{-1}$ & 6.2 &0.94 & 10.09 & 19.2\,GeV$^{-1}$ & 1.0 & $  g_V\lambda_S/\sqrt8$\\
\bottomrule[1.5pt]
\end{tabular}
\end{table}
From these chiral parameters, we obtain the coupling constants associated with the specific Lagrangians used are
\begin{equation}\notag
\begin{aligned}
g_{\bar D^{*0}\bar D^{*-}\pi} &= \frac{2g}{f} M_{\bar D^{*0}} = 17.94, &g_{\Lambda^+_c\Sigma_c^{++}\pi} &= \frac{1}{\sqrt3}(M_{\Lambda_c^+} + M_{\Sigma_c^{++}}) \frac{g_4}{f} = 20.73,\\
g_{\bar D^{*0}\bar D^{*-}\rho}&= \sqrt8 M_{\braket{\bar D^{*0}\bar D^{*-}}} ( g_V\lambda) = 18.47,&
g_{\Lambda_c^+ \Sigma_c^{++}\rho} &=\frac{\sqrt2}{\sqrt3}(M_{\Lambda_c^+} + M_{\Sigma_c^{++}})(\lambda_I g_V)=26.27,\\
g_{\bar D^0 \bar D^{*-}\pi } &= (M_{\bar D^0} M_{\bar D^{*-}})^{\frac12} \frac{2g}{f} = 17.32,&
g_{\bar D^0\bar D^{*-}\rho} & =R_{\braket{\bar D^0\bar D^{*-}}}M_{\bar D^0}\sqrt8 (\lambda g_V) = 17.16, \\
g_{D^-D^{*-}\pi} &= \frac1{\sqrt2}g_{\bar D^0D^{*-}\pi}=12.25,&
g_{\Sigma_c^{++}\Sigma_c^{++}\pi} & =\frac{1}{\sqrt2} (M_{\Sigma_c^{++}} + M_{\Sigma_c^{++}})\frac{g_1}{f}=24.71,\\
g_{D^-D^{*-}\rho} &= \frac1{\sqrt2}g_{\bar D^0D^{*-}\rho}=12.13,&
g_{\Sigma_c^{++}\Sigma_c^{++}\rho} & = \frac{1}{2}  g_V\beta_S=5.05,\\
g_{D^-D^{*-}\omega} &= g_{D^-D^{*-}\rho}, & g_{\Sigma_c^{++}\Sigma_c^{++}\omega} &=g_{\Sigma_c^{++}\Sigma_c^{++}\rho},\\
g_{\bar\Sigma_c^{*++}\Sigma_c^{++}\pi} &= \frac{\sqrt3}{\sqrt8} \frac{g_1}{f}M_{\Sigma_c^*}=10.98,&
g_{\Sigma_c^{*++}\Sigma_c^{++}\rho} &= g_{\Sigma_c^{*++}\Sigma_c^{++}\omega}=\frac{1}{\sqrt{12}} M_{\Sigma_c^*} \lambda_S g_V=13.95.
\end{aligned}
\end{equation}

 Then in the heavy quark limit, from the relationship in \eref{E-g-DDpsi} the coupling constants for the $D^{(*)} \bar D^{(*)}$ pair and charmonia read
\begin{equation}\notag
\begin{aligned}
g_{D\!\bar D\psi}&= 14.89,   ~~g_{D\!\bar D^*\psi}=15.43, ~~g_{D^*\bar D^*\psi}=8.01,
 ~~g_{D\!\bar D^*\eta_c}=7.58, ~~g_{D^*\bar D^*\eta_c}=15.72.
\end{aligned}
\end{equation}
Combined with the total amplitude \eref{E-A}, it can be found that the partial decay widths for $J/\psi(\eta_c)p$ are proportional to $\frac{1}{f_\psi^2}$. The coupling constants for $ND^{(*)}\Sigma_c$ used are\,\cite{Garzon2015} 
\begin{gather}\notag
g_{\!N\!D\Sigma_c}=2.69,~~~~g_{\!N\!D^*\Sigma_c}=4.24.
\end{gather}

Combining above coupling constants and the mass parameters, we finally obtain the dimensionless strong interaction strength  $G_{mn}$ defined for the eight decay channels in the previous section, which are listed in \autoref{T-G} with $G_{54}=G_{64}=G_{52}$ and $G_{74}=G_{84}=G_{72}$.
\begin{table}[ht]
\caption{Dimensionless strong interaction strength $G_{mn}$.}
\label{T-G}
\vspace{0.2em}\centering
\begin{tabular}{ cccccccccccrc }
\toprule[1.5pt]
$J/\psi p$  &$\bar D^*\Lambda_c^+$ &$\eta_cp$  &$\bar D^0\Lambda_c^+$  &$\bar D^0\Sigma_c^+$  &$\bar D^-\Sigma_c^{++}$  &$\bar D^0\Sigma_c^{*+}$ &$\bar D^-\Sigma_c^{*++}$ \\
 \midrule[1.2pt]
$G_{11}$  &$G_{21}$  &$G_{31}$  &$G_{41}$  &$G_{51}$  &$G_{61}$  &$G_{71}$ &$G_{81}$  \\
41.5          & 371.9       & 20.4         & 359.0       & 302.7       &$G_{51}$  & 134.5      & $G_{71}$ \\
 \midrule[1.2pt]
$G_{12}$  &$G_{22}$ &$G_{32}$  &$G_{42}$  &$G_{52}$  &$G_{62}$  &$G_{72}$ &$G_{82}$   \\
34.0          &485.2       &66.7        & 450.8       & 61.3         &$G_{52}$  &169.2       &$G_{72}$     \\
\bottomrule[1.5pt]
\end{tabular}
\end{table}
These values are the standard parameters used in this work. 

It should be pointed out that most of the involved coupling constants are just rough estimations with large uncertainties, and the obtained decay rates make sense for an order-of-magnitude estimations.

\subsection{Numerical results and discussions}
\begin{figure}[h!]
\vspace{0.5em}
\centering
{\includegraphics[width=0.6\textwidth]{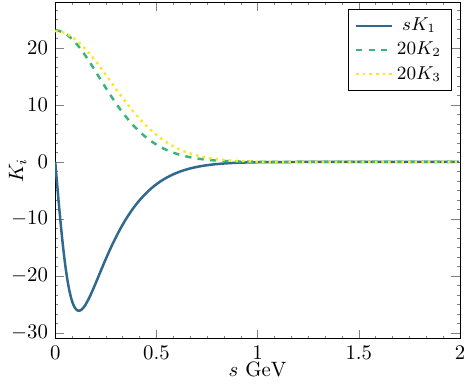} \label{F-V-4440}}
\caption{The interaction kernel  $K_i~(i=1,\cdots,3)$ in the isospin-$\frac12$ with  $m_\Lambda=0.64\,\si{GeV}$.}\label{F-Ki}
\end{figure}
The  introduced free parameters in this work are the relative strength $A_s$ in \eref{E-As} and the regulator $m_\Lambda$ in the form factor $F({s}^2)$ in \eref{E-form-factor}. By solving the relevant BS eigenvalue equations, we find proper values of $A_s$ and $m_\Lambda$ can produce bound states of $\bar D^*\Sigma_c$ based on the one-boson exchange kernel in  isospin-$\frac12$. First by considering all the three pentaquark states $P_\psi^N(4312)^+$, $P_\psi^N(4440)^+$ and $P_\psi^N(4457)^+$, we find that it is difficult to produce all these three bound states when $A_s$ is less than $1.65$, especially for $P_\psi^N(4312)^+$.  The obtained results also show that the mass of the $J^P={\frac{1}{2}}^-$ state is always higher than that of the ${\frac32}^-$ one. By analyzing the eigenvalue equation, it is found that the former one is dominated by the $P$-wave component while the last one is the standard $S$-wave ground state, which is also graphically shown in the following wave functions. Then by fitting the mass splitting of the two $[\bar D^*\Sigma_c]$ bound states to the experimental data $\Delta M\sim 15$\,MeV,  we fix $A_s=1.75$ and $m_\Lambda=0.64$\,GeV. The obtained mass spectra and the corresponding $m_\Lambda$ are also listed in \autoref{T-mA}. 
\begin{table}[ht]
\caption{Mass spectra of the $[\bar D^*\Sigma_c]$ bound states and the corresponding cutoff $m_\Lambda$ defined in \eref{E-form-factor} in units of $\si{GeV}$.}
\label{T-mA}
\vspace{0.2em}\centering
\begin{tabular}{ c|cc|cccc||cccccccc }
\toprule[1.5pt]
$J^P$   &$\frac12^-$	& $\frac32^-$	& $m_\Lambda$  \\
 \midrule[1.2pt]
Mass    &4.457        &	 4.442                    & 0.64   \\
\bottomrule[1.5pt]
\end{tabular}
\end{table}
Under the parameters fixed above, the obtained $K_i$ in the interaction kernel are graphically shown in \autoref{F-Ki}. Notice that the Lorentz structures of $K_1$, $K_2$ and $K_3$ are pseudo-scalar, scalar and vector, respectively.

With $m_\Lambda=0.64\,\si{GeV}$ the obtained masses are $M[P^N_{\psi1/2}]=4.457\,\si{GeV}$ and $M[P_{\psi3/2}^N]=4.442\,\si{GeV}$, which are consistent with the experimental results  $P_\psi^N(4457)^+$ and $P_\psi^N(4440)^+$ reported by LHCb within just a few MeV. Though the obtained bound state mass is dependent on the value of $m_\Lambda$,  the mass splitting between $\frac12^-$ and $\frac32^-$ is always greater than zero when a universal $m_\Lambda$ is set.  Namely, the obtained numerical results show that the mass of $J^P=\frac12^-$ state is larger than that of the $J^P=\frac32^-$ one. This conclusion is consistent with the those in refs.\,\cite{Yamaguchi2020,DuML2020,DuML2021,LiuMZ2021,Yalikun2021,PengFZ2024}. The obtained mass spectra of the $I=\frac12$ $(\bar D^*\Sigma_c)$ molecules also indicate that there does not exist any other bound states below the $\bar D\Sigma_c$ threshold in both $J^P=\frac12^-$ and $J^P=\frac32^-$. The mass ordering that $\frac12^-$ state is higher than the $\frac32^-$ one is mainly caused that the former is dominated by the $P$-wave components while the later one is dominated by the usual $S$-wave components, which can also be see from the wave function shapes.

\begin{figure}[h!]
\vspace{0.5em}
\centering
\includegraphics[width=0.49\textwidth]{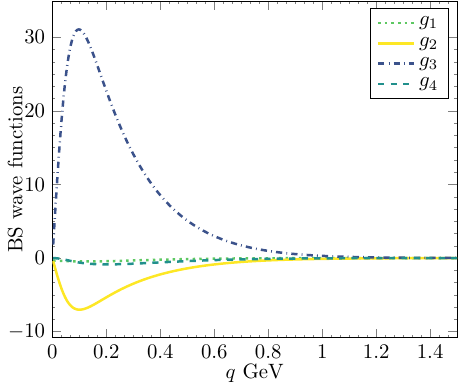} \label{F-wave-J1-2}
\includegraphics[width=0.5\textwidth]{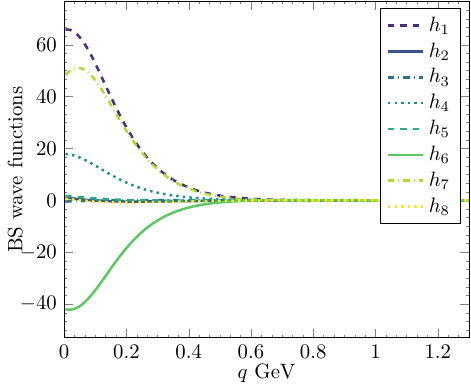} \label{F-wave-J3-2}
\caption{The BS wave functions of  $P_{\psi1/2}^N$ with mass  $4.457\,\si{GeV}$\,(left) and $P_{\psi3/2}^N$ with mass $4.442\,\si{GeV}$\,(right), where two bound states masses are solved with $m_\Lambda=0.64\,\si{GeV}$.}\label{F-wave}
\vspace{0.5em}
\end{figure}
In  \autoref{F-wave} we show the obtained radial BS wave functions $g_i\,(i=1,2,3,4)$ for $P_{\psi1/2}^N(4457)$ (left) and $h_n\,(n=1,\cdots,8)$ for $P_{\psi3/2}^N(4442)$\,(right). When $m_\Lambda$ decreases, the ground state mass increases and the wave functions shifts towards the left. It is clear to see that the $\frac12^-$ bound state is dominated by the $P$-wave $g_3$ and $g_2$ components, while the $\frac32^-$ one is dominated by the $S$-wave $h_1$ and the $D$-wave $h_6$ and $h_7$ with nonnegligible $S$-$D$ mixing effects. The features of the wave functions have important effects in both the mass spectra and decay behaviors, which make the $\frac12^-$ state higher and broader than the $\frac32^-$ one. 

Inserting the numerical wave functions into the expressions of the decay amplitudes, we obtain the numerical form factors for $P_\psi^N$ decays, which are listed in \autoref{T-form-s} and \autoref{T-form-t} with $m_\Lambda=0.64\,\si{GeV}$. The form factors are the same for $\bar D^{0}\Sigma_c^{(*)+}$ and $D^{-}\Sigma_c^{(*)++}$. Since the coupling constants for $\rho$ and  $\omega$ exchange are the same and the mass difference are also quite slight, the form factors for $\rho$ and $\omega$ exchange are almost the same and then are not copied in the tables.
\begin{table}[ht]
\caption{Numerical values of strong decay form factors for $P_{\psi1/2}^N$ with $m=1,\cdots,8$ denoting the 8 decay channels we calculated and $n=1,2$ representing the pseudoscalar and vector meson exchange, respectively.
}\label{T-form-s}
\setlength{\tabcolsep}{3pt}
\vspace{0.2em}\centering
\begin{tabular}{ l|rrrrrrrrr }
\toprule[1.5pt]
\multicolumn{1}{c|}{$s^n_{mo}$} 	&\multicolumn{1}{c}{$J/\psi p$} 				&\multicolumn{1}{c}{$\bar D^{*0}\Lambda_c^+$}  &\multicolumn{1}{c}{$\eta_cp$}			&\multicolumn{1}{c}{$\bar D^{0}\Lambda_c^+$}	&\multicolumn{1}{c}{$\bar D\Sigma_c$}	  &\multicolumn{1}{c}{$\bar D\Sigma_c^*$}\\
\midrule[1.2pt]
$s^1_{m1}$		&$ 1.7\times10^{-5}$ &$-7.8\times10^{-4}$ 		&$2.9\times10^{-4}$ 		&$3.4\times10^{-3}$			&$7.1\times10^{-3}$  	&$-5.5\times10^{-4}$ \\
$s^2_{m1}$		&$3.0\times10^{-4}$	&$7.9\times10^{-3}$ 			&$-5.9\times10^{-5}$ 	&$-2.4\times10^{-4}$			&$-5.4\times 10^{-4}$	&$-2.1\times10^{-3}$ \\
$s^1_{m2}$		&$3.5\times10^{-5}$ 	&$1.7\times10^{-3}$  		&$-$ 					&$-$ 						&$-$ 				&$-$ \\
$s^2_{m2}$		&$-2.7\times10^{-3}$ &$-2.6\times10^{-2}$ 		&$-$ 					&$-$ 						&$-$ 				&$-$ \\
\bottomrule[1.5pt]
\end{tabular}
\end{table}
\begin{table}[ht]
\caption{Numerical values of form factors $t^n_{mo}$ for $P_{\psi3/2}^N$ decays.
}\label{T-form-t}
\vspace{0.2em}\centering
\begin{tabular}{ l|rrrrrrrrr }
\toprule[1.5pt]
\multicolumn{1}{c|}{$t^n_{mo}$} 	&\multicolumn{1}{c}{$J/\psi p$} 				&\multicolumn{1}{c}{$\bar D^{*0}\Lambda_c^+$}  &\multicolumn{1}{c}{$\eta_cp$}			&\multicolumn{1}{c}{$\bar D^{0}\Lambda_c^+$}	&\multicolumn{1}{c}{$\bar D\Sigma_c$}	  &\multicolumn{1}{c}{$\bar D\Sigma_c^*$}\\
\midrule[1.2pt]
$t^1_{m1}$	&$1.2\times10^{-4}$ 		&$1.1\times10^{-2}$ 			&$5.2\times10^{-4}$ 	&$3.7\times10^{-3}$ 			&$2.0\times10^{-2}$   &$-2.1\times10^{-5}$ \\
$t^2_{m1}$	&$-2.3\times10^{-4}$ 	&$4.8\times10^{-4}$ 			&$1.0\times10^{-4}$ 	&$7.4\times10^{-4}$ 			&$2.1\times10^{-3}$	  &$8.2\times10^{-3}$ \\
$t^1_{m2}$	&$-2.3\times10^{-5}$ 	&$-5.1\times10^{-3}$ 		&$-$ 				&$-$ 						&$-$ 		   		  &$3.4\times10^{-4}$ \\
$t^2_{m2}$	&$-1.3\times10^{-3}$ 	&$2.9\times10^{-2}$ 			&$-$ 				&$-$ 						&$-$ 		  		  &$-8.3\times10^{-3}$ \\
$t^1_{m3}$	&$-8.0\times10^{-5}$ 	&$-7.1\times10^{-4}$ 		&$-$ 				&$-$ 						&$-$ 				  &$-4.0\times10^{-2}$ \\
$t^2_{m3}$	&$-9.7\times10^{-4}$ 	&$4.9\times10^{-3}$ 			&$-$ 				&$-$ 						&$-$ 				  &$-8.4\times10^{-2}$ \\
$t^1_{m4}$	&$1.5\times10^{-3}$ 		&$2.1\times10^{-1}$ 			&$-$ 				&$-$ 						&$-$ 				  &$-$ \\
$t^2_{m4}$	&$5.3\times10^{-3}$ 		&$-9.0\times10^{-2}$ 		&$-$ 				&$-$ 						&$-$ 				  &$-$ \\
\bottomrule[1.5pt]
\end{tabular}
\end{table}

Inserting above form factors into the decay width expressions, we obtain the partial decay widths for $P_\psi^N(4440)^+$ and $P_\psi^N(4457)^+$. The obtained partial widths are listed in \autoref{T-width}. We present the results for two different scenarios of the $J^P$ configuration, where $P_\psi^N(4440)^+$ and $P_\psi^N(4457)^+$ are taken as $J^P={\frac32}^-$ and ${\frac12}^-$ respectively in the scenario I, and are just opposite in the scenario II. 
In scenario I, the calculated total widths of $P_\psi^N(4440)^+$ and $P_\psi^N(4457)^+$ are $21.8$\,\si{MeV} and $13.0$\,\si{MeV} respectively, while in scenario II the results are 13.0\,\si{MeV} and $18.5$\,MeV respectively. The calculated  width for $P_\psi^N(4440)^+$ is larger than that for $P_\psi^N(4457)^+$ in scenario I while is smaller for II. On the other hand, the total widths reported by the LHCb are 
\[\textstyle
\Gamma[{P_\psi^N(4440)^+}] =20.6\pm4.9^{+8.7}_{-10.1}\,\si{MeV},~~~~\Gamma[{P_\psi^N(4457)^+}] =6.4\pm2.0^{+5.7}_{-1.9}\,\si{MeV}.
\]
Considering the theoretical errors in the coupling constants of effective Lagrangians and the approximation of heavy and chiral quark symmetry, the obtained results in scenario I are more consistent with the experimental data of LHCb. Namely, our results of the calculated total widths are more favor of the $\frac32^-$ and $\frac12^-$ configuration for $P_\psi^N(4440)^+$ and $P_\psi^N(4457)^+$, which is also consistent with the mass ordering  obtained in this work.  This spin-parity configuration is also favored by the conclusion obtained in refs.\,\cite{Yamaguchi2020,DuML2020,DuML2021,LiuMZ2021,Yalikun2021,PengFZ2024}. 
\begin{table}[ht]
\caption{Comparison of the obtained decay widths for  $P_\psi^N(4440)^+$ and $P_\psi^N(4457)^+$ with other works in units of MeV. The assuming $J^P$ configuration of $P_\psi^N(4440)^+$ and $P_\psi^N(4457)^+$ are $\frac32^-$ and $\frac12^-$ respectively in scenario I, and are exact opposite in scenario II. In refs.\,\cite{YangZY2024,LinYH2019}, the two columns represent the results under  the $\frac12^-$ and $\frac32^-$ spin-party configuration respectively.
}\label{T-width}
\vspace{0.2em}\centering
\setlength{\tabcolsep}{4pt}%
\begin{tabular}{ l|cccccccccc }
\toprule[1.5pt]
\multicolumn{1}{c|}{Channel} 		    									& I						&II					&\cite{YangZY2024} &\cite{LinYH2019} &LHCb\,\cite{LHCb2019-Pc}	  \\
 \midrule[1.2pt]
 $P_\psi^N(4440)^+\!\to\!J/\psi p$ 	  	          		& $6.8\times10^{-3}$ 		&$\sn{1.7}{-2}$		&$4.1-4.1$		&$0.03-0.02$ 		&- \\
 $P_\psi^N(4440)^+\!\to\!\bar D^{*0}\Lambda_c^+$ 	& $\sn{8.8}{-1}$			&$7.6$    		 	&- 		 		&$13.9-6.2$  			&-\\
$P_\psi^N(4440)^+\!\to\!\eta_c p$ 	  	    			& $\sn{1.9}{-1}$    		&$\sn{3.5}{-1}$		&- 				&$0-0$ 	 	 		&-\\
$P_\psi^N(4440)^+\!\to\!\bar D^{0}\Lambda_c^+$		& $20.0$					&$4.6$	   			&$5.98-4.53$ 	&$5.6-1.7$    			&- \\
$P_\psi^N(4440)^+\!\to\!\bar D\Sigma_c$     			& $\sn{7.5}{-2}$		      	&$\sn{5.2}{-2}$ 	 	&$10.43-5.45$	&$3.4-0.5$			&-\\
$P_\psi^N(4440)^+\!\to\!\bar D\Sigma_c^*$ 			& $\sn{6.4}{-1}$          		&$\sn{3.6}{-1}$    		&-				&$0.8-5.4$			&-\\
\midrule[1.2pt]
\multicolumn{1}{c|}{Total} 											& $21.8$                           	&$13.0$ 			 	&$20.52-13.98$	&$23.7-13.9$		 	& $20.6\pm4.9^{+8.7}_{-10.1}$\\
\bottomrule[1.5pt]
$P_\psi^N(4457)^+\!\to\!J/\psi p$ 	  	              	 &  $\sn{1.6}{-2}$    			&$\sn{5.8}{-3}$  		&$1.52-1.52$		 	 &$0.02-0.01$ 	 &-  \\
$P_\psi^N(4457)^+\!\to\!\bar D^{*0}\Lambda_c^+$& $7.8$						&$\sn{8.0}{-1}$		& - 					 &$12.5-6.1$ 		 &-\\
$P_\psi^N(4457)^+\!\to\!\eta_c p$ 	  	    		 &  $\sn{3.2}{-1}$    			&$\sn{1.5}{-1}$ 		&- 				 	 &$0-0$ 		 	 &-  \\
$P_\psi^N(4457)^+\!\to\!\bar D^{0}\Lambda_c^+$ &  $4.3$						&$16.7$				&$2.47-2.15$			 &$3.8-1.5$    	 &- \\
$P_\psi^N(4457)^+\!\to\!\bar D\Sigma_c$     		 &  $\sn{5.5}{-2}$                 		&$\sn{8.1}{-2}$		&$5.60-4.11$			 &$2.6-1.0$		 &-\\
$P_\psi^N(4457)^+\!\to\!\bar D\Sigma_c^*$ 		 &  $\sn{4.7}{-1}$   			&$\sn{7.4}{-1}$  		&-					 &$1.9-6.2$		 &-\\
\midrule[1.2pt]
\multicolumn{1}{c|}{Total} 										 &  $13.0$                   	 		&$18.5$	 	 		&$9.59-7.78$			&$20.7-14.7$	 	& $6.4\pm2.0^{+5.7}_{-1.9}$\\
\bottomrule[1.5pt]
\end{tabular}
\end{table}
The mass ordering and relative widths for $P_\psi^N(4440)$ and $P_\psi^N(4457)$ we obtained stress the importance of the $P$-wave contributions in $P_{\psi1/2}^N$ and the $S$-$D$ mixing effects in $P_{\psi3/2}^N$, which is  consistent with the conclusion  from the spin-dependent tensor force obtained in ref.\,\cite{Yamaguchi2020}.

Our results show that the $\bar D^{0}\Lambda_c^+$ is the dominant decay channel for $P_\psi^N(4440)^+$, which  can amount to more than $90\%$ of the total width. This channel would be a most promising decay channel to be detected in experiments. Besides the $\bar D\Lambda_c^+$, $\bar D^{*0}\Lambda_c^+$ and $\bar D\Sigma_c^*$ are also important decay channels. For $P_\psi^N(4457)^+$, our results show that both the $\bar D^{*0}\Lambda_c^+$ and $\bar D\Sigma_c$  are the more important decay channels, which can amount to $\sim60\%$ and $\sim30\%$ fractions, respectively. A comparison of our results with other works is also listed in \autoref{T-width}, where the first and the second column results in refs.\,\cite{YangZY2024,LinYH2019} represent the $\frac12^-$ and $\frac32^-$ spin-party configuration respectively. Our obtained partial decay widths are roughly consistent with those in refs.\,\cite{LinYH2019,YangZY2024}, while the results in refs.\,\cite{LinYH2019,YangZY2024}  more favor the $\frac12$ and $\frac32$ spin configuration, which are different from our results. The obtained partial decay widths for $J/\psi p$ decay channel are in the order of $10^{-3}$ and $10^{-2}$ which are consistent with the results in ref.\,\cite{LinYH2019} but quite different from the ones in ref.\,\cite{YangZY2024}. We notice that the Belle collaboration determines the upper limits for $B[\Upsilon(1S,2S)\!\to\!(P_\psi^{N+}+X)]\cdot B[P_\psi^{N+}\!\to\! J/\psi p]$ to be at $10^{-6}$ level. 

In conclusion, based on our results, the $\frac32^-$ and $\frac12^-$ spin-parity configuration are favored for $P_\psi^N(4440)^+$ and $P_\psi^N(4457)^+$, respectively; however, more theoretical studies and experimental measurement are important and necessary to determine the essence of the two exotic hadrons.


\subsection{Summary}

In this work, firstly, based on the effective Lagrangians in the heavy quark and chiral limit, we calculate the one-boson-exchange interaction kernel of $\bar D^*\Sigma_c$ in the isospin-$\frac12$, where the light bosons $\sigma$, $\pi$, $\eta$, $\rho$ and $\omega$ are considered in the kernel calculations. Then by solving the Bethe-Salpeter equation for $\bar D^*\Sigma_c$ bound states, we obtain the mass spectra and wave functions of  $P_{\psi1/2}^N$ and $P_{\psi3/2}^N$.  Then by combining the effective Lagrangians and the obtained BS wave function, we calculate the main strong decay widths for the two $P_\psi^N$. 

The mass results favor the $\frac32^-$ and $\frac12^-$ configuration for the experimental $P_\psi^N(4440)^+$ and $P_\psi^N(4457)^+$.  In the favored spin-parity configuration, we obtain the total widths are $21.8$\,\si{MeV} and $13.0$\,\si{MeV} for $P_\psi^N(4440)^+$ and $P_\psi^N(4457)^+$, respectively. The partial decay widths suggest that $\bar D^0\Lambda_c^+$ and $\bar D^{(*)0}\Lambda_c^+$  are the more promising decay channels to detect $P_\psi^N(4440)^+$  and  $P_\psi^N(4457)^+$, respectively.  These results can also serve as important tests for the molecular interpretation of the two $P_\psi^N$ states. Our results are roughly consistent with several other calculations and also the LHCb experimental measurements. Taking into account both the mass spectra and decay widths, our results favor the interpretation of  $P_\psi^N(4440)^+$ and $P_\psi^N(4457)^+$  as the isospin-$\frac12$ $\bar D^*\Sigma_c$ molecular states with $J^P$ configuration $(\frac{3}{2})^-$ and $(\frac12)^-$ respectively. However, more theoretical researches and experimental measurements are necessary to decisively determine the essence of these $P_\psi^N$ states.

\appendix
\section{Effective Lagrangians} \label{A-Ls}

\subsection{Chiral Lagrangians for heavy-light mesons}
Considering the heavy quark spin-flavor symmetry, the heavy field of $\bar D^{(*)}$ in the heavy quark limit can be represented by
\begin{gather} \label{E-HQbar}
H_{\bar Q} = \left(\bar D^{*\alpha}\gamma_\alpha +\i \bar D \gamma_5\right) \frac{1-\slashed v}{2},
\end{gather}
where $\bar D=(\bar D^0, D^-, D_s^-)^\up{T}$ denotes anti-charmed heavy-light meson fields in flavor triplet, and $\bar D^{*\mu}$ is the corresponding vector state; $\bar H_{\bar Q}=\gamma^0H^\dagger_{\bar Q} \gamma_0$ denotes the usual conjunction in Dirac space; and $v$ denotes four-velocity of the anti-heavy-light meson.

Considering the heavy quark spin-flavor symmetry, hidden local symmetry and the light quark chiral symmetry, the involved Lagrangian describing the anti-heavy-light meson and one light meson reads\,\cite{Casalbuoni1997,YangZC2012}
\begin{align} \label{E-LM}
{L}_\up{M} 
&= g_\up{s} \braket{\bar H_{\bar Q} \sigma H_{\bar Q}}+g\braket{\bar H_{\bar Q} \slashed u \gamma_5 H_{\bar Q}} -  \beta  \braket{\bar H_{\bar Q} v_\alpha  {\Omega}^\alpha H_{\bar Q}} -\lambda \braket{ \bar H_{\bar Q} \sigma^{\alpha\beta} F_{\alpha\beta} H_{\bar Q}},
\end{align}
where $\braket{\cdots}$ denotes taking the $4\times4$ Dirac trace; the light scalar meson is represented by $\sigma$. The axial vector current $u_\alpha$ is defined as
\begin{gather}
u_\alpha=\frac12\i\left(\xi^\dagger\partial_\alpha\xi -\xi\partial_\alpha\xi^\dagger\right) = -\frac1f \partial_\alpha \Sigma + \cdots,
\end{gather}
with $f=132\,\si{MeV}$ denoting the pion decay constant, $\xi= \exp(\i \Sigma/f)$. Here we use $\Sigma$ to represent the $3\times3$ traceless hermitian matrix consisting of eight pseudoscalar meson fields, which has been given in \eref{E-Sigma}. 
The tensor field $F_{\alpha\beta}(\Omega)$ is defined as $F_{\alpha\beta}=(\partial_\alpha  {\Omega}_\beta - \partial_\beta  {\Omega}_\alpha)$ with $\Omega = (g_V/\sqrt2) V$, where ${V}$ denotes the $3\times3$ matrix consisting of the 9 light vector meson fields\,\cite{Casalbuoni1997,YangZC2012} and has been presented in \eref{E-V}.
The symbols $\beta$, $\lambda$, and $g_\up{s}$ are the corresponding chiral coupling constants. 

By expanding the second term in \eref{E-LM}, we obtain the Lagrangians between anti-heavy-light $\bar D^{(*)}$ meson and the light pion  as
\begin{equation}\label{E-L-DPD}
\begin{aligned}
{L}_{\bar D^{(*)}\Sigma\bar D^{(*)}} =& +g_{\bar D \Sigma \bar D^* }\i \bar D^\dagger  \partial_\alpha \Sigma \bar D^{*\alpha} \\
&- g_{\bar D^*\Sigma\bar D}\i(\bar D^{*\alpha})^\dagger \partial_\alpha \Sigma \bar D\\
&+g_{\bar D^* \Sigma \bar D^*}\epsilon^{\alpha\beta\mu\nu} \partial_\alpha (\bar D^*_\beta)^\dagger\partial_\mu \Sigma \bar D^*_\nu,
\end{aligned}
\end{equation}
where the coupling constants are determined  within the heavy quark and chiral symmetry as
\begin{equation}
\begin{aligned}
g_{\bar D \Sigma \bar D^* }    &= (M_{\bar D}M_{\bar D^*})^{\frac12}\frac{2g}{f},\\
g_{\bar D^* \Sigma \bar D }    &= (M_{\bar D^*}M_{\bar D})^{\frac12}\frac{2g}{f},\\
g_{\bar D^* \Sigma \bar D^*}  &= R_{\braket{\bar D^{*\dagger}\bar D^*}} \frac{2g}{f},
\end{aligned}
\end{equation}
where the abbreviation $R_{\braket{\bar D^{*\dagger}\bar D^*}}$ denotes the ratio of arithmetic mean to geometric mean of the  corresponding particle masses, namely
\begin{gather}
R_{\braket{\bar D^{*\dagger}\bar D^*}} = \frac{M_{\bar D^{*\dagger}}+M_{\bar D^{*}}}{2\left(M_{\bar D^{*\dagger}}M_{\bar D^{*}}\right)^{\frac12}}.
\end{gather}
Expanding the third and the fourth terms in \eref{E-LM}, we obtain the Lagrangians between anti-heavy-light $\bar D^{(*)}$ meson and light vector mesons as
\begin{equation}\label{E-L-DVD}
\begin{aligned}
L_{\bar  D^{(*)}V\bar D^{(*)}} 
 =& - g_{\bar DV\bar D} \i \partial_\alpha \bar D^\dagger V^\alpha \bar D \\
& + g_{\bar D^*V\bar D}\epsilon^{\alpha\beta\mu\nu}\partial_\alpha \bar D_\beta^{*\dagger} \partial_\mu V_\nu \bar D \\
&- g_{\bar DV\bar D^*} \epsilon^{\alpha\beta\mu\nu} \partial_\alpha \bar D^{\dagger}  \partial_\beta V_\mu \bar D^*_\nu\\
& + g_{\bar D^*V\bar D^*} \i(A_\up{r} \partial_\alpha \bar D_\beta^{*\dagger}V^\alpha \bar D^{*\beta} + \bar D^{*\dagger}_\alpha \partial_\beta V^\alpha \bar D^{*\beta} + \bar D_\alpha^{*\dagger} V_\beta\partial_\alpha \bar D^{*\beta} ),
\end{aligned}
\end{equation}
where the coupling constants are determined under the heavy quark limit as 
\begin{equation}
\begin{aligned}
g_{\bar DV\bar D}    &= R_{\braket{\bar D^{\dagger}\bar D}} \sqrt2(\beta g_V),\\
g_{\bar D^*V\bar D}  &=R_{\braket{\bar D^{*\dagger}\bar D}}  \sqrt8(\lambda g_V),\\
g_{\bar DV\bar D^*}  &= R_{\braket{\bar D^{\dagger}\bar D^*}} \sqrt8(\lambda g_V),\\
g_{\bar D^*V\bar D^*}&=  M_{\braket{\bar D^{*\dagger} \bar D^{*}}} \sqrt8(\lambda g_V),
\end{aligned}
\end{equation}
where we also introduce another abbreviation to denote the mean value for mass
\begin{gather}
M_{\braket{AB}} = \frac{2M_AM_B}{M_A+M_B};
\end{gather}
the coefficient $A_\up{r}$ reads $A_\up{r} =\frac12{\beta}/{\lambda M_{\braket{\bar D^{*\dagger} \bar D^{*}}}}$.
Notice that under the flavor $SU(4)$ symmetry, the coefficient $A_\up{r}$ would be 1 and here we obtain a small value of $A_\up{r}=0.40$.

\subsection{Chiral Lagrangians for heavy-light baryons}
Considering the heavy quark symmetry, hidden local symmetry and chiral symmetry, the effective Lagrangian of the heavy-light baryon and one light meson reads\,\cite{Yan1992,ChengHY2007,LiuYR2012,YangZC2012}
\begin{align} \label{E-LB}
{L}_\up{B} 
&=\frac32 g_1\i \epsilon^{\alpha\beta\mu v} \braket{\bar S_\alpha u_\beta S_\mu} + \beta_S \braket{\bar{ {S}}_\alpha v_\beta   {\rho}^\beta  {S}^\alpha  } + \i \lambda_S   \braket{\bar{ {S}}_\alpha  F^{\alpha\beta}  {S}_\beta} +l_S \braket{\bar{ {S}}_\alpha \sigma  {S}^\alpha}.
\end{align}
where  the symbol $\epsilon^{\mu\nu\alpha\beta}$ denotes the totally antisymmetric  Levi-Civita tensor with $\epsilon^{\mu\nu\alpha\beta}=-\epsilon_{\mu\nu\alpha\beta}$ and the convention $\epsilon^{0123}=1$ is used; the abbreviation $\langle\cdots\rangle$ here denotes taking trace in the $3\times3$ flavor space. The single heavy baryon in spin doublet are incorporated into field as
\begin{gather}
 {S}_\alpha = -\frac{1}{\sqrt3} (\gamma_\alpha + v_\alpha) \gamma^5  {B} +  {B}_{\alpha}^{*},
\end{gather}
where $B$ represents the $3\times3$ matrix for the systematic baryon sextet, 
\begin{eqnarray} \label{E-B6}\renewcommand{\arraystretch}{1.8}
 {B}=\left[
\begin{array}{ccc}
\Sigma_c^{++} &\frac{1}{\sqrt2}\Sigma_c^+   & \frac{1}{\sqrt2}\Xi_c^{\prime+}  \\
\frac{1}{\sqrt2}\Sigma_c^+  & \Sigma_c^{0}  & \frac{1}{\sqrt2}\Xi_c^{\prime0}  \\
\frac{1}{\sqrt2}\Xi_c^{\prime+}  & \frac{1}{\sqrt2}\Xi_c^{\prime0}  &\Omega_c^0
\end{array}\right].
\end{eqnarray}
The conjugation defines as usual for the spinor field $\bar{ {S}}^{mn}_\mu = ( {S}^{mn}_\mu)^\dagger \gamma_0$. An asterisk on the symbol denotes the corresponding spin-$\frac32$ baryon.

Expanding the first item in \eref{E-LB}, we obtain the effective Lagrangians for single heavy baryon and a light pseudoscalar meson as
\begin{equation}
\begin{aligned}
L_{BB\Sigma} =& -g_{\bar BB\Sigma}  \braket{\i\bar B \gamma_5\Sigma B}\\
&+g_{BB^*\Sigma}\braket{(\bar B\partial_\alpha\Sigma B^{*\alpha}+ \bar B^{*\alpha} \partial_\alpha\Sigma B)} \\
&-g_{\bar B^*B^*\Sigma}\epsilon^{\alpha\beta\mu\nu}\braket{  \partial_\beta\bar B^*_\nu \partial_\alpha \Sigma B^*_\mu},
\end{aligned}
\end{equation} 
where $\braket{\cdots}$ here denotes trace in the flavor space; the coupling constants are determined under the heavy quark limit as
\begin{align*}
g_{\bar BB\Sigma}    &=(M_{\bar B}+M_B)\frac{g_1}{f},\\
g_{BB^*\Sigma}  &=\frac{\sqrt3}{2} \frac{g_1}{f},\\
g_{\bar B^*B^*\Sigma}&=\frac32 \frac{1}{M_{\braket{\bar B^* B^*}}}\frac{g_1}{f}.
\end{align*}
Notice that the expressions of $g_{\bar B B\Sigma}$ and $g_{\bar B^* B^*\Sigma}$ are dependent on the masses of the annihilated and created baryons. Also notice that in a specific Lagrangian, there  exists an extra flavor factor since the trace in flavor space had not been explicitly completed.

The second and the third item in \eref{E-LB} show the effective lagrangians for the single heavy baryon and a light vector meson,
\begin{equation}\label{E-L-BVB}
\begin{aligned}
L_{BBV} =& -g_{\bar BBV} \braket{\bar B \gamma_\alpha V^\alpha B} \\
& - g_{\bar BB^*V}\braket{\i \bar B\gamma_\alpha\gamma_5 (\partial^\alpha V^\beta -\partial^\beta V^\alpha)B^*_\beta}\\
& + g_{\bar B^*BV}\braket{\i \bar B^*_\beta \gamma_\alpha\gamma_5 (\partial^\alpha V^\beta -\partial^\beta V^\alpha)B}\\
&+ g_{\bar B^*B^*V} \braket{ \frac{\lambda_S}{\beta_S}\i\bar B^*_\alpha (\partial^\alpha V^\beta - \partial^\beta V^\alpha) B^*_\beta +\bar B^{*\alpha}\gamma_\beta V^\beta B^*_\alpha } ,
\end{aligned}
\end{equation}
where the coupling constants are related to the chiral constants $\beta_S$ and $\lambda_s$ by
\begin{equation}
\begin{aligned}
g_{\bar BBV}    &=\frac1{\sqrt2} g_V\beta_S,\\
g_{\bar BB^*V}  &=g_{\bar B^*BV}=\frac{1}{\sqrt6}g_V\lambda_S,\\
g_{\bar B^*B^*V}&=\frac{1}{\sqrt2}g_V\beta_S.
\end{aligned}
\end{equation}
Notice here the coupling constants are general, and the dimensions are not consistent with each other. A specific coupling constant will also depend on the specific Lagrangian form, the involved baryons\rq{} masses, and the flavor factors of the involved vector mesons. We will listed the specific Lagrangians and dimensionless coupling constants involved when calculating the relevant strong decay amplitudes.  

The single-heavy baryon in flavor anti-triplet can be incorporated in the traceless $3\times3$ matrix as
\begin{gather}
\Lambda = \begin{bmatrix}
0 & \Lambda_c^+ & \Xi_c^+ \\
-\Lambda_c^+ & 0 &\Xi_c^0 \\
-\Xi_c^+ & -\Xi_c^0 & 0
\end{bmatrix}.
\end{gather}
The effective Lagrangian behaves\,\cite{Yan1992,LiuYR2012}
\begin{gather}
L_{\Lambda B} = g_4 \braket{\bar \Lambda u_\alpha S^\alpha} + \lambda_{I} \epsilon^{\alpha\beta\mu\nu} v_\alpha \braket{\bar \Lambda F_{\mu\nu} S_\beta} + \up{H.c.}.
\end{gather}
Expanding above equation we obtain
\begin{equation}\label{E-L-AMB}
\begin{aligned}
L_{\Lambda B} =& -g_{\Lambda\Sigma B} \braket{\i\bar \Lambda \Sigma \gamma_5 B  + \i\bar B \Sigma \gamma_5 \Lambda} \\
& - g_{\Lambda\Sigma B^*}\braket{\bar \Lambda \partial^\alpha\Sigma B^*_\alpha + \bar B^*_\alpha \partial^\alpha\Sigma \Lambda}\\
&+ g_{\Lambda VB} \braket{\bar\Lambda V_\alpha \gamma^\alpha B + \bar B  V_\alpha \gamma^\alpha \Lambda} \\
&- g_{\Lambda VB^*}\epsilon^{\alpha\beta\mu\nu} \braket{\i \partial_\beta \bar \Lambda\partial_\alpha V_\nu B^*_\mu + \i \partial_\beta  \bar B^*_\mu \partial_\alpha V_\nu \Lambda},
\end{aligned}
\end{equation}
where the four coupling constants behaves 
\begin{equation}
\begin{aligned}
g_{\Lambda \Sigma B}    &=\frac{1}{\sqrt3}(M_{\Lambda}+M_B) \frac{g_4}{f},\\
g_{\Lambda \Sigma B^*}  &= \frac{g_4}{f},\\
g_{\Lambda VB}&=\frac{\sqrt2}{\sqrt3} (M_{\Lambda}+M_B)(\lambda_{I}g_V),\\
g_{\Lambda VB^*}&=\sqrt2 \frac{1}{M_{\braket{\Lambda B}}}(\lambda_{I}g_V).
\end{aligned}
\end{equation}

\subsection{Effective Lagrangian for $N\Sigma_cD^{(*)}$ interaction}

The three pentaquark states $P_\psi^N$ are all found in the $J/\psi p$ channel and then the Lagrangians for $N\Sigma_c$ interactions are also needed to calculate this decay width. The nucleon $N=(p,n)^{\up{T}}$ forms a  fundamental representation for $SU(2)$ isospin, and the three $\Sigma_c$ baryons also form a triplet representation of the isospin. Considering the isospin symmetry, the effective Lagrangian for $N\Sigma_cD^{(*)}$ can be expressed as
\begin{equation}
\begin{aligned}
L_{N\Sigma_cD^{(*)}} &= g_{N\Sigma_cD}\i \bar N \gamma_5 \Sigma_c(-\i \sigma_2)\cdot D^\dagger + g_{N\Sigma_cD^*} \bar N\gamma^\alpha \Sigma_c (-\i\sigma_2)\cdot D^{*\dagger}_\alpha+\up{H.c.},
\end{aligned}
\end{equation}
where $\sigma_2$ denotes the usual second Pauli matrix, and the $D^{(*)}$ here only denotes the doublet $(D^{(*)0}, D^{(*)+})$ which also forms a fundamental representation of the isospin, and $\Sigma_c$ represents the 1-2 sector in the symmetric baryon field matrix $B$, namely,
\begin{gather}\renewcommand{\arraystretch}{1.6}
\Sigma_c = \begin{bmatrix}
\Sigma_c^{++} &  \frac{1}{\sqrt2} \Sigma_c^+ \\
\frac{1}{\sqrt2} \Sigma_c^+ & \Sigma_c^{0}
\end{bmatrix}.
\end{gather}
Expanding above Lagrangian, we obtain
\begin{equation}
\begin{aligned}
L_{\bar N\Sigma_cD^{(*)}} =& - g_{N\Sigma_cD}\i \left( \bar p \gamma_5 \Sigma_c^{++} D^{+\dagger} - \bar p  \frac{\gamma_5}{\sqrt2}\Sigma_c^+ D^{0\dagger} + \bar n\frac{\gamma_5}{\sqrt2}\Sigma_c^+D^{+\dagger}  - \bar n\gamma_5\Sigma_c^0 D^{0\dagger}\right) \\
& - g_{N\Sigma_cD^*} \left( \bar p \gamma^\alpha \Sigma_c^{++} D^{*+\dagger}_\alpha -\bar p \frac{\gamma^\alpha}{\sqrt2}\Sigma_c^+ D^{*0\dagger}_\alpha + \bar n  \frac{\gamma^\alpha}{\sqrt2} \Sigma_c^+D^{*+\dagger}_\alpha  - \bar n\gamma^\alpha\Sigma_c^0 D^{*0\dagger}_\alpha\right).
\end{aligned}
\end{equation}

\subsection{Effective Lagrangians for charmonia and heavy-light meson}

The heavy-light charmed mesons in $S$-wave can be represented by\,\cite{Wise1992,Burdman1992,Casalbuoni1997}
\begin{gather}\label{E-HQ}
H_{Q} = \frac{1+\slashed v}{2} (D^{*\alpha} \gamma_\alpha +\i D \gamma_5),
\end{gather}
where $D=(D^0,D^+,D_s^+)$ denotes the fields of the corresponding pseudoscalar charmed mesons, which behaves as a row vector in the light quark flavor space, and then $D^*$ is implied. The anti-heavy-light meson doublet $H_{\bar Q}$ has been presented in \eref{E-HQbar}. 

For heavy quarkonium mesons, the heavy quark flavor symmetry does not hold any longer, while the heavy quark spin symmetry still holds. In the ground states, the charmonium forms a doublet consisting of a pseudoscalar $\eta_c$ and a vector $J/\psi$, which can then be represented by\,\cite{Jenkins1992}
\begin{gather}
R =   \frac{1+\slashed v}{2}   (\psi^\alpha \gamma_\alpha + \i \eta_c \gamma_5)  \frac{1-\slashed v}{2},
\end{gather}
where $\psi^\alpha$ and $\eta_c$  denotes the fields of the corresponding mesons. Noticed that here all the hadron fields in above equations contain a extra factor of $\sqrt{M_H}$ with $M_H$ the corresponding meson mass.

By assuming the invariance under independent rotations of the heavy quark spins, it is possible to write down the effective coupling between the $S$-wave charmonia and the heavy-light mesons as\,\cite{Colangelo2004,GuoFK1008}
\begin{gather}
{L}_{\bar D^{(*)} D^{(*)}R} = \i g_R \up{Tr}\,(\bar R H_{ Q}\gamma_\alpha \partial^\alpha H_{\bar Q} - \bar R \partial^\alpha H_{ Q}\gamma_\alpha H_{\bar Q}) +\up{H.c.},
\end{gather}
which is invariant under independent heavy quark spin symmetry; 
the coupling constant $g_R$ has a dimension of $(-\frac{3}{2})$. Consequently, we obtain the following effective Lagrangians describing the $D^{(*)}D^{(*)}$ coupling to $J/\psi$,
\begin{equation} \label{E-L-psi}
\begin{aligned}
{L}_{ D^{(*)}\bar D^{(*)}\psi^\dagger} =&+ g_{  \bar D D \psi}\i  D {\partial}_\alpha \bar D  \psi^{\dagger\alpha}  \\
& + g_{  D \bar D^*\psi}\frac{1}{M_\psi} \epsilon^{\alpha\beta\mu\nu} \partial_\alpha D \bar D^*_\beta \partial_\mu \psi^\dagger_\nu \\
&-  g_{  D^*\bar D\psi}\frac{1}{M_\psi} \epsilon^{\alpha\beta\mu\nu}\partial_\alpha D^*_\beta \bar D  \partial_\mu \psi^\dagger_\nu\\
& - g_{ D^*\bar D^*\psi} \i({\partial}_\alpha D^{*}_{\beta} \bar D^{*\alpha}    \psi^{\dagger\beta}+ 2   D^{*}_\alpha {\partial}_\beta \bar D^{*\alpha}  \psi^{\dagger\beta} +   D^{*\alpha} \bar D^*_\beta   {\partial}_\alpha \psi^{\dagger\beta}),
\end{aligned}
\end{equation}
where the relavent coupling  constants are related to a single coupling $g_R$, which is determined to be $g_R = {\sqrt{M_\psi}}/({2M_D f_\psi})$ by the VMD method with $f_\psi$ denoting the $J/\psi$ decay constant\,\cite{Colangelo2004}. It is also convenient to express the coupling constants in terms of the $g_{D\bar D\psi}$ as
\begin{equation} \label{E-g-DDpsi}
\begin{aligned}
  g_{ D\bar D\psi} &=    \frac{2M_\psi}{f_\psi},\\
  g_{ D \bar D^*\psi}    &= \left(\frac{M_{\bar D^*}}{M_{\bar D}} \right)^{1/2} \textstyle g_{D \bar D\psi}, \\
  g_{ D^* \bar D\psi}    &= \left(\frac{M_{D^*}}{M_{D}} \right)^{1/2} \textstyle  g_{D \bar D\psi}, \\
  g_{ D^* \bar D^*\psi} &=   \frac12 \left( \frac{M_{D^*}}{M_D} \right)g_{D \bar D\psi}.
\end{aligned}
\end{equation}
The obtained lagrangians for coupling to $\eta_c$ are
\begin{equation} \label{E-L-etac}
\begin{aligned}
{L}_{ D^{(*)}\bar D^{(*)}\eta_c^\dagger} =&+g_{ D \bar D^*\eta_c} \i ( {\partial_\alpha} D\bar D^{*\alpha})\eta_c^\dagger\\
& - g_{ D \bar D^*\eta_c} \i(D^{*\alpha}{\partial_\alpha} \bar D )  \eta^\dagger_c \\
& - g_{ D^*\bar D^*\eta_c} \frac{1}{M_{\eta_c}}\epsilon^{\alpha\beta\mu\nu}  {\partial}_\alpha D^*_\beta \bar D^*_\mu{\partial_\nu \eta^\dagger_c},
\end{aligned}
\end{equation}
where the  coupling constants behaves 
\begin{equation} 
\begin{aligned}
  g_{D \bar D^* \eta_c}&=  \frac12 \left( \frac{M_{\eta_c} M_{\bar D^*}}{M_\psi M_{\bar D}} \right)^{1/2} g_{D \bar D\psi}, \\
  g_{D^* \bar D \eta_c}&=  \frac12 \left( \frac{M_{\eta_c} M_{D^*}}{M_\psi M_{D}} \right)^{1/2} g_{D \bar D\psi}, \\
  g_{D^* \bar D^* \eta_c}&= \left( \frac{M_{\eta_c}}{M_\psi } \right)^{1/2} \frac{M_{D^*}}{M_D} g_{D \bar D\psi}.
\end{aligned}
\end{equation}
In the Lagrangian involved the charmonia, we have already divided a meson mass to keep all the coupling constants dimensionless if necessary, and now the all the meson fields above are dimension 1. 

\section{Calculations of the decay form factors} \label{App}
\subsection{Calculation of amplitude $\mathcal{A}_{12}$}\label{A-2-1}
We can express the decay amplitude $\mathcal{A}_{12}$  as
\begin{align*}
\mathcal{A}_{12;1/2}
=& G_{12}  (e_{1}^{\alpha})^*  \bar u_2 T_{12\alpha} u,
\end{align*}
where we have stripped off the triangle integral over $k$ by defining 
\begin{align}
T_{12\alpha} u(P,r)& =  {O}_{{\alpha\beta\mu\rho}}\gamma_\nu\int \frac{\d^4 k}{(2\pi)^4} [S(k_2) \Gamma_\gamma(k,r)F^2 D^{\gamma\beta}(k_1)] D^{\mu\nu}(k_3) (k_1+k_3)^\rho.
\end{align}
Now we perform the contour integral over $k_P$  as usual and then obtain
\begin{align}
T_{12\alpha } u(P,r) 
&=O_{\alpha\beta\mu\rho}\gamma_\nu \int \frac{\d^3 k_\bot}{(2\pi)^3}\frac{1}{2w_3}F^2\left( a_{3}^{\mu\nu\rho}\Lambda^+ + a_{4}^{\mu\nu \rho}\Lambda^- \right) \gamma_0 \varphi^\beta ,
\end{align}
where 
\begin{equation}
\begin{aligned}
a_{3\mu\nu\rho} 
=\sum_{i=1,3,5}c_i d_{\mu\nu}(k_{3i})(k_{1i}+k_{3i}) , \\
a_{4\mu\nu\rho} = \sum_{i=2,4,6}c_i d_{\mu\nu}(k_{3i})(k_{1i}+k_{3i}),
\end{aligned}
\end{equation}
where we used the notations
\begin{gather}
d_{\mu\nu}(k_{3i}) = -g_{\mu\nu} + \frac{k_{3i\mu} k_{3i\nu}}{m_3^2},\\
k_{3i}=k_{1i}-P_1.
\end{gather}
Notice that contribution of the momentum part in $d_{\mu\nu}$ will be suppressed when the exchanged particle is heavy. 
Now $T_{12\alpha}$ has been expressed by the Salpeter wave function. After performing the three-dimensional integral numerically, we obtain
\begin{gather} \label{E-ff-s12}
{T}_{12\alpha} =  (s^2_{11} \gamma_\alpha  + s^2_{12} \hat P_\alpha ).
\end{gather}

\subsection{Calculation of amplitude $\mathcal{A}_{31}$} \label{A-A31}
For the $\eta_c p$ decay channel by $D$ exchange,  as usual it is convenient to strip off the part involved the integral over $k$ as
\begin{align*}
{T}_{31} u
& = (\gamma_5  P_1^\beta) \int \frac{\d^4 k}{(2\pi)^4}[S(k_2) \Gamma^\gamma(k,r)F^2 D_{\beta\gamma}(k_1)] D(k_3).
\end{align*}
Performing the contour integral over $k_P$, we can express ${T}_3$ by the three-dimensional Salpeter wave function
\begin{align}
{T}_{31} u
&=  (\gamma_5  P_1^\beta)\int \frac{\d^3 k_\bot}{(2\pi)^3}\frac{1}{2w_3} F^2  \left( b_1 \Lambda^+   +b_2 \Lambda^-   \right)\gamma_0\varphi_\beta,
\end{align}
where 
\begin{equation}\label{E-b1-b2}
\begin{gathered}
b_1=c_1+c_3+c_5,\\
b_2=c_2+c_4+c_6.
\end{gathered}
\end{equation}
Inserting the Salpeter wave function \eref{E-wave-1-2N} of $P_{\psi1/2}$, we obtain $T_{31}$ expressed by one form factor,
\begin{gather} \label{E-ff-s31}
T_{31} = s_{31} \gamma_5.
\end{gather}
For $P^N_{\psi3/2} \to \eta_c p$ by $D$ exchange, the triangle amplitude $T_{31\gamma}$ is expressed as
\begin{align}
{T}_{31\gamma} u^\gamma =t_{3}^1 \hat P_{1\gamma}
&=  (\gamma_5  P_1^\beta) \int \frac{\d^3 k_\bot}{(2\pi)^3}\frac{1}{2w_3} F^2 \left( b_1 \Lambda^+   +b_2 \Lambda^-   \right)\gamma_0 A_{\beta\gamma} u^\gamma.
\end{align}

\subsection{Calculation of amplitude $\mathcal{A}_{32}$}\label{A-A32}

The amplitude for $P_{\psi1/2}\to \eta_c p$ by $D^*$ exchange can be simplified as 
\begin{gather}
\i\mathcal{A}[P_{\psi1/2}\to \eta_c p(D^*)] = G_{32}\bar u(P_2,r_2) T_{32} u(P,r),
\end{gather}
where the integral over $k$ has been stripped out, 
\begin{align}
T_{32} u=   - \i\int \frac{\d^4 k}{(2\pi)^4} \gamma^\nu [S(k_2) \Gamma^\gamma(k,r) F^2 D_{\beta\gamma}(k_1)] D_{\mu\nu}(k_3) \frac{\epsilon^{k_3\beta \mu P_1}}{M_1}.
\end{align}
Notice that the contraction with Levi-Civita symbol forces momentum part of the numerator in propagator $D_{\mu\nu}$ to be zero, namely, 
\begin{gather}
d_{\mu\nu}(k_3) = -g_{\mu\nu} + \frac{k_{3\mu} k_{3\nu}}{m_3^2} \to(-g_{\mu\nu}),
\end{gather}
Then above expression is further simplified as
\begin{align}
T_{32} u=   \i\frac{\epsilon^{\alpha\beta \mu P_1}}{M_1} \gamma_\mu \int \frac{\d^4 k}{(2\pi)^4}  [S(k_2) \Gamma^\gamma(k,r) F^2 D_{\beta\gamma}(k_1)] D(k_3) k_{1\alpha} .
\end{align}
Now we perform the contour integral as usual over $k_P$ to express $T$ by the Salpeter wave function
\begin{align}
T_{32} u = \i \frac{\epsilon^{\alpha \beta \mu P_1}}{M_1} \gamma_\mu \int \frac{\d^3 k_\perp}{(2\pi)^3} \frac{1}{2w_3}  (a_{1\alpha}\Lambda^+ + a_{2\alpha} \Lambda^-) \gamma_0 \varphi_\beta .
\end{align}
where the expressions for $a_{1\alpha}$ and $a_{2\alpha}$ have been presented in \eref{E-a1-a2}; the Salpeter wave function  behaves $\varphi_\beta[P_{\psi1/2}^N]=A_\beta u$. Inserting \eref{E-A-alpha} and then calculating the three-dimensional integral, we can express $T_{32}$ by one form factor as
\begin{gather}\label{E-ff-s32}
T_{32}=s_{3}^2 \gamma_5.
\end{gather}

For $P_{\psi3/2}^N \!\to\! \eta_c p$, the triangle amplitude behaves
\begin{align}
T_{32\gamma} u^\gamma =  \i \frac{\epsilon^{\alpha \beta \mu P_1}}{M_1} \gamma_\mu \int \frac{\d^3 k_\perp}{(2\pi)^3} \frac{1}{2w_3} F^2 (a_{1\alpha}\Lambda^+ + a_{2\alpha} \Lambda^-) \gamma_0 A_{\beta\gamma}u^\gamma.
\end{align}
Performing this integral, we obtain
\begin{gather}\label{E-ff-t32}
T_{32\gamma} = t_{3}^2 \hat P_{1\gamma}.
\end{gather}

\subsection{Calculation of amplitude $\mathcal{A}_{71}$}\label{A-A71}
For $\bar D\Sigma_c^*$ decay by $\pi$ exchange, the triangle integral behaves
\begin{gather}
T_{71\alpha} u(P,r) = \frac{P_1^\beta}{M_2}\int \frac{\d^4k}{(2\pi)^4} [S(k_2) \Gamma^\gamma(k,r) F^2D_{\gamma \beta}(k_1)]F^2D(k_3) k_{3\alpha}.
\end{gather}
Performing the contour integral over $k_P$, we obtain
\begin{align} \label{E-S2-D1-D3-k3-kP}
\int \frac{\d k_P}{2\pi} [S(k_2) \Gamma_\gamma(k,r) D_{\gamma \beta}(k_1)] D(k_3)k_{3\alpha} &=  \frac{1}{2w_3}\left( b_{3\alpha} \Lambda^+  + b_{4\alpha} \Lambda^- \right) \gamma_0\varphi_\beta,
\end{align}
where
\begin{equation}\label{E-a3-a4}
\begin{gathered}
b_{3} = c_1 k_{31} + c_3 k_{33} + c_5 k_{35}, \\
b_{4} = c_2 k_{32} + c_4 k_{34} + c_6 k_{36}, 
\end{gathered} 
\end{equation}
where $k_{3i} = k_3(k_P=k_{Pi})$ and $c_i\,(i=1,\cdots 6)$ have been introduced before.
Then the triangle integral can be expressed by the Salpeter wave function as
\begin{align}
T_{71\alpha} u
&=  \frac{P_{1}^{\beta}}{M_2}\int\frac{\d^3 k_\perp}{(2\pi)^3} \frac{1}{2w_3}  F^2\left(b_{3\alpha} \Lambda^+  + b_{4\alpha} \Lambda^- \right) \gamma_0\varphi_\beta,
\end{align}
which can be further simplified as one form factor after finishing above integral numerically, namely,
\begin{gather}
T_{71\alpha} = s_{7}^1  \hat P_\alpha.
\end{gather}

For $P_{\psi3/2}^N \to  D^-\Sigma_c^{*++}$ by $\pi$ exchange, the amplitude behaves
\begin{align}
T_{71\alpha\gamma} u^\gamma
&= \frac{P_{1}^{\beta}}{M_2} \int\frac{\d^3 k_\perp}{(2\pi)^3} \frac{1}{2w_3} F^2\left(b_{3\alpha} \Lambda^+  + b_{4\alpha} \Lambda^- \right) \gamma_0A_{\beta\gamma}u^\gamma,
\end{align}
which can be further simplified as three form factors after finishing above integral over the Salpeter wave function, 
\begin{gather}
T_{71\alpha\beta} = \i t_{71}^1 {\epsilon^{\alpha\beta \hat P\hat P_1}} +  \left( t_{72}^1 g^{\alpha\beta}  +t_{73}^1\hat P^\alpha \hat P_1^\beta  \right) \gamma_5.
\end{gather}

\subsection{Calculation of amplitude $\mathcal{A}_{72}$} \label{A-A72}
For $\bar D-\Sigma_c$ decay channel, the triangle integral behaves
\begin{gather}
T^\sigma_{72} u =\i \frac{\epsilon^{\alpha\beta \mu \hat P_1}}{M_2} \int \frac{\d^4k}{(2\pi)^4}(g^{\sigma\nu}\gamma^\rho-g^{\rho\sigma}\gamma^\nu)\gamma_5 [S(k_2) \Gamma^\gamma(k,r) D_{\gamma \beta}(k_1)]F^2D_{\mu\nu}(k_3)  k_{3\rho} k_{3\alpha}.
\end{gather}
Considering the contraction with Levi-Civita, the above amplitude can be further simplified by $D_{\mu\nu}(k_3)\to (-g_{\mu\nu})D(k_3)$, $k_{3\alpha}\to k_{1\alpha}$, namely,
\begin{gather}
T^\sigma_{72} u =\i \frac{\epsilon^{\alpha\beta \mu \hat P_1}}{M_2} (g^{\rho\sigma}\gamma_\mu-g_{\mu\nu}g^{\sigma\nu}\gamma^\rho)\gamma_5\int \frac{\d^4k}{(2\pi)^4} [S(k_2) \Gamma^\gamma(k,r) D_{\gamma \beta}(k_1)]F^2D(k_3)  k_{3\rho} k_{1\alpha}.
\end{gather}
Then we perform the contour integral over $k_P$ as usual and obtain
\begin{gather}
\int \frac{\d k_P}{2\pi} [S(k_2) \Gamma^\gamma(k,r) D_{\gamma \beta}(k_1)]D (k_3)  k_{3\rho} k_{1\alpha} = \frac{1}{2w_3} \left( b_{5\alpha\rho} \Lambda^+ + b_{6\alpha\rho}\Lambda^- \right)\gamma_0\varphi_\beta,
\end{gather}
where we defined
\begin{equation}
\begin{gathered}
b_{5\alpha \rho} = c_1 k_{11\alpha} k_{31\rho} + c_3 k_{13\alpha}k_{33\rho} + c_5 k_{15\alpha}k_{35\rho}, \\
b_{6\alpha \rho} = c_1 k_{12\alpha} k_{32\rho} + c_4 k_{14\alpha}k_{34\rho} + c_6 k_{16\alpha} k_{36\rho}.
\end{gathered}
\end{equation}
Finally, we can express $T_{72}$ by the three-dimensional integral of the Salpeter wave function as
\begin{gather}
T^\sigma_{72} u =\i \frac{\epsilon^{\alpha\beta \mu \hat P_1}}{M_2} (g^{\rho\sigma}\gamma_\mu-g_{\mu\nu}g^{\sigma\nu}\gamma^\rho)\gamma_5\int \frac{\d^3k_\perp}{(2\pi)^3}  \frac{1}{2w_3} F^2 \left( b_{5\alpha\rho} \Lambda^+ + b_{6\alpha\rho}\Lambda^- \right)\gamma_0\varphi_\beta,
\end{gather}
which can be further simplified as one form factor by finishing the three-dimensional integral, namely,
$T_{72\alpha} = s_{7}^2\hat P_\alpha.$

\subsection{Detailed expressions for form factors} \label{A-2-2}
Since the expressions of the form factors are still quite complicated when expressed by the Salpeter wave functions, here we just list the form factors in \eref{E-s1-s2} for $P_{\psi1/2}^N\!\to\! (V+B)$ as an example to show the calculation details, which are all represented by the integral over the radial Salpeter wave functions $g_1$, $g_2$, $g_3$ and $g_4$. Parts of following expressions are calculated with the help of the \texttt{FeynCalc} package\,\cite{Mertig1990,Shtabovenko2016,Shtabovenko2020}.
The form factors are respectively expressed as
\begin{equation}
\begin{aligned}
s^n_{1o} &=\int \frac{\d^3 k_\perp}{(2\pi)^3} \frac{1}{4w_3} F^2S_{1o}^n
\end{aligned}
\end{equation}
with $n=1,2$ representing the pseudoscalar and vector exchange, respectively, and $o=1,2$ denoting the two form factors,  where $S_{1o}^n$ behaves
\begin{equation}
\begin{aligned}
S^1_{11} &=-\gamma_1x_2 + 2x_1o_8 +x_2o_8+2x_1o_9+x_2o_9,\\
S^1_{12} &= -4x_1o_8-x_2o_8-x_2o_9,\\
S^2_{11}&=y_1+C_1 o_4 y_2-C_1 o_3 o_8 y_2-C_1 o_3 o_9 y_2+C_{22} y_6,\\
S^2_{12}&=-2 C_1 o_4 y_2+2 C_1 o_3 o_8 y_2+y_3-C_1 o_4 y_4+C_1 o_3 o_8 y_4\\
&-C_1 o_3 o_9 y_4-C_1 o_4 y_5-C_{22} y_6+C_{21} o_4^2 y_6-C_{21} o_3 o_4 o_8 y_6+C_{21} o_3 o_4 o_9 y_6,\\
\end{aligned}
\end{equation}
where $x_i$ and $y_i$  are expressed as
\begin{equation}
\begin{aligned}
x_1 & = -C_{22} \left(c_8 g_2  o_2+c_8 g_1  o_1+c_7 g_2 \right), \\
x_2 &= -c_7 C_{21} g_2 o_4^2-c_8 C_{21} g_1 o_1 o_4^2-c_8 C_{21} g_2 o_2 o_4^2+c_9 C_1 g_2 o_4-c_8 C_{21} g_2 o_2
   o_3 o_9 o_4 \\
   &+c_{10} C_1 g_1 o_1 o_4-c_8 C_1 g_2 o_1
   o_4+c_8 C_1 g_1 o_2 o_4+c_{10} C_1 g_2 o_2 o_4+c_7 C_{21} g_2 o_3 o_8
   o_4 \\
   &+c_8 C_{21} g_1 o_1 o_3 o_8 o_4+c_8 C_{21} g_2 o_2 o_3 o_8 o_4-c_7
   C_{21} g_2 o_3 o_9 o_4-c_8 C_{21} g_1 o_1 o_3 o_9 o_4\\
   &+2 c_8 C_{22} g_1 o_1+2 c_8 C_{22} g_2
   o_2-c_{10} C_{22} g_4 o_2-c_9 C_1 g_2 o_3 o_8-c_{10} C_1 g_1 o_1 o_3
   o_8\\
   &-c_{10} C_1 g_2 o_2 o_3 o_8+c_9 C_1 g_2 o_3 o_9+c_{10} C_1 g_1 o_1
   o_3 o_9+c_{10} C_1 g_2 o_2 o_3 o_9+2 c_7 C_{22} g_2\\
   &+c_{10} g_2 o_1-c_{10} g_1 o_2+c_9 g_1-c_7
   C_1 g_1 o_4+c_9 C_{22}g_4-c_{10} C_{22} g_3 o_1;
\end{aligned}
\end{equation}
\begin{equation}
\begin{aligned}
y_1&=-2 c_8 g_2 \iota _3 o_6^3-2 c_9 g_1 o_1 \iota _3 o_6^3-2 c_9 g_2 o_2 \iota _3 o_6^3-2 c_{14} g_1 \iota _3 o_6^2-2 c_{15} g_2 o_1 \iota _3 o_6^2-3 c_{15} g_1 o_2 o_7^2 o_9 \gamma _1 \iota _3\\
&+2 c_{15} g_1 o_2 \iota _3 o_6^2+2 c_8 g_1 o_7 o_8 \iota _3 o_6^2+2 c_9 g_2 o_1 o_7 o_8 \iota _3 o_6^2-2 c_9 g_1 o_2 o_7 o_8 \iota _3 o_6^2+2 c_8 g_1 o_7 o_9 \iota _3 o_6^2\\
&+2 c_9 g_2 o_1 o_7 o_9 \iota _3 o_6^2-2 c_9 g_1 o_2 o_7 o_9 \iota _3 o_6^2+3 c c_8 g_2 o_7 v_1 \iota _3 o_6^2+3 c c_9 g_1 o_1 o_7 v_1 \iota _3 o_6^2+3 c_{14} g_1 o_7^2 o_9 \gamma _1 \iota _3\\
&+3 c c_9 g_2 o_2 o_7 v_1 \iota _3 o_6^2-2 c_8 g_2 o_6-2 c_8 g_3 o_6-2 c_9 g_1 o_1 o_6+2 c_9 g_4 o_1 o_6-2 c_9 g_2 o_2 o_6-2 c_9 g_3 o_2 o_6\\
&+c_9 g_1 o_1 o_7^2 \iota _3 o_6+c_9 g_2 o_2 o_7^2 \iota _3 o_6-2 c_8 g_2 \iota _3 o_6+2 c_{16} g_2 \iota _3 o_6-2 c_9 g_1 o_1 \iota _3 o_6+2 c_{17} g_1 o_1 \iota _3 o_6\\
&-2 c_9 g_2 o_2 \iota _3 o_6+2 c_{17} g_2 o_2 \iota _3 o_6+3 c c_{14} g_1 o_7 v_1 \iota _3 o_6+3 c c_{15} g_2 o_1 o_7 v_1 \iota _3 o_6-3 c c_{15} g_1 o_2 o_7 v_1 \iota _3 o_6\\
&-3 c c_8 g_1 o_7^2 o_8 v_1 \iota _3 o_6-3 c c_9 g_2 o_1 o_7^2 o_8 v_1 \iota _3 o_6+3 c c_9 g_1 o_2 o_7^2 o_8 v_1 \iota _3 o_6-3 c c_8 g_1 o_7^2 o_9 v_1 \iota _3 o_6\\
&-3 c c_9 g_2 o_1 o_7^2 o_9 v_1 \iota _3 o_6+3 c c_9 g_1 o_2 o_7^2 o_9 v_1 \iota _3 o_6-3 c_{14} g_2 o_7 \gamma _1 \iota _3 o_6-3 c_{15} g_1 o_1 o_7 \gamma _1 \iota _3 o_6\\
&-3 c_{15} g_2 o_2 o_7 \gamma _1 \iota _3 o_6-c_8 g_1 o_7 o_8-c_9 g_2 o_1 o_7 o_8+c_9 g_1 o_2 o_7 o_8-c_8 g_1 o_7 o_9-c_9 g_2 o_1 o_7 o_9+c_9 g_1 o_2 o_7 o_9\\
&+2 c c_8 g_2 o_7 v_1+c c_8 g_3 o_7 v_1+2 c c_9 g_1 o_1 o_7 v_1-c c_9 g_4 o_1 o_7 v_1+2 c c_9 g_2 o_2 o_7 v_1+c c_9 g_3 o_2 o_7 v_1+c_8 g_1 o_7 \gamma _1\\
&+c_9 g_2 o_1 o_7 \gamma _1-c_9 g_1 o_2 o_7 \gamma _1+c_{14} g_1 o_7^2 \iota _3+c_{15} g_2 o_1 o_7^2 \iota _3-c_{15} g_1 o_2 o_7^2 \iota _3-2 c_{14} g_1 \iota _3+c_8 g_2 o_7^2 \iota _3 o_6\\
&+2 c_{18} g_1 \iota _3-2 c_{15} g_2 o_1 \iota _3+2 c_{19} g_2 o_1 \iota _3+2 c_{15} g_1 o_2 \iota _3-2 c_{19} g_1 o_2 \iota _3-c_8 g_1 o_7^3 o_8 \iota _3-c_9 g_2 o_1 o_7^3 o_8 \iota _3\\
&+c_9 g_1 o_2 o_7^3 o_8 \iota _3+c_8 g_1 o_7 o_8 \iota _3-2 c_{16} g_1 o_7 o_8 \iota _3+c_9 g_2 o_1 o_7 o_8 \iota _3-2 c_{17} g_2 o_1 o_7 o_8 \iota _3-c_9 g_1 o_2 o_7 o_8 \iota _3\\
&+2 c_{17} g_1 o_2 o_7 o_8 \iota _3-c_8 g_1 o_7^3 o_9 \iota _3-c_9 g_2 o_1 o_7^3 o_9 \iota _3+c_9 g_1 o_2 o_7^3 o_9 \iota _3+c_8 g_1 o_7 o_9 \iota _3-2 c_{16} g_1 o_7 o_9 \iota _3\\
&+c_9 g_2 o_1 o_7 o_9 \iota _3-2 c_{17} g_2 o_1 o_7 o_9 \iota _3-c_9 g_1 o_2 o_7 o_9 \iota _3+2 c_{17} g_1 o_2 o_7 o_9 \iota _3-3 c_{16} g_1 o_7 \gamma _1 \iota _3-3 c_{17} g_2 o_1 o_7 \gamma _1 \iota _3\\
&+3 c_{17} g_1 o_2 o_7 \gamma _1 \iota _3+3 c_{14} g_1 o_7^2 o_8 \gamma _1 \iota _3+3 c_{15} g_2 o_1 o_7^2 o_8 \gamma _1 \iota _3-3 c_{15} g_1 o_2 o_7^2 o_8 \gamma _1 \iota _3+3 c_{15} g_2 o_1 o_7^2 o_9 \gamma _1 \iota _3,
\end{aligned}
\end{equation}

\begin{equation}
\begin{aligned}
y_3&=2 c_8 g_2 \iota _3 o_6^3+2 c_9 g_1 o_1 \iota _3 o_6^3+2 c_9 g_2 o_2 \iota _3 o_6^3+2 c_{14} g_1 o_6^2-2 c_{14} g_4 o_6^2+2 c_{15} g_2 o_1 o_6^2+2 c_{15} g_3 o_1 o_6^2-2 c_{15} g_1 o_2 o_6^2\\
&+2 c_{15} g_4 o_2 o_6^2-c_{14} g_1 \iota _2 o_6^2+c_{14} g_4 \iota _2 o_6^2-c_{15} g_2 o_1 \iota _2 o_6^2-c_{15} g_3 o_1 \iota _2 o_6^2+c_{15} g_1 o_2 \iota _2 o_6^2-c_{15} g_4 o_2 \iota _2 o_6^2\\
&+2 c_{14} g_1 \iota _3 o_6^2+2 c_{15} g_2 o_1 \iota _3 o_6^2-2 c_{15} g_1 o_2 \iota _3 o_6^2-2 c_8 g_1 o_7 o_8 \iota _3 o_6^2-2 c_9 g_2 o_1 o_7 o_8 \iota _3 o_6^2+2 c_9 g_1 o_2 o_7 o_8 \iota _3 o_6^2\\
&-2 c_8 g_1 o_7 o_9 \iota _3 o_6^2-2 c_9 g_2 o_1 o_7 o_9 \iota _3 o_6^2+2 c_9 g_1 o_2 o_7 o_9 \iota _3 o_6^2-3 c c_8 g_2 o_7 v_1 \iota _3 o_6^2-3 c c_9 g_1 o_1 o_7 v_1 \iota _3 o_6^2\\
&-3 c c_9 g_2 o_2 o_7 v_1 \iota _3 o_6^2+2 c_{16} g_2 o_6+2 c_{16} g_3 o_6+2 c_{17} g_1 o_1 o_6-2 c_{17} g_4 o_1 o_6+2 c_{17} g_2 o_2 o_6+2 c_{17} g_3 o_2 o_6\\
&-3 c_{15} g_1 o_1 o_7 o_8 o_6+2 c_{15} g_4 o_1 o_7 o_8 o_6-3 c_{15} g_2 o_2 o_7 o_8 o_6-2 c_{15} g_3 o_2 o_7 o_8 o_6+3 c_{14} g_2 o_7 o_9 o_6+2 c_{14} g_3 o_7 o_9 o_6\\
&-2 c_{15} g_4 o_1 o_7 o_9 o_6+3 c_{15} g_2 o_2 o_7 o_9 o_6+2 c_{15} g_3 o_2 o_7 o_9 o_6-2 c c_{14} g_1 o_7 v_1 o_6+c c_{14} g_4 o_7 v_1 o_6-2 c c_{15} g_2 o_1 o_7 v_1 o_6\\
&+2 c c_{15} g_1 o_2 o_7 v_1 o_6-c c_{15} g_4 o_2 o_7 v_1 o_6+c_{14} g_2 o_7 \gamma _1 o_6+c_{15} g_1 o_1 o_7 \gamma _1 o_6+c_{15} g_2 o_2 o_7 \gamma _1 o_6-c_{16} g_2 \iota _2 o_6\\
&-c_{17} g_1 o_1 \iota _2 o_6+c_{17} g_4 o_1 \iota _2 o_6-c_{17} g_2 o_2 \iota _2 o_6-c_{17} g_3 o_2 \iota _2 o_6+2 c_{14} g_2 o_7 o_8 \iota _2 o_6+c_{14} g_3 o_7 o_8 \iota _2 o_6\\
&-c_{15} g_4 o_1 o_7 o_8 \iota _2 o_6+2 c_{15} g_2 o_2 o_7 o_8 \iota _2 o_6+c_{15} g_3 o_2 o_7 o_8 \iota _2 o_6-2 c_{14} g_2 o_7 o_9 \iota _2 o_6-c_{14} g_3 o_7 o_9 \iota _2 o_6\\
&+c_{15} g_4 o_1 o_7 o_9 \iota _2 o_6-2 c_{15} g_2 o_2 o_7 o_9 \iota _2 o_6-c_{15} g_3 o_2 o_7 o_9 \iota _2 o_6+2 c c_{14} g_1 o_7 v_1 \iota _2 o_6-c c_{14} g_4 o_7 v_1 \iota _2 o_6\\
&+c c_{15} g_3 o_1 o_7 v_1 \iota _2 o_6-2 c c_{15} g_1 o_2 o_7 v_1 \iota _2 o_6+c c_{15} g_4 o_2 o_7 v_1 \iota _2 o_6-c_{14} g_2 o_7 \gamma _1 \iota _2 o_6-c_{15} g_1 o_1 o_7 \gamma _1 \iota _2 o_6\\
&-c_{15} g_2 o_2 o_7 \gamma _1 \iota _2 o_6-c_8 g_2 o_7^2 \iota _3 o_6-c_9 g_1 o_1 o_7^2 \iota _3 o_6-c_9 g_2 o_2 o_7^2 \iota _3 o_6+2 c_8 g_2 \iota _3 o_6-2 c_{16} g_2 \iota _3 o_6+2 c_9 g_1 o_1 \iota _3 o_6\\
&-2 c_{17} g_1 o_1 \iota _3 o_6+2 c_9 g_2 o_2 \iota _3 o_6-2 c_{17} g_2 o_2 \iota _3 o_6-3 c c_{14} g_1 o_7 v_1 \iota _3 o_6-3 c c_{15} g_2 o_1 o_7 v_1 \iota _3 o_6+3 c c_{15} g_1 o_2 o_7 v_1 \iota _3 o_6\\
&+3 c c_8 g_1 o_7^2 o_8 v_1 \iota _3 o_6+3 c c_9 g_2 o_1 o_7^2 o_8 v_1 \iota _3 o_6-3 c c_9 g_1 o_2 o_7^2 o_8 v_1 \iota _3 o_6+3 c c_8 g_1 o_7^2 o_9 v_1 \iota _3 o_6+3 c c_9 g_2 o_1 o_7^2 o_9 v_1 \iota _3 o_6\\
&-3 c c_9 g_1 o_2 o_7^2 o_9 v_1 \iota _3 o_6+3 c_{14} g_2 o_7 \gamma _1 \iota _3 o_6+3 c_{15} g_1 o_1 o_7 \gamma _1 \iota _3 o_6+3 c_{15} g_2 o_2 o_7 \gamma _1 \iota _3 o_6-c_{14} g_1 o_7^2\\
&-c_{15} g_2 o_1 o_7^2+c_{15} g_1 o_2 o_7^2+2 c_8 g_1 o_7 o_8-c_{16} g_1 o_7 o_8+2 c_9 g_2 o_1 o_7 o_8-c_{17} g_2 o_1 o_7 o_8-2 c_9 g_1 o_2 o_7 o_8\\
&+c_{17} g_2 o_1 o_7 o_9-c_{17} g_1 o_2 o_7 o_9-2 c c_{16} g_2 o_7 v_1-c c_{16} g_3 o_7 v_1-2 c c_{17} g_1 o_1 o_7 v_1+c c_{17} g_4 o_1 o_7 v_1-2 c c_{17} g_2 o_2 o_7 v_1\\
&+2 c c_{14} g_2 o_7^2 o_8 v_1+c c_{14} g_3 o_7^2 o_8 v_1+2 c c_{15} g_1 o_1 o_7^2 o_8 v_1-c c_{15} g_4 o_1 o_7^2 o_8 v_1+2 c c_{15} g_2 o_2 o_7^2 o_8 v_1+c c_{15} g_3 o_2 o_7^2 o_8 v_1\\
&-2 c c_{14} g_2 o_7^2 o_9 v_1-c c_{14} g_3 o_7^2 o_9 v_1-2 c c_{15} g_1 o_1 o_7^2 o_9 v_1+c c_{15} g_4 o_1 o_7^2 o_9 v_1-2 c c_{15} g_2 o_2 o_7^2 o_9 v_1-c c_{15} g_3 o_2 o_7^2 o_9 v_1\\
&+c_{16} g_1 o_7 \gamma _1+c_{17} g_2 o_1 o_7 \gamma _1-c_{17} g_1 o_2 o_7 \gamma _1+c_{14} g_1 o_7^2 o_8 \gamma _1+c_{15} g_2 o_1 o_7^2 o_8 \gamma _1-c_{15} g_1 o_2 o_7^2 o_8 \gamma _1\\
&-c_{14} g_1 o_7^2 o_9 \gamma _1-c_{15} g_2 o_1 o_7^2 o_9 \gamma _1+c_{15} g_1 o_2 o_7^2 o_9 \gamma _1+c_{14} g_1 o_7^2 \iota _2+c_{15} g_2 o_1 o_7^2 \iota _2-c_{15} g_1 o_2 o_7^2 \iota _2\\
&-3 c_{14} g_1 \iota _2+c_{14} g_4 \iota _2-3 c_{15} g_2 o_1 \iota _2-c_{15} g_3 o_1 \iota _2+3 c_{15} g_1 o_2 \iota _2-c_{15} g_4 o_2 \iota _2+c_{16} g_1 o_7 o_8 \iota _2+c_{17} g_2 o_1 o_7 o_8 \iota _2\\
&-c_{17} g_1 o_2 o_7 o_8 \iota _2-c_{16} g_1 o_7 o_9 \iota _2-c_{17} g_2 o_1 o_7 o_9 \iota _2+c_{17} g_1 o_2 o_7 o_9 \iota _2+2 c c_{16} g_2 o_7 v_1 \iota _2+c c_{16} g_3 o_7 v_1 \iota _2\\
&+2 c c_{17} g_1 o_1 o_7 v_1 \iota _2-c c_{17} g_4 o_1 o_7 v_1 \iota _2+2 c c_{17} g_2 o_2 o_7 v_1 \iota _2+c c_{17} g_3 o_2 o_7 v_1 \iota _2-2 c c_{14} g_2 o_7^2 o_8 v_1 \iota _2\\
&-c c_{14} g_3 o_7^2 o_8 v_1 \iota _2-2 c c_{15} g_1 o_1 o_7^2 o_8 v_1 \iota _2+c c_{15} g_4 o_1 o_7^2 o_8 v_1 \iota _2-2 c c_{15} g_2 o_2 o_7^2 o_8 v_1 \iota _2-c c_{15} g_3 o_2 o_7^2 o_8 v_1 \iota _2\\
&+2 c c_{14} g_2 o_7^2 o_9 v_1 \iota _2+c c_{14} g_3 o_7^2 o_9 v_1 \iota _2+2 c c_{15} g_1 o_1 o_7^2 o_9 v_1 \iota _2-c c_{15} g_4 o_1 o_7^2 o_9 v_1 \iota _2+2 c c_{15} g_2 o_2 o_7^2 o_9 v_1 \iota _2\\
&+c c_{15} g_3 o_2 o_7^2 o_9 v_1 \iota _2-c_{16} g_1 o_7 \gamma _1 \iota _2-c_{17} g_2 o_1 o_7 \gamma _1 \iota _2+c_{17} g_1 o_2 o_7 \gamma _1 \iota _2-c_{14} g_1 o_7^2 o_8 \gamma _1 \iota _2\\
&-c_{15} g_2 o_1 o_7^2 o_8 \gamma _1 \iota _2+c_{15} g_1 o_2 o_7^2 o_8 \gamma _1 \iota _2+c_{14} g_1 o_7^2 o_9 \gamma _1 \iota _2+c_{15} g_2 o_1 o_7^2 o_9 \gamma _1 \iota _2\\
&-c_{15} g_1 o_2 o_7^2 o_9 \gamma _1 \iota _2-c_{14} g_1 o_7^2 \iota _3-c_{15} g_2 o_1 o_7^2 \iota _3+c_{15} g_1 o_2 o_7^2 \iota _3+2 c_{14} g_1 \iota _3-2 c_{18} g_1 \iota _3+2 c_{15} g_2 o_1 \iota _3\\
&-2 c_{19} g_2 o_1 \iota _3-2 c_{15} g_1 o_2 \iota _3+2 c_{19} g_1 o_2 \iota _3+c_8 g_1 o_7^3 o_8 \iota _3+c_9 g_2 o_1 o_7^3 o_8 \iota _3-c_9 g_1 o_2 o_7^3 o_8 \iota _3-c_8 g_1 o_7 o_8 \iota _3\\
&+2 c_{16} g_1 o_7 o_8 \iota _3-c_9 g_2 o_1 o_7 o_8 \iota _3+2 c_{17} g_2 o_1 o_7 o_8 \iota _3+c_9 g_1 o_2 o_7 o_8 \iota _3-2 c_{17} g_1 o_2 o_7 o_8 \iota _3+c_8 g_1 o_7^3 o_9 \iota _3\\
&+c_9 g_2 o_1 o_7^3 o_9 \iota _3-c_9 g_1 o_2 o_7^3 o_9 \iota _3-c_8 g_1 o_7 o_9 \iota _3+2 c_{16} g_1 o_7 o_9 \iota _3-c_9 g_2 o_1 o_7 o_9 \iota _3+2 c_{17} g_2 o_1 o_7 o_9 \iota _3\\
&-2 c_{17} g_1 o_2 o_7 o_9 \iota _3+3 c_{16} g_1 o_7 \gamma _1 \iota _3+3 c_{17} g_2 o_1 o_7 \gamma _1 \iota _3-3 c_{17} g_1 o_2 o_7 \gamma _1 \iota _3-3 c_{14} g_1 o_7^2 o_8 \gamma _1 \iota _3\\
&-3 c_{15} g_2 o_1 o_7^2 o_8 \gamma _1 \iota _3+3 c_{15} g_1 o_2 o_7^2 o_8 \gamma _1 \iota _3-3 c_{14} g_1 o_7^2 o_9 \gamma _1 \iota _3-3 c_{15} g_2 o_1 o_7^2 o_9 \gamma _1 \iota _3+3 c_{15} g_1 o_2 o_7^2 o_9 \gamma _1 \iota _3,\\
&-2 c_{14} g_3 o_7 o_8 o_6+3 c_{15} g_1 o_1 o_7 o_9 o_6-c c_{15} g_3 o_1 o_7 v_1 o_6-2 c_{15} g_1 o_1 o_7 o_9 \iota _2 o_6+2 c_{15} g_1 o_1 o_7 o_8 \iota _2 o_6-c_{16} g_3 \iota _2 o_6\\
&-3 c_{14} g_2 o_7 o_8 o_6+c_9 g_1 o_2 o_7 o_9 \iota _3-c c_{17} g_3 o_2 o_7 v_1+2 c c_{15} g_2 o_1 o_7 v_1 \iota _2 o_6+c_{17} g_1 o_2 o_7 o_8+c_{16} g_1 o_7 o_9,
\end{aligned}
\end{equation}
\begin{equation}
\begin{aligned}
y_4&=-2 c_8 g_1 \iota _3 o_6^3-2 c_9 g_2 o_1 \iota _3 o_6^3+2 c_9 g_1 o_2 \iota _3 o_6^3+2 c_{14} g_2 o_6^2+2 c_{14} g_3 o_6^2+2 c_{15} g_1 o_1 o_6^2-2 c_{15} g_4 o_1 o_6^2\\
&+2 c_{15} g_2 o_2 o_6^2+2 c_{15} g_3 o_2 o_6^2-c_{14} g_2 \iota _2 o_6^2-c_{14} g_3 \iota _2 o_6^2-c_{15} g_1 o_1 \iota _2 o_6^2+c_{15} g_4 o_1 \iota _2 o_6^2-c_{15} g_2 o_2 \iota _2 o_6^2\\
&-c_{15} g_3 o_2 \iota _2 o_6^2-2 c_{14} g_2 \iota _3 o_6^2-2 c_{15} g_1 o_1 \iota _3 o_6^2-2 c_{15} g_2 o_2 \iota _3 o_6^2+2 c_8 g_2 o_7 o_8 \iota _3 o_6^2+2 c_9 g_1 o_1 o_7 o_8 \iota _3 o_6^2\\
&+2 c_9 g_2 o_2 o_7 o_8 \iota _3 o_6^2-2 c_8 g_2 o_7 o_9 \iota _3 o_6^2-2 c_9 g_1 o_1 o_7 o_9 \iota _3 o_6^2-2 c_9 g_2 o_2 o_7 o_9 \iota _3 o_6^2+3 c c_8 g_1 o_7 v_1 \iota _3 o_6^2\\
&+3 c c_9 g_2 o_1 o_7 v_1 \iota _3 o_6^2-3 c c_9 g_1 o_2 o_7 v_1 \iota _3 o_6^2+2 c_{16} g_1 o_6-2 c_{16} g_4 o_6+2 c_{17} g_2 o_1 o_6+2 c_{17} g_3 o_1 o_6-2 c_{17} g_1 o_2 o_6\\
&+2 c_{17} g_4 o_2 o_6-c_{14} g_1 o_7 o_8 o_6+2 c_{14} g_4 o_7 o_8 o_6-c_{15} g_2 o_1 o_7 o_8 o_6-2 c_{15} g_3 o_1 o_7 o_8 o_6+c_{15} g_1 o_2 o_7 o_8 o_6\\
&-c_{14} g_1 o_7 o_9 o_6+2 c_{14} g_4 o_7 o_9 o_6-c_{15} g_2 o_1 o_7 o_9 o_6-2 c_{15} g_3 o_1 o_7 o_9 o_6+c_{15} g_1 o_2 o_7 o_9 o_6-2 c_{15} g_4 o_2 o_7 o_9 o_6\\
&+c c_{15} g_4 o_1 o_7 v_1 o_6-c c_{15} g_3 o_2 o_7 v_1 o_6-c_{14} g_1 o_7 \gamma _1 o_6-c_{15} g_2 o_1 o_7 \gamma _1 o_6+c_{15} g_1 o_2 o_7 \gamma _1 o_6-c_{16} g_1 \iota _2 o_6\\
&+c_{16} g_4 \iota _2 o_6-c_{17} g_2 o_1 \iota _2 o_6-c_{17} g_3 o_1 \iota _2 o_6+c_{17} g_1 o_2 \iota _2 o_6-c_{17} g_4 o_2 \iota _2 o_6-c_{14} g_4 o_7 o_8 \iota _2 o_6\\
&+c_{15} g_3 o_1 o_7 o_8 \iota _2 o_6+c_{15} g_4 o_2 o_7 o_8 \iota _2 o_6-c_{14} g_4 o_7 o_9 \iota _2 o_6+c_{15} g_3 o_1 o_7 o_9 \iota _2 o_6+c_{15} g_4 o_2 o_7 o_9 \iota _2 o_6\\
&+c c_{14} g_3 o_7 v_1 \iota _2 o_6-c c_{15} g_4 o_1 o_7 v_1 \iota _2 o_6+c c_{15} g_3 o_2 o_7 v_1 \iota _2 o_6+c_{14} g_1 o_7 \gamma _1 \iota _2 o_6+c_{15} g_2 o_1 o_7 \gamma _1 \iota _2 o_6\\
&-c_{15} g_1 o_2 o_7 \gamma _1 \iota _2 o_6+c_8 g_1 o_7^2 \iota _3 o_6+c_9 g_2 o_1 o_7^2 \iota _3 o_6-c_9 g_1 o_2 o_7^2 \iota _3 o_6-2 c_8 g_1 \iota _3 o_6+2 c_{16} g_1 \iota _3 o_6\\
&-2 c_9 g_2 o_1 \iota _3 o_6+2 c_{17} g_2 o_1 \iota _3 o_6+2 c_9 g_1 o_2 \iota _3 o_6-2 c_{17} g_1 o_2 \iota _3 o_6+3 c c_{14} g_2 o_7 v_1 \iota _3 o_6+3 c c_{15} g_1 o_1 o_7 v_1 \iota _3 o_6\\
&+3 c c_{15} g_2 o_2 o_7 v_1 \iota _3 o_6-3 c c_8 g_2 o_7^2 o_8 v_1 \iota _3 o_6-3 c c_9 g_1 o_1 o_7^2 o_8 v_1 \iota _3 o_6-3 c c_9 g_2 o_2 o_7^2 o_8 v_1 \iota _3 o_6\\
&+3 c c_8 g_2 o_7^2 o_9 v_1 \iota _3 o_6+3 c c_9 g_1 o_1 o_7^2 o_9 v_1 \iota _3 o_6+3 c c_9 g_2 o_2 o_7^2 o_9 v_1 \iota _3 o_6-3 c_{14} g_1 o_7 \gamma _1 \iota _3 o_6\\
&-3 c_{15} g_2 o_1 o_7 \gamma _1 \iota _3 o_6+3 c_{15} g_1 o_2 o_7 \gamma _1 \iota _3 o_6+c_{14} g_2 o_7^2+c_{15} g_1 o_1 o_7^2+c_{15} g_2 o_2 o_7^2-2 c_8 g_2 o_7 o_8\\
&+c_{16} g_2 o_7 o_8-2 c_9 g_1 o_1 o_7 o_8+c_{17} g_1 o_1 o_7 o_8-2 c_9 g_2 o_2 o_7 o_8+c_{17} g_2 o_2 o_7 o_8+c_{16} g_2 o_7 o_9+c_{17} g_1 o_1 o_7 o_9\\
&+c c_{16} g_4 o_7 v_1-c c_{17} g_3 o_1 o_7 v_1-c c_{17} g_4 o_2 o_7 v_1-c c_{14} g_4 o_7^2 o_8 v_1+c c_{15} g_3 o_1 o_7^2 o_8 v_1+c c_{15} g_4 o_2 o_7^2 o_8 v_1\\
&+c c_{15} g_3 o_1 o_7^2 o_9 v_1+c c_{15} g_4 o_2 o_7^2 o_9 v_1-c_{16} g_2 o_7 \gamma _1-c_{17} g_1 o_1 o_7 \gamma _1-c_{17} g_2 o_2 o_7 \gamma _1-c_{14} g_2 o_7^2 o_8 \gamma _1\\
&-c_{15} g_1 o_1 o_7^2 o_8 \gamma _1-c_{15} g_2 o_2 o_7^2 o_8 \gamma _1-c_{14} g_2 o_7^2 o_9 \gamma _1-c_{15} g_1 o_1 o_7^2 o_9 \gamma _1-c_{15} g_2 o_2 o_7^2 o_9 \gamma _1\\
&-c_{14} g_2 o_7^2 \iota _2-c_{15} g_1 o_1 o_7^2 \iota _2-c_{15} g_2 o_2 o_7^2 \iota _2+c_{14} g_2 \iota _2-c_{14} g_3 \iota _2+c_{15} g_1 o_1 \iota _2+c_{15} g_4 o_1 \iota _2\\
&+c_{15} g_2 o_2 \iota _2-c_{15} g_3 o_2 \iota _2-c_{16} g_2 o_7 o_8 \iota _2-c_{17} g_1 o_1 o_7 o_8 \iota _2-c_{17} g_2 o_2 o_7 o_8 \iota _2-c_{16} g_2 o_7 o_9 \iota _2-c_{17} g_1 o_1 o_7 o_9 \iota _2\\
&-c_{17} g_2 o_2 o_7 o_9 \iota _2-c c_{16} g_4 o_7 v_1 \iota _2+c c_{17} g_3 o_1 o_7 v_1 \iota _2+c c_{17} g_4 o_2 o_7 v_1 \iota _2+c c_{14} g_4 o_7^2 o_8 v_1 \iota _2\\
&-c c_{15} g_3 o_1 o_7^2 o_8 v_1 \iota _2-c c_{15} g_4 o_2 o_7^2 o_8 v_1 \iota _2+c c_{14} g_4 o_7^2 o_9 v_1 \iota _2-c c_{15} g_3 o_1 o_7^2 o_9 v_1 \iota _2-c c_{15} g_4 o_2 o_7^2 o_9 v_1 \iota _2\\
&+c_{16} g_2 o_7 \gamma _1 \iota _2+c_{17} g_1 o_1 o_7 \gamma _1 \iota _2+c_{17} g_2 o_2 o_7 \gamma _1 \iota _2+c_{14} g_2 o_7^2 o_8 \gamma _1 \iota _2+c_{15} g_1 o_1 o_7^2 o_8 \gamma _1 \iota _2\\
&+c_{15} g_2 o_2 o_7^2 o_8 \gamma _1 \iota _2+c_{14} g_2 o_7^2 o_9 \gamma _1 \iota _2+c_{15} g_1 o_1 o_7^2 o_9 \gamma _1 \iota _2+c_{15} g_2 o_2 o_7^2 o_9 \gamma _1 \iota _2\\
&+c_{14} g_2 o_7^2 \iota _3+c_{15} g_1 o_1 o_7^2 \iota _3+c_{15} g_2 o_2 o_7^2 \iota _3-2 c_{14} g_2 \iota _3+2 c_{18} g_2 \iota _3-2 c_{15} g_1 o_1 \iota _3+2 c_{19} g_1 o_1 \iota _3\\
&-2 c_{15} g_2 o_2 \iota _3+2 c_{19} g_2 o_2 \iota _3-c_8 g_2 o_7^3 o_8 \iota _3-c_9 g_1 o_1 o_7^3 o_8 \iota _3-c_9 g_2 o_2 o_7^3 o_8 \iota _3+c_8 g_2 o_7 o_8 \iota _3-2 c_{16} g_2 o_7 o_8 \iota _3\\
&+c_9 g_1 o_1 o_7 o_8 \iota _3-2 c_{17} g_1 o_1 o_7 o_8 \iota _3+c_9 g_2 o_2 o_7 o_8 \iota _3-2 c_{17} g_2 o_2 o_7 o_8 \iota _3+c_8 g_2 o_7^3 o_9 \iota _3+c_9 g_1 o_1 o_7^3 o_9 \iota _3\\
&+c_9 g_2 o_2 o_7^3 o_9 \iota _3-c_8 g_2 o_7 o_9 \iota _3+2 c_{16} g_2 o_7 o_9 \iota _3-c_9 g_1 o_1 o_7 o_9 \iota _3+2 c_{17} g_1 o_1 o_7 o_9 \iota _3-c_9 g_2 o_2 o_7 o_9 \iota _3\\
&+2 c_{17} g_2 o_2 o_7 o_9 \iota _3-3 c_{16} g_2 o_7 \gamma _1 \iota _3-3 c_{17} g_1 o_1 o_7 \gamma _1 \iota _3-3 c_{17} g_2 o_2 o_7 \gamma _1 \iota _3+3 c_{14} g_2 o_7^2 o_8 \gamma _1 \iota _3\\
&+3 c_{15} g_1 o_1 o_7^2 o_8 \gamma _1 \iota _3+3 c_{15} g_2 o_2 o_7^2 o_8 \gamma _1 \iota _3-3 c_{14} g_2 o_7^2 o_9 \gamma _1 \iota _3-3 c_{15} g_1 o_1 o_7^2 o_9 \gamma _1 \iota _3\\
&-3 c_{15} g_2 o_2 o_7^2 o_9 \gamma _1 \iota _3-c c_{14} g_3 o_7 v_1 o_6-2 c_{15} g_4 o_2 o_7 o_8 o_6-c c_{14} g_4 o_7^2 o_9 v_1+c_{17} g_2 o_2 o_7 o_9,\\
\end{aligned}
\end{equation}
\begin{equation}
\begin{aligned}
y_6&=2 c_8 g_2 o_6^3+2 c_8 g_3 o_6^3+2 c_9 g_1 o_1 o_6^3-2 c_9 g_4 o_1 o_6^3+2 c_9 g_2 o_2 o_6^3+2 c_9 g_3 o_2 o_6^3-c_8 g_2 \iota _2 o_6^3-c_8 g_3 \iota _2 o_6^3\\
&-c_9 g_1 o_1 \iota _2 o_6^3+c_9 g_4 o_1 \iota _2 o_6^3-c_9 g_2 o_2 \iota _2 o_6^3-c_9 g_3 o_2 \iota _2 o_6^3+2 c_8 g_3 \iota _3 o_6^3-2 c_9 g_4 o_1 \iota _3 o_6^3+2 c_9 g_3 o_2 \iota _3 o_6^3\\
&+2 c_{14} g_1 o_6^2-2 c_{14} g_4 o_6^2+2 c_{15} g_2 o_1 o_6^2+2 c_{15} g_3 o_1 o_6^2-2 c_{15} g_1 o_2 o_6^2+2 c_{15} g_4 o_2 o_6^2-c_8 g_1 o_7 o_8 o_6^2\\
&-c_9 g_2 o_1 o_7 o_8 o_6^2-2 c_9 g_3 o_1 o_7 o_8 o_6^2+c_9 g_1 o_2 o_7 o_8 o_6^2-2 c_9 g_4 o_2 o_7 o_8 o_6^2-c_8 g_1 o_7 o_9 o_6^2+2 c_8 g_4 o_7 o_9 o_6^2\\
&-2 c_9 g_3 o_1 o_7 o_9 o_6^2+c_9 g_1 o_2 o_7 o_9 o_6^2-2 c_9 g_4 o_2 o_7 o_9 o_6^2-c c_8 g_3 o_7 v_1 o_6^2+c c_9 g_4 o_1 o_7 v_1 o_6^2-c c_9 g_3 o_2 o_7 v_1 o_6^2\\
&-c_8 g_1 o_7 \gamma _1 o_6^2-c_9 g_2 o_1 o_7 \gamma _1 o_6^2+c_9 g_1 o_2 o_7 \gamma _1 o_6^2-c_{14} g_1 \iota _2 o_6^2+c_{14} g_4 \iota _2 o_6^2-c_{15} g_2 o_1 \iota _2 o_6^2\\
&-c_{15} g_3 o_1 \iota _2 o_6^2+c_{15} g_1 o_2 \iota _2 o_6^2-c_{15} g_4 o_2 \iota _2 o_6^2-c_8 g_4 o_7 o_8 \iota _2 o_6^2+c_9 g_3 o_1 o_7 o_8 \iota _2 o_6^2+c_9 g_4 o_2 o_7 o_8 \iota _2 o_6^2\\
&-c_8 g_4 o_7 o_9 \iota _2 o_6^2+c_9 g_3 o_1 o_7 o_9 \iota _2 o_6^2+c_9 g_4 o_2 o_7 o_9 \iota _2 o_6^2+c c_8 g_3 o_7 v_1 \iota _2 o_6^2-c c_9 g_4 o_1 o_7 v_1 \iota _2 o_6^2\\
&+c c_9 g_3 o_2 o_7 v_1 \iota _2 o_6^2+c_8 g_1 o_7 \gamma _1 \iota _2 o_6^2+c_9 g_2 o_1 o_7 \gamma _1 \iota _2 o_6^2-c_9 g_1 o_2 o_7 \gamma _1 \iota _2 o_6^2\\
&-2 c_{14} g_4 \iota _3 o_6^2+2 c_{15} g_3 o_1 \iota _3 o_6^2+2 c_{15} g_4 o_2 \iota _3 o_6^2+2 c_8 g_4 o_7 o_8 \iota _3 o_6^2-2 c_9 g_3 o_1 o_7 o_8 \iota _3 o_6^2\\
&-2 c_9 g_4 o_2 o_7 o_8 \iota _3 o_6^2+2 c_8 g_4 o_7 o_9 \iota _3 o_6^2-2 c_9 g_3 o_1 o_7 o_9 \iota _3 o_6^2-2 c_9 g_4 o_2 o_7 o_9 \iota _3 o_6^2-3 c c_8 g_3 o_7 v_1 \iota _3 o_6^2\\
&+3 c c_9 g_4 o_1 o_7 v_1 \iota _3 o_6^2-3 c c_9 g_3 o_2 o_7 v_1 \iota _3 o_6^2+c_8 g_2 o_7^2 o_6+c_9 g_1 o_1 o_7^2 o_6+c_9 g_2 o_2 o_7^2 o_6+c_{14} g_2 o_7 o_8 o_6\\
&+c_{15} g_1 o_1 o_7 o_8 o_6+c_{15} g_2 o_2 o_7 o_8 o_6+c_{14} g_2 o_7 o_9 o_6+c_{15} g_1 o_1 o_7 o_9 o_6+c_{15} g_2 o_2 o_7 o_9 o_6+c c_{14} g_4 o_7 v_1 o_6\\
&-c c_{15} g_4 o_2 o_7 v_1 o_6-c c_8 g_4 o_7^2 o_8 v_1 o_6+c c_9 g_3 o_1 o_7^2 o_8 v_1 o_6+c c_9 g_4 o_2 o_7^2 o_8 v_1 o_6-c c_8 g_4 o_7^2 o_9 v_1 o_6\\
&+c c_9 g_4 o_2 o_7^2 o_9 v_1 o_6-c_{14} g_2 o_7 \gamma _1 o_6-c_{15} g_1 o_1 o_7 \gamma _1 o_6-c_{15} g_2 o_2 o_7 \gamma _1 o_6-c_8 g_2 o_7^2 o_8 \gamma _1 o_6\\
&-c_9 g_1 o_1 o_7^2 o_8 \gamma _1 o_6-c_9 g_2 o_2 o_7^2 o_8 \gamma _1 o_6-c_8 g_2 o_7^2 o_9 \gamma _1 o_6-c_9 g_1 o_1 o_7^2 o_9 \gamma _1 o_6-c_9 g_2 o_2 o_7^2 o_9 \gamma _1 o_6\\
&-c_8 g_2 o_7^2 \iota _2 o_6-c_9 g_1 o_1 o_7^2 \iota _2 o_6-c_9 g_2 o_2 o_7^2 \iota _2 o_6+c_8 g_2 \iota _2 o_6-c_8 g_3 \iota _2 o_6+c_9 g_1 o_1 \iota _2 o_6+c_9 g_4 o_1 \iota _2 o_6\\
&+c_9 g_2 o_2 \iota _2 o_6-c_9 g_3 o_2 \iota _2 o_6-c_{14} g_2 o_7 o_8 \iota _2 o_6-c_{15} g_1 o_1 o_7 o_8 \iota _2 o_6-c_{15} g_2 o_2 o_7 o_8 \iota _2 o_6-c_{14} g_2 o_7 o_9 \iota _2 o_6\\
&-c_{15} g_1 o_1 o_7 o_9 \iota _2 o_6-c_{15} g_2 o_2 o_7 o_9 \iota _2 o_6-c c_{14} g_4 o_7 v_1 \iota _2 o_6+c c_{15} g_3 o_1 o_7 v_1 \iota _2 o_6+c c_{15} g_4 o_2 o_7 v_1 \iota _2 o_6\\
&+c c_8 g_4 o_7^2 o_8 v_1 \iota _2 o_6-c c_9 g_3 o_1 o_7^2 o_8 v_1 \iota _2 o_6-c c_9 g_4 o_2 o_7^2 o_8 v_1 \iota _2 o_6+c c_8 g_4 o_7^2 o_9 v_1 \iota _2 o_6\\
&-c c_9 g_3 o_1 o_7^2 o_9 v_1 \iota _2 o_6-c c_9 g_4 o_2 o_7^2 o_9 v_1 \iota _2 o_6+c_{14} g_2 o_7 \gamma _1 \iota _2 o_6+c_{15} g_1 o_1 o_7 \gamma _1 \iota _2 o_6\\
&+c_{15} g_2 o_2 o_7 \gamma _1 \iota _2 o_6+c_8 g_2 o_7^2 o_8 \gamma _1 \iota _2 o_6+c_9 g_1 o_1 o_7^2 o_8 \gamma _1 \iota _2 o_6+c_9 g_2 o_2 o_7^2 o_8 \gamma _1 \iota _2 o_6\\
&+c_8 g_2 o_7^2 o_9 \gamma _1 \iota _2 o_6+c_9 g_1 o_1 o_7^2 o_9 \gamma _1 \iota _2 o_6+c_9 g_2 o_2 o_7^2 o_9 \gamma _1 \iota _2 o_6-c_8 g_3 o_7^2 \iota _3 o_6\\
&+c_9 g_4 o_1 o_7^2 \iota _3 o_6-c_9 g_3 o_2 o_7^2 \iota _3 o_6+2 c_8 g_3 \iota _3 o_6-2 c_{16} g_3 \iota _3 o_6-2 c_9 g_4 o_1 \iota _3 o_6+2 c_{17} g_4 o_1 \iota _3 o_6\\
&+2 c_9 g_3 o_2 \iota _3 o_6-2 c_{17} g_3 o_2 \iota _3 o_6+3 c c_{14} g_4 o_7 v_1 \iota _3 o_6-3 c c_{15} g_3 o_1 o_7 v_1 \iota _3 o_6-3 c c_{15} g_4 o_2 o_7 v_1 \iota _3 o_6\\
&-3 c c_8 g_4 o_7^2 o_8 v_1 \iota _3 o_6+3 c c_9 g_3 o_1 o_7^2 o_8 v_1 \iota _3 o_6+3 c c_9 g_4 o_2 o_7^2 o_8 v_1 \iota _3 o_6-3 c c_8 g_4 o_7^2 o_9 v_1 \iota _3 o_6\\
&+3 c c_9 g_3 o_1 o_7^2 o_9 v_1 \iota _3 o_6+3 c c_9 g_4 o_2 o_7^2 o_9 v_1 \iota _3 o_6+3 c_{14} g_3 o_7 \gamma _1 \iota _3 o_6-3 c_{15} g_4 o_1 o_7 \gamma _1 \iota _3 o_6\\
&+3 c_{15} g_3 o_2 o_7 \gamma _1 \iota _3 o_6+c_{14} g_4 o_7^2 \iota _3-c_{15} g_3 o_1 o_7^2 \iota _3-c_{15} g_4 o_2 o_7^2 \iota _3-2 c_{14} g_4 \iota _3+2 c_{18} g_4 \iota _3\\
&+2 c_{15} g_3 o_1 \iota _3-2 c_{19} g_3 o_1 \iota _3+2 c_{15} g_4 o_2 \iota _3-2 c_{19} g_4 o_2 \iota _3-c_8 g_4 o_7^3 o_8 \iota _3+c_9 g_3 o_1 o_7^3 o_8 \iota _3+c_9 g_4 o_2 o_7^3 o_8 \iota _3\\
&+c_8 g_4 o_7 o_8 \iota _3-2 c_{16} g_4 o_7 o_8 \iota _3-c_9 g_3 o_1 o_7 o_8 \iota _3+2 c_{17} g_3 o_1 o_7 o_8 \iota _3-c_9 g_4 o_2 o_7 o_8 \iota _3+2 c_{17} g_4 o_2 o_7 o_8 \iota _3\\
&-c_8 g_4 o_7^3 o_9 \iota _3+c_9 g_3 o_1 o_7^3 o_9 \iota _3+c_9 g_4 o_2 o_7^3 o_9 \iota _3+c_8 g_4 o_7 o_9 \iota _3-2 c_{16} g_4 o_7 o_9 \iota _3-c_9 g_3 o_1 o_7 o_9 \iota _3\\
&+2 c_{17} g_3 o_1 o_7 o_9 \iota _3-c_9 g_4 o_2 o_7 o_9 \iota _3+2 c_{17} g_4 o_2 o_7 o_9 \iota _3-3 c_{16} g_4 o_7 \gamma _1 \iota _3+3 c_{17} g_3 o_1 o_7 \gamma _1 \iota _3\\
&+3 c_{17} g_4 o_2 o_7 \gamma _1 \iota _3+3 c_{14} g_4 o_7^2 o_8 \gamma _1 \iota _3-3 c_{15} g_3 o_1 o_7^2 o_8 \gamma _1 \iota _3-3 c_{15} g_4 o_2 o_7^2 o_8 \gamma _1 \iota _3\\
&+3 c_{14} g_4 o_7^2 o_9 \gamma _1 \iota _3-3 c_{15} g_3 o_1 o_7^2 o_9 \gamma _1 \iota _3-3 c_{15} g_4 o_2 o_7^2 o_9 \gamma _1 \iota _3-c_9 g_2 o_1 o_7 o_9 o_6^2+2 c_8 g_4 o_7 o_8 o_6^2,\\
&+c c_9 g_3 o_1 o_7^2 o_9 v_1 o_6-c c_{15} g_3 o_1 o_7 v_1 o_6,
\end{aligned}
\end{equation}
\begin{equation}
\begin{aligned}
y_5&=2 c_8 g_1 o_6^3-2 c_8 g_4 o_6^3+2 c_9 g_2 o_1 o_6^3+2 c_9 g_3 o_1 o_6^3-2 c_9 g_1 o_2 o_6^3+2 c_9 g_4 o_2 o_6^3-c_8 g_1 \iota _2 o_6^3+c_8 g_4 \iota _2 o_6^3\\
&-c_9 g_2 o_1 \iota _2 o_6^3-c_9 g_3 o_1 \iota _2 o_6^3+c_9 g_1 o_2 \iota _2 o_6^3-c_9 g_4 o_2 \iota _2 o_6^3-2 c_8 g_4 \iota _3 o_6^3+2 c_9 g_3 o_1 \iota _3 o_6^3\\
&+2 c_9 g_4 o_2 \iota _3 o_6^3+2 c_{14} g_2 o_6^2+2 c_{14} g_3 o_6^2+2 c_{15} g_1 o_1 o_6^2-2 c_{15} g_4 o_1 o_6^2+2 c_{15} g_2 o_2 o_6^2+2 c_{15} g_3 o_2 o_6^2\\
&-3 c_8 g_2 o_7 o_8 o_6^2-2 c_8 g_3 o_7 o_8 o_6^2-3 c_9 g_1 o_1 o_7 o_8 o_6^2+2 c_9 g_4 o_1 o_7 o_8 o_6^2-3 c_9 g_2 o_2 o_7 o_8 o_6^2-2 c_9 g_3 o_2 o_7 o_8 o_6^2\\
&+3 c_8 g_2 o_7 o_9 o_6^2+2 c_8 g_3 o_7 o_9 o_6^2+3 c_9 g_1 o_1 o_7 o_9 o_6^2-2 c_9 g_4 o_1 o_7 o_9 o_6^2+3 c_9 g_2 o_2 o_7 o_9 o_6^2+2 c_9 g_3 o_2 o_7 o_9 o_6^2\\
&-2 c c_8 g_1 o_7 v_1 o_6^2+c c_8 g_4 o_7 v_1 o_6^2-2 c c_9 g_2 o_1 o_7 v_1 o_6^2-c c_9 g_3 o_1 o_7 v_1 o_6^2+2 c c_9 g_1 o_2 o_7 v_1 o_6^2-c c_9 g_4 o_2 o_7 v_1 o_6^2\\
&+c_8 g_2 o_7 \gamma _1 o_6^2+c_9 g_1 o_1 o_7 \gamma _1 o_6^2+c_9 g_2 o_2 o_7 \gamma _1 o_6^2-c_{14} g_2 \iota _2 o_6^2-c_{14} g_3 \iota _2 o_6^2\\
&-c_{15} g_1 o_1 \iota _2 o_6^2+c_{15} g_4 o_1 \iota _2 o_6^2-c_{15} g_2 o_2 \iota _2 o_6^2-c_{15} g_3 o_2 \iota _2 o_6^2+2 c_8 g_2 o_7 o_8 \iota _2 o_6^2\\
&+c_8 g_3 o_7 o_8 \iota _2 o_6^2+2 c_9 g_1 o_1 o_7 o_8 \iota _2 o_6^2-c_9 g_4 o_1 o_7 o_8 \iota _2 o_6^2+2 c_9 g_2 o_2 o_7 o_8 \iota _2 o_6^2+c_9 g_3 o_2 o_7 o_8 \iota _2 o_6^2\\
&-2 c_8 g_2 o_7 o_9 \iota _2 o_6^2-c_8 g_3 o_7 o_9 \iota _2 o_6^2-2 c_9 g_1 o_1 o_7 o_9 \iota _2 o_6^2+c_9 g_4 o_1 o_7 o_9 \iota _2 o_6^2-2 c_9 g_2 o_2 o_7 o_9 \iota _2 o_6^2\\
&-c_9 g_3 o_2 o_7 o_9 \iota _2 o_6^2+2 c c_8 g_1 o_7 v_1 \iota _2 o_6^2-c c_8 g_4 o_7 v_1 \iota _2 o_6^2+2 c c_9 g_2 o_1 o_7 v_1 \iota _2 o_6^2\\
&+c c_9 g_3 o_1 o_7 v_1 \iota _2 o_6^2-2 c c_9 g_1 o_2 o_7 v_1 \iota _2 o_6^2+c c_9 g_4 o_2 o_7 v_1 \iota _2 o_6^2-c_8 g_2 o_7 \gamma _1 \iota _2 o_6^2\\
&-c_9 g_1 o_1 o_7 \gamma _1 \iota _2 o_6^2-c_9 g_2 o_2 o_7 \gamma _1 \iota _2 o_6^2+2 c_{14} g_3 \iota _3 o_6^2-2 c_{15} g_4 o_1 \iota _3 o_6^2+2 c_{15} g_3 o_2 \iota _3 o_6^2\\
&-2 c_8 g_3 o_7 o_8 \iota _3 o_6^2+2 c_9 g_4 o_1 o_7 o_8 \iota _3 o_6^2-2 c_9 g_3 o_2 o_7 o_8 \iota _3 o_6^2+2 c_8 g_3 o_7 o_9 \iota _3 o_6^2\\
&-2 c_9 g_4 o_1 o_7 o_9 \iota _3 o_6^2+2 c_9 g_3 o_2 o_7 o_9 \iota _3 o_6^2+3 c c_8 g_4 o_7 v_1 \iota _3 o_6^2-3 c c_9 g_3 o_1 o_7 v_1 \iota _3 o_6^2\\
&-3 c c_9 g_4 o_2 o_7 v_1 \iota _3 o_6^2-c_8 g_1 o_7^2 o_6-c_9 g_2 o_1 o_7^2 o_6+c_9 g_1 o_2 o_7^2 o_6+4 c_8 g_1 o_6-4 c_8 g_4 o_6+4 c_9 g_2 o_1 o_6+4 c_9 g_3 o_1 o_6\\
&-4 c_9 g_1 o_2 o_6+4 c_9 g_4 o_2 o_6-c_{14} g_1 o_7 o_8 o_6-c_{15} g_2 o_1 o_7 o_8 o_6+c_{15} g_1 o_2 o_7 o_8 o_6+c_{14} g_1 o_7 o_9 o_6+c_{15} g_2 o_1 o_7 o_9 o_6\\
&-c_{15} g_1 o_2 o_7 o_9 o_6-2 c c_{14} g_2 o_7 v_1 o_6-c c_{14} g_3 o_7 v_1 o_6-2 c c_{15} g_1 o_1 o_7 v_1 o_6+c c_{15} g_4 o_1 o_7 v_1 o_6-2 c c_{15} g_2 o_2 o_7 v_1 o_6\\
&-c c_{15} g_3 o_2 o_7 v_1 o_6+2 c c_8 g_2 o_7^2 o_8 v_1 o_6+c c_8 g_3 o_7^2 o_8 v_1 o_6+2 c c_9 g_1 o_1 o_7^2 o_8 v_1 o_6-c c_9 g_4 o_1 o_7^2 o_8 v_1 o_6\\
&+2 c c_9 g_2 o_2 o_7^2 o_8 v_1 o_6+c c_9 g_3 o_2 o_7^2 o_8 v_1 o_6-2 c c_8 g_2 o_7^2 o_9 v_1 o_6-c c_8 g_3 o_7^2 o_9 v_1 o_6-2 c c_9 g_1 o_1 o_7^2 o_9 v_1 o_6\\
&+c c_9 g_4 o_1 o_7^2 o_9 v_1 o_6-2 c c_9 g_2 o_2 o_7^2 o_9 v_1 o_6-c c_9 g_3 o_2 o_7^2 o_9 v_1 o_6+c_{14} g_1 o_7 \gamma _1 o_6+c_{15} g_2 o_1 o_7 \gamma _1 o_6\\
&-c_{15} g_1 o_2 o_7 \gamma _1 o_6+c_8 g_1 o_7^2 o_8 \gamma _1 o_6+c_9 g_2 o_1 o_7^2 o_8 \gamma _1 o_6-c_9 g_1 o_2 o_7^2 o_8 \gamma _1 o_6\\
&-c_8 g_1 o_7^2 o_9 \gamma _1 o_6-c_9 g_2 o_1 o_7^2 o_9 \gamma _1 o_6+c_9 g_1 o_2 o_7^2 o_9 \gamma _1 o_6+c_8 g_1 o_7^2 \iota _2 o_6+c_9 g_2 o_1 o_7^2 \iota _2 o_6\\
&-c_9 g_1 o_2 o_7^2 \iota _2 o_6-3 c_8 g_1 \iota _2 o_6+c_8 g_4 \iota _2 o_6-3 c_9 g_2 o_1 \iota _2 o_6-c_9 g_3 o_1 \iota _2 o_6+3 c_9 g_1 o_2 \iota _2 o_6\\
&-c_9 g_4 o_2 \iota _2 o_6+c_{14} g_1 o_7 o_8 \iota _2 o_6+c_{15} g_2 o_1 o_7 o_8 \iota _2 o_6-c_{15} g_1 o_2 o_7 o_8 \iota _2 o_6-c_{14} g_1 o_7 o_9 \iota _2 o_6\\
&-c_{15} g_2 o_1 o_7 o_9 \iota _2 o_6+c_{15} g_1 o_2 o_7 o_9 \iota _2 o_6+2 c c_{14} g_2 o_7 v_1 \iota _2 o_6+c c_{14} g_3 o_7 v_1 \iota _2 o_6+2 c c_{15} g_1 o_1 o_7 v_1 \iota _2 o_6\\
&-c c_{15} g_4 o_1 o_7 v_1 \iota _2 o_6+2 c c_{15} g_2 o_2 o_7 v_1 \iota _2 o_6+c c_{15} g_3 o_2 o_7 v_1 \iota _2 o_6-2 c c_8 g_2 o_7^2 o_8 v_1 \iota _2 o_6-c c_8 g_3 o_7^2 o_8 v_1 \iota _2 o_6\\
&-2 c c_9 g_1 o_1 o_7^2 o_8 v_1 \iota _2 o_6+c c_9 g_4 o_1 o_7^2 o_8 v_1 \iota _2 o_6-2 c c_9 g_2 o_2 o_7^2 o_8 v_1 \iota _2 o_6-c c_9 g_3 o_2 o_7^2 o_8 v_1 \iota _2 o_6\\
&+2 c c_8 g_2 o_7^2 o_9 v_1 \iota _2 o_6+c c_8 g_3 o_7^2 o_9 v_1 \iota _2 o_6+2 c c_9 g_1 o_1 o_7^2 o_9 v_1 \iota _2 o_6-c c_9 g_4 o_1 o_7^2 o_9 v_1 \iota _2 o_6\\
&+2 c c_9 g_2 o_2 o_7^2 o_9 v_1 \iota _2 o_6+c c_9 g_3 o_2 o_7^2 o_9 v_1 \iota _2 o_6-c_{14} g_1 o_7 \gamma _1 \iota _2 o_6-c_{15} g_2 o_1 o_7 \gamma _1 \iota _2 o_6\\
&+c_{15} g_1 o_2 o_7 \gamma _1 \iota _2 o_6-c_8 g_1 o_7^2 o_8 \gamma _1 \iota _2 o_6-c_9 g_2 o_1 o_7^2 o_8 \gamma _1 \iota _2 o_6+c_9 g_1 o_2 o_7^2 o_8 \gamma _1 \iota _2 o_6\\
&+c_8 g_1 o_7^2 o_9 \gamma _1 \iota _2 o_6+c_9 g_2 o_1 o_7^2 o_9 \gamma _1 \iota _2 o_6-c_9 g_1 o_2 o_7^2 o_9 \gamma _1 \iota _2 o_6+c_8 g_4 o_7^2 \iota _3 o_6\\
&-c_9 g_3 o_1 o_7^2 \iota _3 o_6-c_9 g_4 o_2 o_7^2 \iota _3 o_6-2 c_8 g_4 \iota _3 o_6+2 c_{16} g_4 \iota _3 o_6+2 c_9 g_3 o_1 \iota _3 o_6-2 c_{17} g_3 o_1 \iota _3 o_6\\
&+2 c_9 g_4 o_2 \iota _3 o_6-2 c_{17} g_4 o_2 \iota _3 o_6-3 c c_{14} g_3 o_7 v_1 \iota _3 o_6+3 c c_{15} g_4 o_1 o_7 v_1 \iota _3 o_6-3 c c_{15} g_3 o_2 o_7 v_1 \iota _3 o_6\\
&+3 c c_8 g_3 o_7^2 o_8 v_1 \iota _3 o_6-3 c c_9 g_4 o_1 o_7^2 o_8 v_1 \iota _3 o_6+3 c c_9 g_3 o_2 o_7^2 o_8 v_1 \iota _3 o_6-3 c c_8 g_3 o_7^2 o_9 v_1 \iota _3 o_6\\
&+3 c c_9 g_4 o_1 o_7^2 o_9 v_1 \iota _3 o_6-3 c c_9 g_3 o_2 o_7^2 o_9 v_1 \iota _3 o_6-3 c_{14} g_4 o_7 \gamma _1 \iota _3 o_6+3 c_{15} g_3 o_1 o_7 \gamma _1 \iota _3 o_6
\end{aligned}
\end{equation}
\begin{equation}\notag
\begin{aligned}
\cdots&+3 c_{15} g_4 o_2 o_7 \gamma _1 \iota _3 o_6-2 c_8 g_2 o_7 o_8-2 c_9 g_1 o_1 o_7 o_8-2 c_9 g_2 o_2 o_7 o_8+2 c_8 g_2 o_7 o_9+2 c_9 g_1 o_1 o_7 o_9\\
&+2 c c_8 g_4 o_7 v_1-2 c c_9 g_3 o_1 o_7 v_1-2 c c_9 g_4 o_2 o_7 v_1+2 c_8 g_2 o_7 \gamma _1+2 c_9 g_1 o_1 o_7 \gamma _1+2 c_9 g_2 o_2 o_7 o_9\\
&-c_{14} g_3 o_7^2 \iota _3+c_{15} g_4 o_1 o_7^2 \iota _3-c_{15} g_3 o_2 o_7^2 \iota _3+2 c_{14} g_3 \iota _3-2 c_{18} g_3 \iota _3-2 c_{15} g_4 o_1 \iota _3+2 c_{19} g_4 o_1 \iota _3\\
&+2 c_{15} g_3 o_2 \iota _3-2 c_{19} g_3 o_2 \iota _3+c_8 g_3 o_7^3 o_8 \iota _3-c_9 g_4 o_1 o_7^3 o_8 \iota _3+c_9 g_3 o_2 o_7^3 o_8 \iota _3-c_8 g_3 o_7 o_8 \iota _3\\
&+2 c_{16} g_3 o_7 o_8 \iota _3+c_9 g_4 o_1 o_7 o_8 \iota _3-2 c_{17} g_4 o_1 o_7 o_8 \iota _3-c_9 g_3 o_2 o_7 o_8 \iota _3+2 c_{17} g_3 o_2 o_7 o_8 \iota _3\\
&-c_8 g_3 o_7^3 o_9 \iota _3+c_9 g_4 o_1 o_7^3 o_9 \iota _3-c_9 g_3 o_2 o_7^3 o_9 \iota _3+c_8 g_3 o_7 o_9 \iota _3-2 c_{16} g_3 o_7 o_9 \iota _3-c_9 g_4 o_1 o_7 o_9 \iota _3\\
&+2 c_{17} g_4 o_1 o_7 o_9 \iota _3+c_9 g_3 o_2 o_7 o_9 \iota _3-2 c_{17} g_3 o_2 o_7 o_9 \iota _3+3 c_{16} g_3 o_7 \gamma _1 \iota _3-3 c_{17} g_4 o_1 o_7 \gamma _1 \iota _3\\
&+3 c_{17} g_3 o_2 o_7 \gamma _1 \iota _3-3 c_{14} g_3 o_7^2 o_8 \gamma _1 \iota _3+3 c_{15} g_4 o_1 o_7^2 o_8 \gamma _1 \iota _3-3 c_{15} g_3 o_2 o_7^2 o_8 \gamma _1 \iota _3\\
&+3 c_{14} g_3 o_7^2 o_9 \gamma _1 \iota _3-3 c_{15} g_4 o_1 o_7^2 o_9 \gamma _1 \iota _3+3 c_{15} g_3 o_2 o_7^2 o_9 \gamma _1 \iota _3+2 c_9 g_2 o_2 o_7 \gamma _1,
\end{aligned}
\end{equation}
\begin{equation}
\begin{aligned}
y_2&=2 c_8 g_1 \iota _3 o_6^3+2 c_9 g_2 o_1 \iota _3 o_6^3-2 c_9 g_1 o_2 \iota _3 o_6^3+2 c_{14} g_2 \iota _3 o_6^2+2 c_{15} g_1 o_1 \iota _3 o_6^2+2 c_{15} g_2 o_2 \iota _3 o_6^2-2 c_8 g_2 o_7 o_8 \iota _3 o_6^2 \\
&-2 c_9 g_1 o_1 o_7 o_8 \iota _3 o_6^2-2 c_9 g_2 o_2 o_7 o_8 \iota _3 o_6^2+2 c_8 g_2 o_7 o_9 \iota _3 o_6^2+2 c_9 g_1 o_1 o_7 o_9 \iota _3 o_6^2+2 c_9 g_2 o_2 o_7 o_9 \iota _3 o_6^2-2 c_9 g_4 o_2 o_6\\
&-3 c c_9 g_2 o_1 o_7 v_1 \iota _3 o_6^2+3 c c_9 g_1 o_2 o_7 v_1 \iota _3 o_6^2-2 c_8 g_1 o_6+2 c_8 g_4 o_6-2 c_9 g_2 o_1 o_6-2 c_9 g_3 o_1 o_6+2 c_9 g_1 o_2 o_6\\
&-c_9 g_2 o_1 o_7^2 \iota _3 o_6+c_9 g_1 o_2 o_7^2 \iota _3 o_6+2 c_8 g_1 \iota _3 o_6-2 c_{16} g_1 \iota _3 o_6+2 c_9 g_2 o_1 \iota _3 o_6-2 c_{17} g_2 o_1 \iota _3 o_6-2 c_9 g_1 o_2 \iota _3 o_6\\
&-3 c c_{14} g_2 o_7 v_1 \iota _3 o_6-3 c c_{15} g_1 o_1 o_7 v_1 \iota _3 o_6-3 c c_{15} g_2 o_2 o_7 v_1 \iota _3 o_6+3 c c_8 g_2 o_7^2 o_8 v_1 \iota _3 o_6+3 c c_9 g_1 o_1 o_7^2 o_8 v_1 \iota _3 o_6 \\
&+3 c c_9 g_2 o_2 o_7^2 o_8 v_1 \iota _3 o_6-3 c c_8 g_2 o_7^2 o_9 v_1 \iota _3 o_6-3 c c_9 g_1 o_1 o_7^2 o_9 v_1 \iota _3 o_6-3 c c_9 g_2 o_2 o_7^2 o_9 v_1 \iota _3 o_6+3 c_{14} g_1 o_7 \gamma _1 \iota _3 o_6\\
&+3 c_{15} g_2 o_1 o_7 \gamma _1 \iota _3 o_6-3 c_{15} g_1 o_2 o_7 \gamma _1 \iota _3 o_6+c_8 g_2 o_7 o_8+c_9 g_1 o_1 o_7 o_8+c_9 g_2 o_2 o_7 o_8-c_8 g_2 o_7 o_9-c_9 g_1 o_1 o_7 o_9\\
&-c c_8 g_4 o_7 v_1+c c_9 g_3 o_1 o_7 v_1+c c_9 g_4 o_2 o_7 v_1-c_8 g_2 o_7 \gamma _1-c_9 g_1 o_1 o_7 \gamma _1-c_9 g_2 o_2 o_7 \gamma _1-c_{14} g_2 o_7^2 \iota _3-c_{15} g_1 o_1 o_7^2 \iota _3\\
&+2 c_{14} g_2 \iota _3-2 c_{18} g_2 \iota _3+2 c_{15} g_1 o_1 \iota _3-2 c_{19} g_1 o_1 \iota _3+2 c_{15} g_2 o_2 \iota _3-2 c_{19} g_2 o_2 \iota _3+c_8 g_2 o_7^3 o_8 \iota _3+c_9 g_1 o_1 o_7^3 o_8 \iota _3\\
&+c_9 g_2 o_2 o_7^3 o_8 \iota _3-c_8 g_2 o_7 o_8 \iota _3+2 c_{16} g_2 o_7 o_8 \iota _3-c_9 g_1 o_1 o_7 o_8 \iota _3+2 c_{17} g_1 o_1 o_7 o_8 \iota _3-c_9 g_2 o_2 o_7 o_8 \iota _3\\
&-c_8 g_2 o_7^3 o_9 \iota _3-c_9 g_1 o_1 o_7^3 o_9 \iota _3-c_9 g_2 o_2 o_7^3 o_9 \iota _3+c_8 g_2 o_7 o_9 \iota _3-2 c_{16} g_2 o_7 o_9 \iota _3+c_9 g_1 o_1 o_7 o_9 \iota _3-2 c_{17} g_1 o_1 o_7 o_9 \iota _3\\
&+c_9 g_2 o_2 o_7 o_9 \iota _3-2 c_{17} g_2 o_2 o_7 o_9 \iota _3+3 c_{16} g_2 o_7 \gamma _1 \iota _3+3 c_{17} g_1 o_1 o_7 \gamma _1 \iota _3+3 c_{17} g_2 o_2 o_7 \gamma _1 \iota _3-3 c_{14} g_2 o_7^2 o_8 \gamma _1 \iota _3\\
&-3 c_{15} g_1 o_1 o_7^2 o_8 \gamma _1 \iota _3-3 c_{15} g_2 o_2 o_7^2 o_8 \gamma _1 \iota _3+3 c_{14} g_2 o_7^2 o_9 \gamma _1 \iota _3+3 c_{15} g_1 o_1 o_7^2 o_9 \gamma _1 \iota _3+3 c_{15} g_2 o_2 o_7^2 o_9 \gamma _1 \iota _3,\\
&-3 c c_8 g_1 o_7 v_1 \iota _3 o_6^2-c_8 g_1 o_7^2 \iota _3 o_6+2 c_{17} g_1 o_2 \iota _3 o_6-c_9 g_2 o_2 o_7 o_9-c_{15} g_2 o_2 o_7^2 \iota _3+2 c_{17} g_2 o_2 o_7 o_8 \iota _3.
\end{aligned}
\end{equation}

In above expressions,  $v_1=\beta_1\gamma_1$ with $\beta_1=\frac{P_1}{M_1}$ and $\gamma_1=\frac{E_1}{M_1}$;  the expressions of $o_i$s are
\begin{equation}
\begin{aligned}
o_1 & = \frac{k}{w_2},&o_2 &=\frac{m_2}{w_2},&o_3 &= \frac{M_1}{P_1},&o_4 &= \frac{1}{\beta_1} = \frac{E_1}{P_1},\\
o_6 &=\frac{k}{m_3},& o_7 & = \frac{M_1}{m_3},&o_8 &=\frac{M}{M_1},&o_9 &= \frac{M_2}{M_1};
\end{aligned}
\end{equation}
 and we define
\begin{align}
C_1 = c=\cos\theta,~~C_{21}= \frac{1}{2} (3\cos^2\theta -1),~~C_{22}= \frac{1}{2} (\cos^2\theta-1),
\end{align}
where $\theta$ denotes the angle between $\bm k$ and $\bm{P}_1$. We also define $c_{8-19}$ to denote the following relevant variables 
\begin{equation}
\begin{aligned}
c_{8}&=(c_1 + c_2+c_3+c_4+c_5+c_6),\\
c_{9}&=(c_1 - c_2 + c_3-c_4+c_5-c_6),\\
c_{10} &= kc_8,\\
c_{11} &= kc_9,\\
c_{12} & =  (c_1k_{11} + c_2k_{12}+c_3k_{13}+c_4k_{14}+c_5k_{15}+c_6k_{16}),\\
c_{13} & =  (c_1k_{11} - c_2k_{12}+c_3k_{13}-c_4k_{14}+c_5k_{15}-c_6k_{16}),\\
c_{14}& = c_{12}/m_3,\\
c_{15} &= c_{13}/m_3,\\
c_{16} & =  (c_1k_{11}^2 + c_2k_{12}^2+c_3k_{13}^2+c_4k_{14}^2+c_5k_{15}^2+c_6k_{16}^2)/m_3^2,\\
c_{17} & =  (c_1k_{11}^2 - c_2k_{12}^2+c_3k_{13}^2-c_4k_{14}^2+c_5k_{15}^2-c_6k_{16}^2)/m_3^2,\\
c_{18} & =  (c_1k_{11}^3 + c_2k_{12}^3+c_3k_{13}^3+c_4k_{14}^3+c_5k_{15}^3+c_6k_{16}^3)/m_3^3,\\
c_{19} & =  (c_1k_{11}^3 - c_2k_{12}^3+c_3k_{13}^3-c_4k_{14}^3+c_5k_{15}^3-c_6k_{16}^3)/m_3^3,
\end{aligned}
\end{equation}
where the expressions of $c_i(i=1,\cdots,6)$ has already been listed in \eref{E-cn}.  

\section{Squared amplitude} \label{A-3}

For the decay mode of $P_{\psi1/2}^N\to (V+B)$, we obtain 
\begin{gather}
\sum |\mathcal{A}|^2 =  \left( -g^{\alpha\beta}+ {\hat P_1^\alpha \hat P_1^\beta} \right)\up{Tr}  \left(\slashed P_{\!2}+M_2\right)  T_{m\alpha} \left( \slashed P+M \right) \bar{T}_{m\beta},
\end{gather} 
where the summation is over all the spin states of the involved initial and final particles, namely, $r_1,r_2$ and $r$;  $T_{m\alpha}$  is expressed as
\begin{gather}
T_{m\alpha} = \left(   s_{m1} \gamma_\alpha + s_{m2}\hat P_\alpha \right)
\end{gather}
with with $m=1$ and 2 representing the form factor for $J/\psi p$ and $\bar D^{*0}\Lambda^+_c$ channel, respectively;
and $\bar{T}_{\beta} = \gamma^0 T_{\beta}^\dagger \gamma_0$ is defined as the usual conjugation variable;  and we also used the relationship of the summation over the vector polarization states $r_1$,
\begin{gather}
 \sum_{r_1} e_{(r_1)}^\alpha e_{(r_1)}^{*\beta} = -g^{\alpha\beta}+{\hat P_1^\alpha \hat P_1^\beta};
\end{gather} 
and the summation over the polarization states of the spinors
\begin{gather}
 \sum_{r_2} u(P_2,r_2) \bar u(P_2,r_2) =  (\slash P_2 + M_2),\\
  \sum_r u(P,r) \bar u(P,r) =  (\slashed P + M).
\end{gather}
For the mode of $P_{\psi1/2}^N\!\to\! (P+B)$, the squared amplitude behaves
\begin{gather}
\sum |\mathcal{A}|^2 = \up{Tr}  \left(\slashed P_{\!2}+M_2\right)  T_m \left( \slashed P+M \right) \bar T_m = 4M(E_2-M_2)s^2_{m},
\end{gather} 
where $T_m= s_{m} \gamma_5$ with $m=3,4,5$ representing the form factor for $\eta_c p$, $\bar D^0\Lambda^+_c$ and $\bar D\Sigma_c$ channels, respectively.
The squared amplitude is proportional to the kinetic energy of the final baryon. 
For the decay mode of $P_{\psi1/2}^N\to (P+B^*)$, the squared amplitude behaves
\begin{gather}
\sum |\mathcal{A}|^2 = \up{Tr} \, u^{\alpha_1}_2 \bar u^{\alpha}_2 T_{6\alpha}  (\slash P+M) \bar{T}_{6\alpha_1}
\end{gather} 
with $T_{6\alpha} = s_6 \hat P_\alpha$ denoting the form factor for $\bar D\Sigma_c^*$ channel, where we need the following relationship for Rarita-Schwinger spinor
\begin{equation}
u_2^\alpha\bar u_2^{\alpha_1}=(\sl P_2+M_2)\left[ -g^{\alpha{\alpha_1}}+\frac{1}{3}\gamma^\alpha\gamma^{\alpha_1} - \frac{P^\alpha_2\gamma^{\alpha_1}-P^{\alpha_1}_2\gamma^\alpha}{3M_2}+\frac{2P^\alpha_2 P^{\alpha_1}_2}{3M_2^2} \right].
\end{equation}

For the decay mode of $P_{\psi3/2}^N\!\to\! (V+B)$, we obtain
\begin{gather}
\sum |\mathcal{A}|^2 =  \left( -g^{\alpha\alpha_1}+\hat P_1^\alpha  \hat P_1^{\alpha_1}  \right)\up{Tr}  \left(\slashed P_{\!2}+M_2\right)  T_{i\alpha\beta}  u^{\beta} (P,r)\bar u^{\beta_1}(P,r) \bar{T}_{i\alpha_1\beta_1},
\end{gather} 
where 
\begin{gather}
T_{i\alpha\beta} = \i t_{i1} \epsilon_{\alpha\beta \hat P \hat P_1} +  \left(t_{i2}g_{\alpha\beta}  + t_{i3} \hat P_{1\beta} \gamma_\alpha + t_{i4} \hat P_{\alpha}\hat P_{1\beta} \right)\gamma_5 
\end{gather}
with $i=1$ and $2$ denotes the form factors for the $J/\psi p$ and   $\bar D^{*0}\Lambda_c^+$ channel, respectively; and we also need the relationship \eref{E-RS} for Rarita-Schwinger spinor $u^\alpha$.
For the decay mode $P_{\psi3/2}^N\to(P+B)$, we obtain
\begin{gather}
\sum |\mathcal{A}|^2 = \up{Tr}  \left(\slashed P_{\!2}+M_2\right)  T_{m\alpha}  u^{\alpha} (P,r)\bar u^{\beta}(P,r) \bar{T}_{m\beta},
\end{gather} 
where $T_{m\alpha} = s_m\hat P_{1\alpha}$ with $m=3,4,5$ denotes the form factor for $\eta_cp$, $\bar D^0\Lambda^+_c$ and $\bar D\Sigma_c$ channel, respectively. For the $P_{\psi3/2}^N\to (P+B^*)$ decay mode, the squared amplitude behaves
\begin{gather}
\sum |\mathcal{A}|^2 = \up{Tr} \, u^{\alpha_1}(P_2,r_2) \bar u^{\alpha}(P_2,r_2) T_{6\alpha\beta}  u^{\beta} (P,r)\bar u^{\beta_1}(P,r) \bar{T}_{7\alpha_1\beta_1},
\end{gather} 
where  the form factor for $\bar D\Sigma_c^*$ decay channel behaves
\begin{gather}
T_{7\alpha\beta} =\i t_{61}  \epsilon^{\alpha\beta \hat P\hat P_1}  +  \left( t_{62} g^{\alpha\beta} +t_{63}\hat P^\alpha \hat P_1^\beta  \right) \gamma_5.
\end{gather}

\acknowledgments
The authors thank Prof.\,Feng-Kun Guo, Xu-Chang Zheng, and Hao Xu for  helpful suggestions and discussions. This work is supported by the National Key R\&D Program of China\,(2022YFA1604803), the Natural Science Basic Research Program of Shaanxi\,(No.\,2025JC-YBMS-020), and  is also supported by the National Natural Science Foundation of China\,(NSFC) under Grant Nos.\,12047503, 12005169, 12075301, 11821505, 12047503, and 12075073.  X.Z. Tan also acknowledges the support from Helmholtz-OCPC Postdoctoral Fellowship Program. 


%

\end{document}